\documentclass[11pt, authoryear]{elsarticle}

\usepackage{amssymb}
\usepackage{amsthm}
\usepackage{amsmath}
\usepackage{xcolor}
\usepackage{multirow}
\usepackage{subcaption}
\usepackage{tabularx}
\usepackage{float}
\usepackage{caption}
\usepackage{hyperref}
\usepackage{lscape}
\usepackage{adjustbox}
\usepackage{upgreek}
\usepackage{rotating}

\usepackage{tablefootnote}
\usepackage{footnote}

\newcolumntype{Y}{>{\centering\arraybackslash}X}

\usepackage{booktabs}
\usepackage[font=small,labelfont=bf,tableposition=top]{caption}

\usepackage{color}
\usepackage{fullpage}
\usepackage{makecell}
\usepackage{xfrac}
\usepackage{manyfoot}
\usepackage{comment}
\usepackage{tabularx}
\usepackage{dcolumn}
\usepackage{tablefootnote}
\usepackage{xltabular}

\usepackage{multirow}
\usepackage{booktabs}
\usepackage{ragged2e}

\usepackage{tikz}
\usepackage{tikz-cd}
\usepackage{tikz-layers}
\usetikzlibrary{arrows}
\usetikzlibrary{automata}
\usetikzlibrary{matrix,fit}
\usetikzlibrary{calc}
\usetikzlibrary{decorations.text}
\usetikzlibrary{decorations.pathmorphing}
\usetikzlibrary{decorations.markings}
\usetikzlibrary{arrows.meta}
\usetikzlibrary{positioning}

\usepackage[symbol*]{footmisc}

\newcommand{\tablefontsize}{\normalsize}

\newcommand{\eM}{\mathcal{M}}
\newcommand{\dee}[1]{\mathrm{d}{#1}}
\newcommand{\lgrl}[1]{
    \lambda_{\rm GRL}(#1)}

\usepackage{array}\newcolumntype{C}[1]{>{\centering\arraybackslash}p{#1}}

\newcommand{\STAB}[1]{\begin{tabular}{@{}c@{}}#1\end{tabular}}

\NewDocumentCommand{\rot}{O{90} O{1em} m}{\makebox[#2][l]{\rotatebox{#1}{#3}}}%

\newcommand{\x}{x}

\newcolumntype{b}{X}
\newcolumntype{s}{>{\hsize=.055\hsize}X}
\newcolumntype{m}{>{\hsize=.3\hsize}X}
\newcolumntype{k}{>{\hsize=.5\hsize}X}
\newcolumntype{v}{>{\hsize=0.75\hsize}X}

\usepackage[left=2cm, right=2cm, top=2cm, bottom=2cm, footskip=0.5cm]{geometry}

\newcommand{\abm}{\textsc{abm}}

\newcommand{\npe}{\textsc{npe}}
\newcommand{\nre}{\textsc{nre}}

\makeatletter
\newcommand*{\myfnsymbolsingle}[1]{%
  \ensuremath{%
    \ifcase#1
    \or 
      \star %
    \or 
      \dagger
    \or 
      \ddagger
    \or 
      \mathsection
    \or 
      \mathparagraph
    \else 
      \@ctrerr  
    \fi
  }%
}   
\makeatother

\newcommand*{\myfnsymbol}[1]{%
  \myfnsymbolsingle{\value{#1}}%
}

\journal{ }

\begin{document}

\begin{frontmatter}

\title{{\bf Forecasting Macroeconomic Dynamics using a Calibrated Data-Driven Agent-based Model$^{\rm \star}$}
\vskip -0.17in}
\date{\today}

\author[]{\textsc{Samuel Wiese}$^{\rm a,b,\dagger}$}
\author[]{\textsc{Jagoda Kaszowska-Mojsa}$^{\rm a,c,d}$}
\author[]{\textsc{Joel Dyer}$^{\rm a,b}$}
\author[]{\textsc{Jos\'e Moran}$^{\rm e, a}$}
\author[]{\textsc{Marco Pangallo}$^{\rm a,f}$}
\author[]{\textsc{Fran\c{c}ois Lafond}$^{\rm a,g}$}
\author[]{\textsc{John Muellbauer}$^{\rm a,h}$}
\author[]{\textsc{Anisoara Calinescu}$^{\rm a,b}$}
\author[]{\textsc{J. Doyne Farmer}$^{\rm a,e,g,i}$\vspace{-0.05cm}}

\address[1]{Institute for New Economic Thinking, University of Oxford}
\address[2]{Department of Computer Science, University of Oxford}
\address[3]{Institute of Economics, Polish Academy of Sciences}
\address[4]{Mathematical Institute, University of Oxford}
\address[5]{Macrocosm Inc.}
\address[6]{CENTAI Institute}
\address[7]{Smith School of Enterprise and the Environment, University of Oxford}
\address[8]{Department of Economics, University of Oxford}
\address[9]{Santa Fe Institute\vspace{-0.9cm}}


\begin{abstract}
In the last few years, economic agent-based models have made the transition from qualitative models calibrated to match stylised facts to quantitative models for time series forecasting, and in some cases, their predictions have performed as well or better than those of standard models (see, e.g. \cite{poledna2023economic,hommesCANVAS2022,pichler2022forecasting}). Here, we build on the model of Poledna et al., adding several new features such as housing markets, realistic synthetic populations of individuals with income, wealth and consumption heterogeneity, enhanced behavioural rules and market mechanisms, and an enhanced credit market. We calibrate our model for all 38 OECD member countries using state-of-the-art approximate Bayesian inference methods and test it by making out-of-sample forecasts. It outperforms both the Poledna and AR(1) time series models by a highly statistically significant margin. Our model is built within a platform we have developed, making it easy to build, run, and evaluate alternative models, which we hope will encourage future work in this area.  
\end{abstract}

\end{frontmatter}

\renewcommand{\thefootnote}{\myfnsymbol{footnote}}
\footnotetext[1]{Acknowledgements: We are grateful to Sebastian Poledna for many helpful suggestions during model development and for his support in replicating the IIASA model. We would also like to thank the participants of the 1$^\text{st}$ IIASA-MacroABM Workshop for their comments, which enriched our work. The funders of this work include Baillie Gifford, UKRI AI World Leading Researcher Fellowship (grant EP/W002949/1), ESRC grant PRINZ (ES/W010356/1), Marie Skłodowska-Curie grant No 101023445 and the James S. McDonnell Foundation.}

\footnotetext[2]{Corresponding author: samuel.wiese@cs.ox.ac.uk}
\renewcommand*{\thefootnote}{\arabic{footnote}}
\setcounter{footnote}{0}

\section{Introduction}
Agent-Based Models (\abm s) have recently emerged as a credible alternative to mainstream macroeconomic models, offering a fundamentally different approach to understanding economic dynamics \citep{doynesOverview}.  Although macroeconomic \abm s have been around for twenty years \citep{dawidAgentbasedMacroeconomics2018}, they have so far mostly been successful at generating qualitative insights, for instance, about policy \citep{fagioloMacroeconomicPolicyDSGE2016}. Typical macroeconomic \abm s are validated by showing that they can reproduce stylised facts or summary statistics of real-world data. To achieve this validation target, researchers calibrate the most important parameters of the model while initialising the state variables of individual agents at random. This is because standard macroeconomic \abm s do not aim at reproducing a specific economy at a specific time but rather try to reproduce a plausible generic economy replicating key stylised facts.

In the last few years, \abm s have become more data-driven, striving to faithfully represent a specific economy at a specific point in time \citep{pangallo2024datadriven}. To do so, it is not enough to calibrate a few parameters; it is also necessary to initialise the state variables of individual agents in a way that is compatible with real-world data. The main advantage of this approach is that data-driven \abm s can track and forecast empirical time series, making it possible to explicitly show that they are a faithful representation of the economy and enhancing trust in counterfactuals and policy recommendations. Most of the applications of data-driven \abm s have been in specific sectors of the economy, such as housing \citep{geanakoplosGettingSystemicRisk2012} and labour markets \citep{del2021occupational}, and on the economic impact of major disasters \citep{hallegatte2008adaptive,pichler2022forecasting}. However, data-driven macroeconomic \abm s also exist \citep{papadopoulos2019income,kaszowska2020macroprudential}.

A breakthrough in using data-driven \abm s for macroeconomics was achieved by \cite{polednaEconomicForecastingAgentbased2020}. This model, which we call the \textit{IIASA model}\footnote{The model of Poledna et al. was developed at the International Institute for Applied Systems Analysis (IIASA).} and use as our benchmark, was the first to attempt to compare the out-of-sample forecasting performance of a macroeconomic \abm~ to that of statistical and DSGE models, demonstrating that \abm s could match and potentially surpass the performance of traditional models in analyzing and predicting economic dynamics. The IIASA model quantitatively reproduced the economy of Austria, initialising firm and household variables so that they were consistent with national accounts and macroeconomic statistics in a given year. All of the parameters in the model are pinned down directly in the data or estimated from time series, making any parameter calibration unnecessary. However, this suggests a more flexible model specification and calibration approach could improve its forecasting power.

In this paper, we introduce a novel macroeconomic \abm, and show that calibrating its free parameters leads to statistically significant improvements over simple statistical models and the IIASA model. We rely on state-of-the-art Bayesian calibration techniques, namely neural posterior estimation and neural density ratio estimation, which, along with other neural network methods for approximate Bayesian inference for agent-based models \citep{dyer2023gradient}, have been seen to be far faster and more accurate than alternative methods employed in the agent-based modelling literature \citep{dyer2024black}.

Although inspired by the IIASA model, our model departs with several key specifications designed to improve its realism and scope. Our main innovations include:
\begin{itemize}
   \item \textbf{Housing Market}: Households buy, sell, let and rent properties. Because houses are a substantial fraction of household wealth, slowdowns in the housing market can have ripple effects throughout the economy.
   \item \textbf{Credit Market}: Banks make loans to firms and households, considering realistic supply-side loan requirements and setting interest rates using a statistical model used by private banks. Our more detailed modelling of the credit market makes our model a better fit for studying macroprudential policy.
   \item \textbf{Realistic Synthetic Populations}: Individuals are grouped into households and differ by several socio-economic characteristics, including age, income, wealth and occupation status. Our model considers sophisticated synthetic populations that accurately match the demographic and economic characteristics of the simulated countries. This allows us to better capture the economic conditions and the different behaviours of the agents within the model.
   \item \textbf{Enhanced Behavioural Rules and Market Mechanisms}: We have incorporated more realistic behavioural rules and mechanisms that are not present in the IIASA model. For instance, in the IIASA model, firms set production targets that are limited by past resource constraints, which unrealistically constrains aggregate growth. Here, we let firms set production targets that only partly consider past resource constraints, and that involve more forward-thinking, enabling firms to set more ambitious targets. 
   %
\end{itemize}

Out of the seven parameters we found to have the most impact on calibration, five influence firms' production targets, and two influence firms' prices. Our key result is that these improvements lead to better performance in forecasting economic aggregates. We compare our model's three-year predictions of GDP, inflation, household consumption, government consumption, and investment against the IIASA model and a benchmark AR(1). We perform these forecasts across different initialisation times and for 38 countries to show statistical significance. Our calibrated model outperforms the benchmark models overall, providing a lower forecasting error for all considered economic aggregates for 33 out of 38 countries. Additionally, we estimate a Bayes factor for each time of initialisation and country. Of all estimated Bayes factors, 65\% support our model over the IIASA model.

This paper is structured as follows: Section \ref{sec:literature} briefly summarises the literature on agent-based models in macroeconomics, Section \ref{sec:novel_abm} gives a high-level overview of our model, Section \ref{sec:calibration} discusses the methods for calibration, Section \ref{sec:forecasting} compares the forecasting performance of our model to popular benchmark models, and Section \ref{sec:conclusion} concludes, highlighting future work.

\section{Literature}
\label{sec:literature}
We now place our model in the context of the agent-based macroeconomics literature. We mostly follow \cite{dawidAgentbasedMacroeconomics2018} in choosing the models we compare to and the dimensions we consider. Specifically, we consider the types of agents (banks, firms, households, etc.) and the markets in which they interact (goods, labour, credit, and housing). We add another particularly relevant dimension here, namely how models relate to data \citep{pangallo2024datadriven}. We first consider which parts of the \abm~ are taken from data, including the calibration of parameters and the initialisation of individual agents' state variables. We also consider what property of the data is targeted, that is, if the models only aim at reproducing summary statistics or stylised facts, or if they attempt to forecast empirical time series. Our comparison is summarized in Table \ref{table:table_simple_model_comp}.

We consider ten models or families of models. First, we consider the so-called CATS model, whose baseline version is described in the book by \cite{delli2011macroeconomics}. We also consider the extension of the CATS model that includes capital goods \citep{assenzaEmergentDynamicsMacroeconomic2015}, as it has been calibrated \citep{gatti2020rising,glielmo2023reinforcement} and has been used as the building block of \cite{poledna2023economic}. Conversely, we also include a simplified version of the CATS model, named Mark-0 \citep{tipping_points}, which was the object of rich theoretical studies that showed that contrary to the common criticism of \abm s, a good mathematical understanding of their behaviour was possible~\citep{Gualdi2015}, or that advanced numerical techniques enabled an efficient exploration of their parameter space~\citep{karl_expl}. Next, we compare our model to the two offsprings of the EURACE project, respectively developed at the University of Bielefeld (UNIBI) and first described in \cite{deissenbergEURACEMassivelyParallel2008} and at the University of Genoa (UNIGE) and introduced in \cite{cincotti2012eurace}. Furthermore, we compare our model to the family of Keynes meets Schumpeter (K+S) \abm s developed at Sant'Anna Pisa (see for instance \citealt{dosiSchumpeterMeetingKeynes2010,dosiFiscalMonetaryPolicies2015}), to the framework developed by \cite{ashrafHowInflationAffects2016,ashrafBanksMarketOrganization2017}, and to the baseline macroeconomic \abm~ by \cite{lengnickAgentbasedMacroeconomicsBaseline2013}. Finally, we consider our main baseline, namely the IIASA model \citep{polednaEconomicForecastingAgentbased2020} (for a more detailed comparison, see Section \ref{sec:diffIIASA}), and its extension by the Bank of Canada, CANVAS \citep{hommesCANVAS2022}.

In terms of types of agents, all models feature one or more banks, firms and households, at least in some versions.\footnote{For instance, \cite{dosiSchumpeterMeetingKeynes2010} do not include banks, but \cite{dosiFiscalMonetaryPolicies2015} do.} Out of the chosen models, only the EURACE models \cite{deissenbergEURACEMassivelyParallel2008,cincotti2012eurace} and our model explicitly distinguish between households and individuals. Every model features markets for goods, labour and credit; again, only the EURACE models and ours feature a housing market.

In terms of relation to real-world data, most models aim to replicate stylised facts by calibrating parameters. Most papers calibrate parameters heuristically by setting reasonable values or taking them from other studies. A formal parameter calibration procedure is only followed for the CC-MABM \citep{gatti2020rising,glielmo2023reinforcement}, for the K+S model \citep{martinoli2022calibration}, in \cite{ashrafBanksMarketOrganization2017}, and in our model. Even without a formal calibration procedure, most models succeed at replicating stylised facts such as cross-correlations between macroeconomic variables or distributions of firm size.\footnote{The IIASA model and our model do not aim at replicating stylised facts, but we conjecture that they would easily replicate most of the other models' stylised facts if they were tested.} As discussed in the introduction, only the IIASA model and our model initialise variables of individual agents so that they are compatible with real-world data, and this enables them to compete at out-of-sample forecasting.\footnote{See also \cite{gatti2020rising} for an approach to forecasting with \abm s that relies on a sort of Bayesian surrogate model.}

As highlighted in Table~\ref{table:table_simple_model_comp}, our model distinguishes itself by its comprehensive scope. It covers more economic sectors than other models, making it more complete, and introduces robust Bayesian calibration for empirical validation, improving on the benchmarks by~\cite{polednaEconomicForecastingAgentbased2020} and~\cite{hommesCANVAS2022}. Furthermore, we are applying this model to a substantial number of countries -- 38 -- increasing the statistical validity of our approach. These innovations position our model as a significant advancement in the field of macroeconomic ABMs. 

\begin{table}[H]
\center
\renewcommand{\arraystretch}{1.1}
\begin{minipage}{\textwidth}
\setlength{\tabcolsep}{3.5pt}
\begin{tabularx}{\textwidth}{X|ccccccc|cccc|cccc}
\toprule
\multicolumn{1}{l}{\textbf{Model}} & \multicolumn{7}{c}{\textbf{Agents}} & \multicolumn{4}{c}{\textbf{Markets}} & \multicolumn{4}{c}{\textbf{Rel. to Data}}\\
 & \rot{Banks} & \rot{Central Bank} & \rot{Firms} & \rot{Central Gov.} & \rot{Gov. Entities} & \rot{Households} & \rot{Individuals} & \rot{Goods} & \rot{Labour} & \rot{Credit} & \rot{Housing} & \rot{Calibration} & \rot{Initialization} & \rot{Stylised facts} & \rot{Forecasting} \\
\cmidrule(l){2-16}
\emph{CATS} \citep{delli2011macroeconomics} & \x & & \x & & & \x & & \x & \x & \x & & & & \x & \\
\emph{CC-MABM} \citep{assenzaEmergentDynamicsMacroeconomic2015} & \x & \x & \x & & & \x & & \x & \x & \x & & \x & & \x & \x \\
\emph{Mark-0} \citep{tipping_points}  & \x & & \x & & & \x & & \x & \x & \x & & & & & \\
\emph{EURACE UNIBI} \citep{deissenbergEURACEMassivelyParallel2008} & \x & \x & \x & \x & \x & \x & \x & \x & \x & \x & \x & & & \x & \\
\emph{EURACE UNIGE} \citep{cincotti2012eurace}  & \x & \x & \x & \x & \x & \x & \x & \x & \x & \x & \x & & & \x & \\
\emph{K+S} \citep{dosiSchumpeterMeetingKeynes2010,dosiFiscalMonetaryPolicies2015}  & \x & \x & \x & \x & & \x & & \x & \x & \x & & \x & & \x & \\
\cite{ashrafHowInflationAffects2016,ashrafBanksMarketOrganization2017}  & \x & \x & \x & \x & \x & \x & & \x & \x & \x & & \x & & \x & \\
\cite{lengnickAgentbasedMacroeconomicsBaseline2013}  & \x & & \x & & & \x & & \x & \x & \x & & & & \x & \\
CANVAS Model \citep{hommesCANVAS2022}  & \x & \x & \x & \x & \x & \x & & \x & \x & \x & &  & \x & & \x \\
IIASA Model \citep{polednaEconomicForecastingAgentbased2020}  & \x & \x & \x & \x & \x & \x & & \x & \x & \x & &  & \x & & \x \\
\textbf{Our Model}  & \x & \x & \x & \x & \x & \x & \x & \x & \x & \x & \x & \x & \x &  & \x \\
\bottomrule
\end{tabularx}
\end{minipage}
\vspace{0.05cm}
\caption{Comparison of macroeconomic agent-based models.}
\label{table:table_simple_model_comp}
\end{table}

\section{A Novel Large-Scale Economic Agent-based Model}
\label{sec:novel_abm}
This section provides an overview of our agent-based model, which, while inspired by the foundational work of~\cite{polednaEconomicForecastingAgentbased2020} that we set as our benchmark, diverges significantly in terms of implementation, scope and mechanisms. We describe the different types of agents and markets it incorporates, the sequence of events within an iteration of the simulation, and outline the key distinctions with the benchmark by~\cite{polednaEconomicForecastingAgentbased2020} and its extension by~\cite{hommesCANVAS2022}.

\subsection{Model Overview}
Our model is designed with a modular structure that enables the simulation of multiple countries simultaneously, providing a closed model where the entire world economy is portrayed. The modeller has the choice of which countries to model explicitly, while all of the others are aggregated into a collective entity we call the ``Rest of the World'' (RoW). For example, one could simulate Germany and the United States. The simulation would then be organised into three main blocks: one for Germany, one for the United States, and one for the RoW. In this work, however, we restrict ourselves to a simple configuration where we only run one country at a time, paired with the RoW. This systematic approach will then be applied to 38 countries.

\begin{figure}[H]
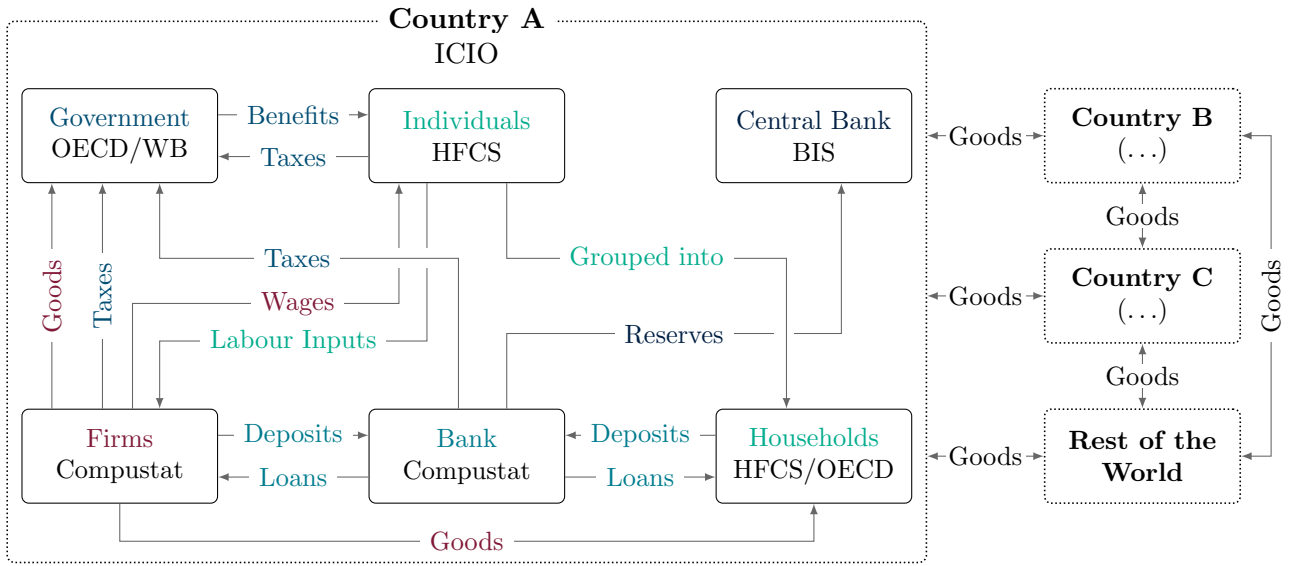

\begin{adjustbox}{max totalsize={1.0\textwidth}{.7\textheight},center}
\tikz[draw=black!60]{
\tikzstyle{block}=[draw=black,rounded corners=.9mm,inner sep=0pt,minimum width=2.6cm,minimum height=1.25cm,align=center,font=\small]
\tikzstyle{label}=[fill=white,align=center,font=\small]

\definecolor{oxfordb}{HTML}{801937}
\definecolor{gov_green}{HTML}{004D74}
\definecolor{bank_blue}{HTML}{007C8F}
\definecolor{ind_green}{HTML}{00AC8D}
\definecolor{firm_red}{HTML}{801937}
\definecolor{cb_blue}{HTML}{002044}

\node[block] (firms) {\color{oxfordb}Firms\\Compustat};
\node[block,right=2cm of firms] (bank) {\color{bank_blue}Bank\\Compustat};
\node[block,right=2cm of bank] (households) {\color{ind_green}Households\\HFCS/OECD};

\node[block,above=3cm of firms] (govt) {\color{gov_green}Government\\OECD/WB};
\node[block,right=2cm of govt] (individuals) {\color{ind_green}Individuals\\HFCS};
\node[block,right=2cm of individuals] (centbank) {\color{cb_blue}Central Bank\\BIS};

\draw[-latex] (firms.12)--node[label]{\color{bank_blue}Deposits} (bank.168);
\draw[latex-] (firms.-12)--node[label]{\color{bank_blue}Loans} (bank.-168);
\draw[latex-] (bank.12)--node[label]{\color{bank_blue}Deposits} (households.168);
\draw[-latex] (bank.-12)--node[label]{\color{bank_blue}Loans} (households.-168);
\draw[-latex] (firms.-90) --++(0,-.5) -| node[pos=.25,label]{\color{oxfordb}Goods} (households.-90);
\draw[-latex] (firms.75) --++(0,1.4) -| node[label,pos=.69]{} node[pos=.31,label]{\color{firm_red}Wages} (individuals.-145);
\draw[latex-] (firms.50) --++(0,.9) -| node[label,pos=.763]{} node[pos=.25,label]{\color{ind_green}Labour Inputs} (individuals.-130);
\draw[-latex] (firms.110)--node[label,sloped]{\color{gov_green}Taxes} (govt.-110);
\draw[-latex] (firms.145)--node[label,sloped]{\color{oxfordb} Goods} (govt.-145);
\draw[-latex] (govt.12)--node[label]{\color{gov_green}Benefits} (individuals.168);
\draw[latex-] (govt.-12)--node[label]{\color{gov_green}Taxes} (individuals.-168);
\draw[-latex] (bank.50) --++(0,1) -| node[label,pos=.415]{} node[pos=.25,label]{\color{cb_blue}Reserves} (centbank.-60);
\draw[-latex] (individuals.-50) --++(0,-1) -| node[pos=.25,label]{\color{ind_green}Grouped into} (households.120);

\draw[-latex] (bank.100) --++(0,2) -| node[pos=.27,label]{\color{gov_green}Taxes} (govt.-50);

\node[rounded corners=1mm,draw=black,anchor=north west,densely dotted,semithick,minimum width=12.22cm,minimum height=7.2cm] (r) at ($(govt.north west)+(-2mm,9mm)$) {};

\node[block,densely dotted,semithick,right=1.75cm of centbank] (b1) {\textbf{Country B}\\($\ldots$)};
\node[block,densely dotted,semithick,right=1.75cm of households] (b3) {\textbf{Rest of the}\\\textbf{World}};
\node[block,densely dotted,semithick] (b2) at ($(b1)!.5!(b3)$) {\textbf{Country C}\\($\ldots$)};

\draw[latex-latex] (b1.west)--node[label,inner xsep=1pt]{Goods} (b1.west-|r.east);
\draw[latex-latex] (b2.west)--node[label,inner xsep=1pt]{Goods} (b2.west-|r.east);
\draw[latex-latex] (b3.west)--node[label,inner xsep=1pt]{Goods} (b3.west-|r.east);
\draw[latex-latex] (b1)--node[label,inner ysep=1pt]{Goods} (b2);
\draw[latex-latex] (b2)--node[label,inner ysep=1pt]{Goods} (b3);
\draw[latex-latex] (b3.east)--++(.4,0) |- node[pos=.25,label,sloped]{Goods} (b1.east);

\node[fill=white] (country_main_label) at (r.north) {\textbf{Country A}};
\node[fill=white, below=-0.14cm of country_main_label] {ICIO};

}
\end{adjustbox}
\caption{The basic structure of the agent-based model with the most important datasets used to calibrate each agent. These include the inter-country input-output tables (ICIO) from the OECD, World Bank (WB) data, Compustat microdata for firms and banks, Household Finance and Consumption Survey (HFCS) data for households and individuals, and policy rates provided by the Bank for International Settlements (BIS).}
\label{fig:model_structure}
\end{figure}

Our model is calibrated to represent the economic structure of countries at a given point in time. Within each country, different economic agents -- firms, individuals, households, banks, a central bank and the government -- play distinct roles and interact with each other. Each country is initialised with 18 sectors\footnote{See Table \ref{table:nace2}.}. To manage computational complexity, we apply a scale factor: one agent (firms or households, etc.) in the simulation represents 1000 real agents. Every agent is calibrated using microdata, allowing for heterogeneity. As in the IIASA model, we ensure full stock-flow consistency. Figure \ref{fig:model_structure} schematizes the basic structure of the agent-based model.

\subsection{Description of the Agents and their Interactions}

\paragraph{Individuals and Households}
Individuals in the model represent the inhabitants of a country, with demographic characteristics such as age mirroring the real population distribution. These individuals are grouped into households. Each household's income and wealth are derived from the economic activities of its members, shaping their consumption habits and decisions. Households participate in the housing market, where they can rent, let or buy homes; in the credit market, where they can apply for loans; and in the goods market, where they purchase what they consume. Individuals can also be hired and fired in the labour market.

\paragraph{Firms}
Firms are categorized by industry and are the primary production units in the economy. They interact by purchasing inputs from local and international markets and transforming them into goods and services. Firms hire workers and apply for loans.

\paragraph{Banks}
Banks play a crucial role by providing credit to agents and holding their deposits. They operate within the credit market, where loans are issued with varying maturities and interest rates that are contingent on the type of loan and on the agent to whom it is given -- for example, consumption loans or mortgages for households or investment loans for firms. The borrower's financial health and the broader economic context influence these terms, particularly the interest rate. The interest rate charged by banks is typically a markup over the policy rate, which is set by the central bank.

\paragraph{The Government}
The government is responsible for a wide range of activities. It collects taxes from individuals and firms, which constitutes its main source of revenue, and has the freedom to set fiscal policy. These funds are used to purchase goods in the goods market, fueling public investment. The government also provides individuals with benefits based on specific criteria, such as their income, employment status or other demographic characteristics. 

\paragraph{The Rest of the World}
This agent, external to all countries, models all of the global economy's non-explicit components. It interacts with domestic economies exclusively via the goods market, encapsulating import/export activities. This allows for a mathematically closed model.

\subsection{Data Sources}
At its core, our simulation is structured around the detailed description of the agents' balance sheets and income statements described above. These balance sheets and their relationships are initialised using a comprehensive array of data sources that are crucial to represent a country's characteristics accurately. For example, the \emph{Inter-Country Input-Output Tables} (ICIO) from the OECD clearly describe the input dependencies of industries within and across national economies. Additional data for initialising the model comes from the IMF, the World Bank, Compustat, the European Central Bank (ECB), and the Bank for International Settlements (BIS). Table \ref{table:data_sources} summarizes these data sources.

\begin{table}[H]
\centering
\renewcommand{\arraystretch}{1.1}
\begin{tabularx}{\textwidth}{l|X}
\toprule
\textbf{Dataset} & \textbf{Usage} \\
\cmidrule(l){1-2}
OECD & Input-Output tables, balance sheets, business demography, labour statistics \\
IMF & Quarterly time series of economic aggregates \\
World Bank & Taxes, government spending, government revenue \\
BIS & Policy rates \\
ECB & Microdata on households and individuals with details on consumption, wealth, and debt \\
Compustat & Microdata on firms and banks with balance sheets and income statements \\
\bottomrule
\end{tabularx}
\vspace{0.1cm}
\caption{Data sources used for model initialisation and calibration.}
\label{table:data_sources}
\end{table}

Aside from initialising balance sheets, the comprehensive data sources we use in our model help establish certain fixed parameters. For example, the tax rates applied within the model are directly derived from World Bank data. However, it is important to distinguish them from the \emph{free} parameters of the model, which are not directly observable and require careful calibration. Free parameters typically include behavioural factors determining how agents extrapolate past data to predict future conditions and make decisions. Since these parameters fundamentally influence the model's dynamics, they must be inferred by systematically comparing the model's time series output against actual economic data. This is the goal of the calibration.

The operation of the simulation consists of updating these balance sheets sequentially. Agents in the model are fully aware of their financial position and have access to historical data regarding their activity and certain economic aggregates. They generally have no detailed economic information about other agents and their intentions, as is common in \abm s. Based on this partial information, they form expectations and make decisions that will influence their balance sheets and those of other agents. Therefore, the agents' balance sheets are updated sequentially at each iteration of the model. The following section describes the different steps in each of these iterations.

\subsection{Sequence of Events}
In our model, each iteration represents a discrete step of the evolution of our economic system. The timeframe corresponding to each iteration is one quarter, a choice that was made based on the time-frequency of the data sources we use and that strikes a good balance with computational efficiency. During these periods, agents make decisions based on partial information and using heuristic ``rules of thumb''. Our implementation allows these rules to be flexible and generic, allowing for different implementation strategies that allow, e.g., reproducing the rules laid out in the model of~\cite{polednaEconomicForecastingAgentbased2020} or~\cite{hommesCANVAS2022}, but also using different sets of rules. 

The type of heuristics includes, for example, those used by firms to determine their target production levels by considering factors such as inflation, historical demand or input costs. The specifics of the rules we use are abstracted into functions that transform inputs -- like market conditions and agent states -- into outputs, such as production targets or hiring decisions. Similarly, market mechanisms, like those governing the labour or goods market, are modelled through matching functions that try to align supply with demand but generally fail to do so exactly. The specifics of these functions are generic enough to be adapted or replaced. We outline the different intermediate steps of each iteration below. The full model description is provided in \ref{sec:model_description}.

\subsubsection{Estimation Phase}
\label{ssub:estimation_phase}
This step lays the foundation for each iteration of our simulations. It involves creating forecasts that will be used by the agents in our simulation in their decision-making process.

\paragraph{Central Economic Forecasts}
This step constitutes the core of this phase, as it includes forecasts for various economic conditions, such as inflation across different markets (consumer prices, producer prices, and housing prices) and overall economic growth rates. These forecasts are based on historical data and provide a common set of expectations about the future state of the economy. This provides all agents with a centralised forecast of these key indicators, which make up a uniform basis from which they form their own, possibly idiosyncratic, predictions and plans.

\paragraph{Firm-Specific Forecasts}
Following the establishment of central economic forecasts, individual firms create their own expectations about the economy's future. Each firm takes into account the past history of supply availability, demand for its products, and price movements to project its own future growth and forecast future demand for its products.

By the end of this phase, all agents have expectations about future economic conditions, which guide their decisions in the \emph{Target Setting Phase}.  

\subsubsection{Target Setting Phase}
\label{ssub:target_setting_phase}
During this phase, agents make strategic decisions based on the above forecasts.

\paragraph{Firm Decisions} Firms begin by setting their production level objective. They do so based on several factors, such as future market conditions established during the \emph{Estimation Phase}, and in particular, their estimate of future demand and their current finished goods inventory. Each firm also assesses its current financial health, including its levels of debt and equity. This determines how much a firm can afford to expand. Once target production is determined, firms set objectives for the labour market and decide on the hiring strategy and the wages they are willing to offer. They also determine their procurement needs based on their current intermediate inputs stock. 

\paragraph{Individual Decisions}
Individuals in the economy set their reservation wages -- the minimum wage they are willing to accept from a potential employer. Their personal wage history influences this decision.

This phase sets the stage for later market interactions. It determines part of the demand that will be considered in the Goods Market, but it also crucially sets the stage for the Labour Market phase.

\subsubsection{Labour Market Phase}
\label{ssub:labour_market_phase}
This phase simply consists of a clearinghouse where the labour supply from individuals is matched with labour demand from firms, considering offered and reservation wages. In parallel to this, firms may fire individuals if their objectives require them to do so, or individuals may spontaneously quit their employment with a given probability. Following this, the different states of the agents are updated so that the firms' employee rolls and the individual's employment status and income match the new situation. 

This step is crucial in setting firms' production capacity, which depends on their workforce, and in setting the income available to each household.

\subsubsection{Planning Metrics Phase}
\label{ssub:planning_metrics_phase}
In this phase, agents refine their expectations and set further targets in response to changes in unemployment.

\paragraph{Government Decisions}
The government updates its social welfare expenditures in response to changes in unemployment and inflation expectations. The government also sets its target level of consumption for goods and services depending on economic conditions and forecasts and its historic spending trajectory.\footnote{Note, however, that in our implementation, this last step is done in the Goods Market Phase without impacting the ordering of the steps described here.}

\paragraph{Central Bank Decisions} The Central Bank sets its policy rate based on current inflation trends.

\paragraph{Household Decisions} Households update their information about their current income based on labour status changes and on changes in social benefits. With updated income estimates, they can determine their spending capacities and adjust their demand for housing and goods and services.

\paragraph{Firm Production} Knowing their output capacity, which is determined by their workforce, their intermediate inputs stock and their capital stock, and guided by the target level described above, firms carry out their production. They set their selling price depending on inflation forecasts and past market conditions. This sets the available supply of each good.

This section is, therefore, crucial in setting the stage for the housing, goods and credit markets.

\subsubsection{Housing Market Phase}
\label{ssub:housing_market_phase}
This phase consists of matching sellers and buyers of housing and matching households seeking to rent property with landlords. It is essentially a reimplementation of the model described by~\cite{geanakoplosGettingSystemicRisk2012} and ~\cite{carro2023heterogeneous} and consists of two distinct phases:

\paragraph{Preparation Phase}
First, the valuation of each property is readjusted depending on market conditions. Households also update their priors on the amount they are willing to buy or rent property and decide whether they want to own or rent. Landlords also decide whether to update the rent on their property depending on inflation.

\paragraph{Clearing Phase}
With the volume and prices set for both supply and demand of the housing market, the housing market is cleared by matching properties up for sale with prospective buyers and properties up for rent with renters. The wealth and housing expenses of households are updated accordingly. Note that a household attempting to buy a property may need to obtain a mortgage in the next phase for the transaction to happen.

\subsubsection{Credit Market Phase}
\label{ssub:credit_market_phase}
This phase is similar to the one described above and consists of two steps. 

\paragraph{Preparation Phase}
Firms and households set their credit targets by studying their current financial health and economic forecasts. Banks set the interest rates they offer on different loan types (mortgages and consumption loans to households, loans for firms) as a markup on the current Central Bank rate. This markup may also depend on the borrower's financial health.

\paragraph{Clearing Phase}
The demand for loans from households and firms is matched with the loans offered by banks. The transactions of households that applied for and obtained mortgages are executed, and the corresponding real estate is updated to be in their name. Mortgages that do not go through result in no transaction registered.

\subsubsection{Goods Market Phase}
In contrast to the phases described above, which all happen within a country, this phase involves the global economy and matches the global supply and demand of goods. It again involves two steps.

\paragraph{Preparation Phase}
The supply and demand of agents is aggregated into an order book in USD, requiring demand in local currencies to be converted to demand for goods in USD using the current exchange rates.

\paragraph{Clearing Phase}
The order book defined above is cleared by matching buyers and sellers. Note that the degree to which agents prefer to buy goods domestically rather than in the international market is a parameter that can be adjusted. We use international trade data to realistically match flows of goods and services in international markets.

\subsubsection{Realised Metrics Phase}
\label{ssub:updates_phase}
Here, agents update the actual outcomes of their operations and financial activities over the past quarter. This phase is crucial for comparing predicted and realised outcomes and making the necessary updates for future planning.

\paragraph{Firms}
After subtracting the amounts they sold, firms update their records to reflect their inventory levels. They also register the realised demand, which will be used to predict demand for the next quarter. Firms also review their financial health and analyse their solvency after updating their financial data with proceeds from sales. The financial records of firms and their bank accounts are also updated. Firms under stress may file for bankruptcy.

\paragraph{Banks}
Banks compute their profits, liabilities, and reserves after all the operations described above, including the consequences of possible borrowers' defaults. Banks under stress may also go bankrupt.

\paragraph{Households}
Households update their wealth and income after operations in the goods and credit market.

\subsection{Key Differences to the IIASA Model}
\label{sec:diffIIASA}
The key differences between the IIASA model \citep{poledna2023economic} and ours are summarized in Table \ref{table:comp_to_iiasa}. While there are many similarities in terms of agents and structure, the main differences are that our model can be calibrated to many countries, features a housing market\footnote{See \ref{sec:housing_market}.}, and explicitly distinguishes between households and individuals.




\begin{table}[H]
\center
\renewcommand{\arraystretch}{0.97}
\begin{minipage}{\textwidth}
\begin{tabularx}{\textwidth}{sk|vX}
\toprule
 & \textbf{Category} & \textbf{IIASA Model} & \textbf{Our Model} \\
\cmidrule(l){1-4}
 & Scope & One country & Multiple countries \\
\cmidrule(l){2-4}
 & Markets & Goods, Labour, Credit & Goods, Labour, Credit, Housing \\
\cmidrule(l){1-4}
\multirow{14}{*}{\centering\STAB{\rotatebox[origin=c]{90}{\,\,\textbf{Individuals \& Households}}}}
& Matching & One-to-one & Multiple individuals living in the same household \\
\cmidrule(l){2-4}
& Income & Wages, benefits, dividends & Wages, benefits, rental income, investment profits \\
\cmidrule(l){2-4}
& Wealth & Deposits, properties & Deposits, properties, other real assets, other financial assets \\
\cmidrule(l){2-4}
& Debt & Overdrafts & Consumption loans, mortgages, overdrafts \\
\cmidrule(l){2-4}
& Consumption & Fixed fraction of income & Heterogeneous fraction of income, consumption smoothing, minimal consumption \\
\cmidrule(l){2-4}
& Reservation Wage & No & Yes \\
\cmidrule(l){1-4}
\multirow{5}{*}{\centering\STAB{\rotatebox[origin=c]{90}{\textbf{Firms}}}} 
& Production & Nested Leontief that allows substitution between intermediate goods and capital goods & Nested Leontief \\
\cmidrule(l){2-4}
& Target setting & Only buffer & Active management \\
\cmidrule(l){2-4}
\cmidrule(l){2-4}
\cmidrule(l){1-4}
\multirow{13}{*}{\centering\STAB{\rotatebox[origin=c]{90}{\textbf{Banks}}}}
& Number of banks & 1 & Multiple \\
\cmidrule(l){2-4}
& Loans recipients & Firms, households (only overdrafts) & Firms, households (overdrafts, consumption loans, mortgages) \\
\cmidrule(l){2-4}
& Bankruptcy & Cannot fail & Bail-in mechanism \\
\cmidrule(l){2-4}
& Supply-side loan requirements & Capital requirement & Capital requirement and risk diversification \\
\cmidrule(l){2-4}
& Demand-side loan requirements & Loan-to-value & Loan-to-value, return on equity/assets (firms), loan-to-income/debt-service-to-income (households) \\
\cmidrule(l){2-4}
& Interest rates & Fixed markup on the central bank policy rate & Statistical model that considers pass-through from the central bank, inflation and non-performing loans \\
\cmidrule(l){1-4}
\multirow{2}{*}{\centering\STAB{\rotatebox[origin=c]{90}{\textbf{Gov.}}}}
& Benefits & Unemployment benefits grow with the economy & Countercyclical unemployment benefits \\
\bottomrule
\end{tabularx}
\end{minipage}
\vspace{0.05cm}
\caption{Comparison of the IIASA Model to our model.}
\label{table:comp_to_iiasa}
\end{table}

\section{Model Calibration}
\label{sec:calibration}
Our model, denoted $\eM$, contains free parameters $\theta$ that cannot be \textit{directly} estimated from aggregate time series or microdata. Our goal is to estimate them using the calibration data $y$ to which we have access. To do so, we use approximate, simulation-based Bayesian inference procedures to obtain a posterior density $\pi(\theta \mid y, \eM)$ for the parameters given the observed calibration data. Approximate, simulation-based inference techniques allow us to overcome the difficulty of conducting inference for complex models such as ours, which stems from the fact that the model's likelihood function -- the distribution $p(y \mid \theta, \eM)$ of simulation data given a set of parameters -- is unknown and generally difficult to derive or evaluate numerically \citep{dyer2022calibrating, dyer2024black}.

Exact Bayesian inference proceeds by posing a prior distribution $\pi(\theta \mid \eM)$ for the parameters, which captures the modeller's \textit{a priori} beliefs about the credibility of different parameter values. This belief distribution is subsequently updated to a posterior distribution using Bayes's theorem:
\begin{equation}\label{eq:bayes}
    \pi(\theta \mid y, \eM) = \frac{p(y\mid \theta, \eM)}{m(y \mid \eM)}\, \pi(\theta \mid \eM),
\end{equation}
where
\begin{equation}
    m(y \mid \eM) = \int p(y \mid \theta, \eM) \, \pi(\theta \mid \eM)\, \dee{\theta}
\end{equation}
is the model's marginal likelihood function. This requires, in particular, being able to evaluate the likelihood function $p(y \mid \theta, \eM)$. 
In contrast, simulation-based Bayesian inference procedures approximate this density using only the modeler's ability to simulate the model. This circumvents the difficulty posed above of not being able to evaluate the likelihood function, and consequently the posterior density, via Bayes's theorem. 

In this work, we employ two different simulation-based inference procedures: \emph{neural posterior estimation} ({\npe}) and \emph{neural density ratio estimation} ({\nre}), which have been seen to be faster and more accurate~\citep{dyer2024black}  than alternative methods employed in the agent-based modelling literature\footnote{E.g., the approaches discussed in~\cite{grazziniBayesian2017}.}. We briefly describe both approaches below.

\subsection{Approximate Bayesian Inference with Neural Posterior Estimation}
\npe{} models the posterior density $\pi(\theta \mid y, \eM)$ directly using a neural network $q_\phi$, with $\phi$ the weights of the network. The trained weights $\hat{\phi}$ are obtained by minimising the Kullback-Leibler divergence between the joint density over parameters and the data, i.e., by minimising
\begin{align}
    \mathbb{E}_{p(x \mid \theta, \eM)\,\pi(\theta \mid \eM)}\left[\log\frac{p(x \mid \theta, \eM)\, \pi(\theta \mid \eM)}{q_{\phi}(\theta \mid x, \eM)\, m(x \mid \eM)}\right]
    = 
    \mathbb{E}_{p(x \mid \theta, \eM)\,\pi(\theta \mid \eM)}\left[-\log q_{\phi}(\theta \mid x, \eM)\right] + \text{constants in }\phi,
\end{align}
which motivates using the following loss function
\begin{equation}\label{eq:loss}
    \mathcal{L}(\phi) = \mathbb{E}_{p(x \mid \theta, \eM)\,\pi(\theta \mid \eM)}\left[-\log q_{\phi}(\theta \mid x, \eM)\right].
\end{equation}

In practice, we estimate \eqref{eq:loss} using a finite Monte Carlo sample of $N$ data-parameter pairs $\left(x^{(i)}, \theta^{(i)}\right)_{i=1}^{N}$ drawn from the distribution $p(x \mid \theta, \eM)\,\pi(\theta \mid \eM)$ by simulating from the prior of the parameters and the \abm. The posterior $\pi(\theta \mid y, \eM)$ is then approximated as $q_{\hat{\phi}}(\theta \mid y, \eM)$. In the current work, we take $q_{\phi}$ to be a normalising flow \citep{tabak2010density} and train the network parameters $\phi$ through gradient-based optimisation procedures; further details on the neural network architecture and training procedure are provided in \ref{app:training_npe}.

\subsection{Approximate Bayesian Inference with Neural Density Ratio Estimation}
An alternative approach, which again uses neural networks, is \nre. This approach consists of training a neural network $r_{\varphi}$ with weights $\varphi$ to approximate the log ratio between the likelihood $p(y \mid \theta, \eM)$ and the marginal likelihood $m(y \mid \eM)$. While this can be achieved in various ways, we do so by searching for neural network weights $\hat{\varphi}$ that minimise the loss function given by
\begin{equation}
    \mathcal{L}(\varphi) = -\frac{1}{B} \sum_{b=1}^{B} \frac{\exp\left(r_{\varphi}(x^{(b)}, \theta^{(b)})\right)}{\exp\left(r_{\varphi}(x^{(b)}, \theta^{(b)})\right) + \sum_{k=1}^K \exp\left(r_{\varphi}(x^{(b)}, {\theta}^{(b,k)})\right)},
\end{equation}
where a batch of $B > 1$ data-parameter pairs $\left(x^{(b)}, \theta^{(b)}\right)_{b=1}^B$ are drawn from the true joint density, $p(x \mid \theta, \eM)\,\pi(\theta \mid \eM)$, and each of the $\theta^{(b,k)}$ is drawn from the prior $ \pi(\theta \mid \eM)$, $K > 1$.

It can be shown \citep{durkan2020contrastive} that this induces the neural network to approximate, up to some additive constant $c(x)$, the log-ratio of the likelihood and marginal likelihood at $x$, i.e.,
\begin{equation}
    r_{\hat{\varphi}}(x, \theta) \approx \log \frac{p(x\mid \theta, \eM)}{m(x \mid \eM)} + c(x).
\end{equation}
This can then be used to recover approximate posterior samples for $y$, for example, by Markov chain Monte Carlo procedures. Further details on the training procedure are provided in \ref{app:training_nre}.

\subsection{Prior and Posterior Estimation}
We choose seven parameters among the full set of free model parameters\footnote{For the full model description, including parameters, see \ref{sec:model_description}.} for estimation. This choice was made since they appear to have the strongest positive influence on forecasting performance\footnote{\label{fn:forecasting} Our chosen metric is described in Section \ref{sec:comp_gdp}. While there exists a variety of alternative model selection rules we could employ here instead, such as selecting the maximum a posteriori model through a procedure mirroring the Bayes factor computations presented below, we opt for model selection based on this RMSE metric for simplicity.}. Table \ref{table:prior} summarises the prior and the estimated posterior using \npe{} for Austria as a representative example, as it is the country the IIASA model is calibrated to. We calibrate these parameters to match the first moment of growth rates of five time series between 1990-Q1 and 2013-Q1: real gross domestic product, inflation, household consumption, government consumption, and investment. In Table \ref{table:prior}, U$(\{0,1\})$ refers to binary parameters chosen with equal likelihood and U$([0,1])$ to the continuous uniform distribution on the closed interval $[0,1]$.

The first three rows in Table \ref{table:prior} are not calibrated using approximate Bayesian inference methods. Instead, for each of the 8 possible combinations, corresponding to $\left(\phi^Q_F, \phi^\text{DP}, \phi^\text{CP}\right)=\left(\pm1, \pm1, \pm1\right)$, we ran the calibration procedure using $10^6$ samples and a uniform prior for the remaining four rows of the table. For these rows, we show the first two moments of the estimated posterior corresponding to $\phi^Q_F=\phi^\text{DP}=\phi^\text{CP}=0$, which is the configuration that minimises the forecasting error\footnote{See the previous footnote \ref{fn:forecasting}.} to the true nominal gross domestic product growth rates between 1990-Q1 and 2013-Q1.

Summary statistics describing the estimated posterior are shown in Table \ref{table:prior}. In the rest of this section, we discuss the economic meaning of the parameters we chose to estimate.

\begin{table}[H]
\centering
\renewcommand{\arraystretch}{1.1}
\begin{tabularx}{\textwidth}{Xc|c|cc}
\toprule
\multirow{2}{*}{\textbf{Description}} & \multirow{2}{*}{\textbf{Notation}} & \multirow{2}{*}{\textbf{Prior}} & \multicolumn{2}{c}{\textbf{Posterior}} \\
 & & & \textbf{Mean} & \textbf{Variance} \\
\cmidrule(l){1-5}
Demand adjustment speed on firm growth & $\phi^Q_F$ & U$(\{0,1\})$ & 0 & - \\
Influence of demand-pull inflation on prices & $\phi^\text{DP}$ & U$(\{0,1\})$ & 0 & - \\
Influence of cost-push inflation on prices & $\phi^\text{CP}$ & U$(\{0,1\})$ & 0 & - \\
\cmidrule(l){1-5}
Target inventory to production fraction & $\phi^\text{StY}$ & U$([0,1])$ & 0.10 & 0.0 \\
Influence of labour inputs on target production & $\chi^\text{H}$ & U$([0,1])$ & 0.53 & 0.02 \\
Influence of intermediate inputs on target production & $\chi^\text{M}$ & U$([0, 1])$ & 0.03 & 0.0 \\
Influence of capital inputs on target production & $\chi^\text{K}$ & U$([0, 1])$ & 0.18 & 0.03 \\
\bottomrule
\end{tabularx}
\vspace{0.5cm}
\caption{The prior and the corresponding posterior moments estimated by \npe{} for Austria between 1990-Q1 and 2013-Q1. U$(\{0,1\})$ refers to binary parameters chosen with equal likelihood, U$([0,1])$ to the continuous uniform distribution on the closed interval $[0,1]$.}
\label{table:prior}
\end{table}

\paragraph{Predicted Firm Demand}
In our model, growth is driven by expectations. Specifically, firms set target production based on an expectation for future demand. The predicted future demand of a firm $f$ operating in sector $s$ is set as a markup on previously observed demand $Q_f(t-1)$,
\begin{equation}
    \label{eq:firm_exp_demand_main}
    \overline{Q}_f(t) = \underbrace{\left(1 + \overline{\gamma}_s(t)\right)}_\text{Predicted sectoral growth} \times \underbrace{\left(1 + \phi^Q_F \overline{\gamma}_f(t)\right)}_\text{Predicted idiosyncratic growth} \times \underbrace{Q_f(t-1).}_\text{Previous demand}
\end{equation}
The first term in equation \eqref{eq:firm_exp_demand_main} corresponds to a global sectoral forecast $\overline{\gamma}_s(t)$, and the second term corresponds to a firm-specific idiosyncratic growth forecast $\overline{\gamma}_f(t)$\footnote{The firm-specific growth forecast was introduced in the CANVAS model \citep{hommesCANVAS2022}. It is driven by demand-pull and cost-push inflation, see \eqref{eq:firm_exp_growth}.}. A parameter $\phi^Q_F$ governs to what extent that forecast is considered when predicting future demand. For \npe{} and for \nre{}, $\phi^Q_F=0$ for all 38 countries, thereby excluding firm-specific growth forecasts.

\paragraph{Firm Prices}
Firms set prices $P_f(t)$ as a markup on previous prices $P_f(t-1)$,
\begin{equation}
    \label{eq:firm_prices_main}
    P_f(t) = \underbrace{\left(1+  \overline{\pi}^\text{PPI}(t)\right)}_\text{Predicted PPI inflation} \times  \underbrace{\left(1+\phi^\text{DP} \overline{\pi}_f^\text{DP}(t)\right)}_\text{Demand-pull inflation} \times \underbrace{\left(1 + \phi^\text{CP} \overline{\pi}_f^\text{CP}(t)\right)}_\text{Cost-push inflation} \times \underbrace{P_f(t-1).}_\text{Previous price}
\end{equation}
The first term in equation \eqref{eq:firm_prices_main} corresponds to a global forecast for producer price index (PPI) inflation $\overline{\pi}^\text{PPI}(t)$. The second term corresponds to (firm-specific) demand-pull inflation $\overline{\pi}_f^\text{DP}(t)$, driven by observed demand, whose influence is governed by a parameter $\phi^\text{DP}$. The third term corresponds to (firm-specific) cost-push inflation $\overline{\pi}_f^\text{CP}(t)$, driven by observed unit costs, whose influence is governed by a parameter $\phi^\text{CP}$.\footnote{The firm-specific drivers of inflation were introduced in the CANVAS model \cite{hommesCANVAS2022}, see Eq. \eqref{eq:dp_inflation} and Eq. \eqref{eq:cp_inflation}.} For \npe{} and for \nre{}, $\phi^\text{DP}=\phi^\text{CP}=0$ for all 38 countries, thereby excluding firm-specific inflation from price setting.

\paragraph{Firm Target Production}
Firm target production $\hat{Y}_f(t)$ is set based on predicted demand $\overline{Q}_f(t)$, current inventory, and considers the current financial situation of the firm. It may also be limited by the firm's current workforce, stock of intermediate inputs, and stock of capital inputs. Specifically,
\begin{equation}
\label{eq:firm_target_production_main}
\begin{aligned}
    \hat{Y}_f(t) = \min\Bigg(&\underbrace{\overline{Q}_f(t) + \phi^\text{StY} Y_f(t-1) - S_f(t-1)}_\text{Predicted demand given current inventory},\, \underbrace{\overline{Q}_f(t) + \chi^H \left(H_f(t) - \overline{Q}_f(t)\right)}_\text{Labour inputs},\\
    & \underbrace{\overline{Q}_f(t) + \chi^M \left(M_f(t) - \overline{Q}_f(t)\right)}_\text{Intermediate inputs},\,\underbrace{\overline{Q}_f(t) + \chi^K \left(K_f(t) - \overline{Q}_f(t)\right)}_\text{Capital inputs}
    \Bigg),
\end{aligned}
\end{equation}
where $Y_f(t-1)$ is real output of firm $f$ in the previous time step, $\phi^\text{StY}$ is the target inventory to production fraction and $S_f(t-1)$ are inventories kept from the previous time step. For \npe, we get an mean of $\phi^\text{StY}=0.06$ across all 38 countries, for \nre, we get $\phi^\text{StY}=0.04$, implying that the role of inventories in setting target production is small. Moreover, $H_f(t)$, $M_f(t)$ and $K_f(t)$ denote maximum production given the firm's labour, intermediate, and capital inputs, respectively.

The firms' labour inputs $H_f(t)$ are set as
\begin{equation}
    H_f(t) = \underbrace{h_f(t)}_\text{Work effort} \,\,\underbrace{\sum_{i\in\mathcal{I}_f(t)} H_i(t)}_\text{Labour supply}
\end{equation}
where $\mathcal{I}_f(t)$ is the set of individuals employed by firm $f$, $H_i(t)$ are labour inputs from individual $i$, and $h_f(t)$ is a firm-specific factor denoting work effort. This factor follows \cite{poledna2023economic} and corresponds to increased or decreased work effort due to overtime or part-time employment (see \eqref{eq:we}).

The firm's production is also limited by its level of intermediate and capital inputs, set as
\begin{align}
    M_f(t) ={}& \min_{s'\in\mathcal{S}}\left(\frac{M_{fs'}(t-1)}{m_{s's}}\right) \\
    K_f(t) ={}& \min_{s'\in\mathcal{S}}\left(\frac{K_{fs'}(t-1)}{k_{s's}}\right)
\end{align}
where $\mathcal{S}$ is the set of sectors, $M_{fs}(t-1)$ is the firms' previous stock of intermediate inputs, and $K_{fs}(t-1)$ is the firms' previous stock of capital inputs of sector $s$. The matrix $m_{s's}$ denotes the real amount of intermediate inputs of sector $s'$ necessary to produce one real unit of output of sector $s$, while the matrix $k_{s's}$ plays the same role for the real amount of capital inputs required for production.

These functional forms are substantially different to those used in \cite{poledna2023economic}. Indeed, \cite{poledna2023economic} aggregates all intermediate inputs into a single intermediate composite and all capital inputs into a single capital composite, essentially assuming a linear production function in each intermediate and capital input. Here, we assume instead stricter substitution possibilities in the economy by using a Leontief production function for each input. 

The parameters $\chi^\text{H},\chi^\text{M},\chi^\text{K}\in [0, 1]$, determining the influence of resource constraints on target production, are another key innovation in our framework. We think it unreasonable to assume that firms limit target production from the constraints on labour, intermediate and capital inputs. Instead, it is likely that firms have imperfect knowledge of these constraints and wish to produce more than they could --- although they would naturally fail to produce as much as planned if such constraints bind.

Thus, for instance, if $\chi^\text{H}=0$, firms would ignore labour inputs in determining target production, while if $\chi^\text{H}=1$, they would fully consider labour input constraints. Target production directly determines the firm's demand for labour, intermediate inputs, and capital, and so it is clear that limiting it ex-ante creates substantial problems for macroeconomic dynamics. Across all 38 countries, and for both \npe{} and \nre{}, the role of intermediate inputs (determined by $\chi^\text{M}$) is small, suggesting that firms would substantially expand production if the constraints were not binding. Labour inputs and capital inputs are partially binding (averaged across all 38 countries, $\chi^\text{H}=0.54$ and $\chi^\text{K}=0.56$ for \npe, $\chi^\text{H}=0.45$ and $\chi^\text{K}=0.55$ for \nre), and taken into account by firms when setting their production targets.

\section{Results}
\label{sec:forecasting}
This section summarises the forecasting performance of our model in out-of-sample prediction of economic aggregates.
%
%
We highlight the improvement in root mean squared error from using more realistic modelling rules, micro-data, and systematic calibration over the IIASA model. Compared to the literature \citep{hommesCANVAS2022,poledna2023economic}, our validation is more meaningful since we perform a Bayes factor estimation and include multiple (38) countries and individual Monte-Carlo trajectories.

\begin{figure}[H]
\includegraphics[width=\textwidth]{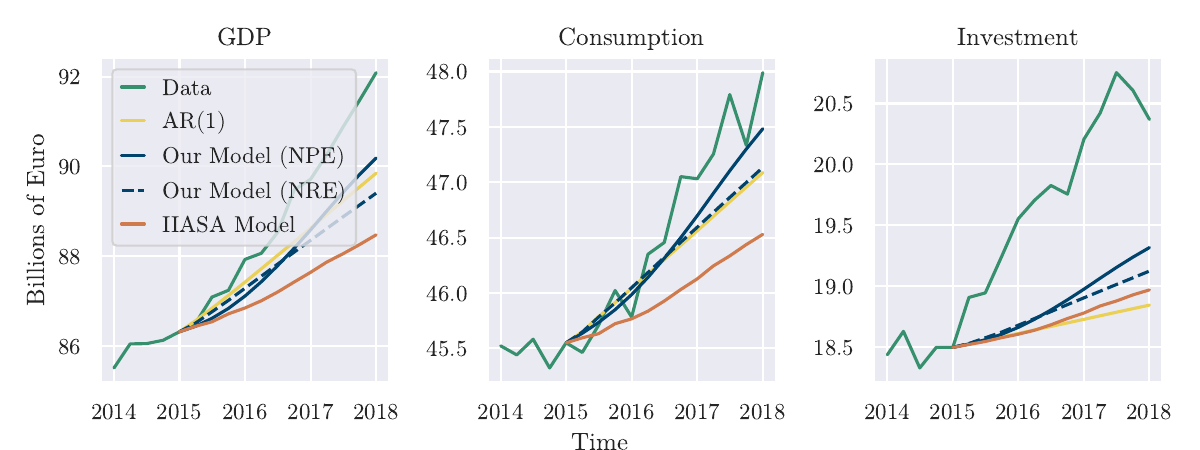}
\centering
\caption{Forecasts for real GDP, consumption, and investment for Austria from 2015-Q1 to 2018-Q1.}
\label{fig:basic_comp}
\end{figure}

Figure \ref{fig:basic_comp} shows forecasts from 2015-Q1 to 2018-Q1 for real GDP, real consumption, and real investment for Austria. The green line is true data, and the yellow line corresponds to an AR(1) iterated multiperiod forecast, fit on quarterly first differences of the log nominal gross domestic product between 1990-Q1 and 2015-Q1. The blue lines correspond to the average of $1000$ simulation runs of our model, sampled from the posterior predictive distribution associated with the estimated posterior. The solid line corresponds to \npe, and the dashed line to \nre, in both cases targeting moments of economic aggregates between 1990-Q1 and 2010-Q1. The orange line corresponds to the average of 1000 simulation runs of the IIASA model.\footnote{We run the IIASA model for all 38 OECD member countries; see \ref{sec:appendix_sp_reproduction}.} For this particular country and period of time, our model with \npe{} performs well for predicting consumption; all models fail to predict investment well.

\subsection{Bayesian Model Selection}
\label{sec:bayesian_ms}
To further test the forecasting capabilities of our model relative to the benchmark \abm~in \citet{poledna2023economic}, we construct an approximate \emph{Bayes factor}  \citep{jeffreys1998theory} for our model, $\eM$, against the IIASA benchmark model, $\eM'$. 
The Bayes factor $B(\eM, \eM', y)$ for model $\eM$ against model $\eM'$ and for data $y$ is the ratio of the two models' marginal likelihood functions:
\begin{equation}
    B(\eM, \eM', y) = \frac{m(y \mid \eM)}{m(y \mid \eM')}.
\end{equation}

For our model, the marginal likelihood function is the likelihood function at data $y$ integrated over the prior distribution over the model's free parameters, $\theta$:
\begin{equation}\label{eq:marginal_likelihood}
    m(y \mid \eM) = \int p(y \mid \theta, \eM)\, \pi(\theta \mid \eM) \, \dee{\theta}.
\end{equation}
In contrast, the IIASA benchmark $\eM'$ has no free parameters; $m(y \mid \eM')$ is, therefore, simply the distribution over model output resulting from the model's use of pseudorandom numbers internally. 

Given the use of the prior (rather than a posterior) density in Equation~\ref{eq:marginal_likelihood}, Bayes factors assess the relative predictive performance of the two models $\eM$ and $\eM'$ treating all data as out-of-sample; that is, Bayes factors can be interpreted as the predictive probability (density) for the data before data becomes available for use in parameter estimation tasks, and a summary of the evidence provided by unseen data in favour of one probabilistic model over another \citep{kass1995bayes}. Further, Bayes factors naturally penalise a model for any increased complexity introduced by the appearance of additional model parameters, resulting from the fact that it integrates the model's likelihood function over the prior. Finally, in contrast to alternative approaches to model comparison (e.g., non-Bayesian likelihood ratio tests), Bayes factors do not require the models being compared to be nested\footnote{Model $A$ is nested within model $B$ with parameterisation $\xi = (\alpha, \beta)$ (with $\alpha$ and $\beta$ scalars or vectors) if model $A$ is attained from model $B$ when $\beta = \beta_0$ for some $\beta_0$.}. Bayes factors, therefore, provide a suitable vehicle for comparing the predictive performance of our more highly parameterised model against the IIASA baseline.

In our Bayes factor computations, we choose a uniform parameter prior $\pi(\theta \mid \eM)$ over the (bounded) support of the parameters for our model\footnote{We discuss in \ref{app:stat_sig} that this prior is not entirely suitable, and is likely to lead to Bayes factors that are less favourable for our model than other more sensible priors.}. Given that the models' likelihood functions (and therefore marginal likelihood functions) are difficult to derive, given the complexity of $\eM$ and $\eM'$, we estimate Bayes factors using density ratio estimation: we train a real-valued neural network $b_{\upphi}$ to minimise (Monte Carlo estimates of) the loss function
\begin{equation}
    \mathcal{L}(\upphi) = -\mathbb{E}_{x \sim m(\cdot \mid \eM)}\left[\log\frac{\exp(b_{\upphi}(x))}{1 + \exp(b_{\upphi}(x))}\right] - \mathbb{E}_{x' \sim m(\cdot \mid \eM')}\left[\log \frac{1}{1 + \exp(b_{\upphi}(x'))}\right],
\end{equation}
which induces the network to learn the log-ratio of the marginal likelihoods \citep{gutmann2012noise}, i.e.,
\begin{equation}
    b_{\hat{\upphi}}(x) \approx \log \frac{m(x \mid \eM)}{m(x \mid \eM')} = \log B(\eM, \eM', x)
\end{equation}
for any given data $x$, where $\hat{\upphi}$ are the trained neural network weights. We then estimate $B(\eM, \eM', y)$ as $\exp(b_{\hat{\upphi}}(y))$. We provide details on the training procedure and neural architecture in \ref{app:training_bf}.

We estimate Bayes factors for the 38 OECD countries at 20 times of initialisation between 2013-Q1 and 2017-Q4. Each ratio estimator is trained using 1000 Monte Carlo samples, each of length three years. In particular, the ratio estimators are trained on summary statistics derived from five macroeconomic aggregates\footnote{See Table \ref{table:forecasting_gdp}.} simulated over these three-year periods. Further details on the experimental setup and results are shown in, respectively, \ref{app:training_bf} and \ref{sec:appendix_bayes_factors}. Of all estimated Bayes factors, approximately $65\%$ support our model over the IIASA model. 

\subsection{Forecasting Economic Aggregates}
\label{sec:comp_gdp}
We measure the forecasting performance for different economic aggregates of a chosen country and a chosen time of initialisation $T_1$ by calculating the median of the root mean squared error (RMSE) of 1000 Monte Carlo trajectories,
\begin{equation}
\label{eq:mmse}
    \text{RMSE} = \underset{1\leq i \leq 1000}{\operatorname{median}}\,\,\sqrt{\frac{1}{h} \sum_{t=T_1 + 1}^{T_1+1+h} \left(X_i(t) - \widehat{X}(t)\right)^2}
\end{equation}
where $h$ is the length of the forecasting window, $X_i(t)$ is simulation output of the $i$-th Monte-Carlo run, and $\widehat{X}(t)$ is the corresponding true data at time $t$. For each time $T_1$ and each country for which we can initialise the model, we calculate the median of the quantity in Equation \eqref{eq:mmse}.

Specifically, we calculate the median across 38 OECD member countries and 20 initialisation times between 2013-Q1 and 2017-Q4. We have chosen the median in the previous steps as it is a robust metric that is not swayed by outliers. We compare the performance of our model to a benchmark AR(1) iterated multiperiod forecast, fit on the first differences of the logarithm, and to the IIASA model by its relative improvement in median RMSE.

The first super-column of Table \ref{table:forecasting_gdp} (``AR1 RMSE'') shows the median RMSE of an AR(1) model predicting five different economic aggregates for six different forecasting horizons. As described above, we take the median over the full set of 38 countries, 20 initialisation times, and 1000 trajectories. The remaining rows of Table \ref{table:forecasting_gdp} compare the performance of the agent-based models (the IIASA model and our calibrated models) to the AR(1) by showing the relative improvement\footnote{Measured by subtracting the median RMSE of the agent-based model from the median RMSE of the AR(1), and then dividing that difference by the median RMSE of the AR(1).} in forecasting error (measured as a median across countries and initialisation times) compared to the AR(1). Overall, our model performs significantly better than both the AR(1) and the IIASA model. On a country level (see \ref{sec:appendix_economic_forecasting_country}), our model outperforms the IIASA model across all initialisation times and economic aggregates for 33 out of 38 countries.

\begin{table}[h]
\centering
\renewcommand{\arraystretch}{1.1}
\begin{tabularx}{\textwidth}{s|X|cccccc}
\toprule
 & \multirow{2}{*}{Aggregate} & \multicolumn{6}{c}{Horizon} \\
 & & 1 Quarter & 2 Quarters & 3 Quarters & 1 Year & 2 Years & 3 Years \\
\cmidrule(l){1-8}
\multirow{5}{*}{\centering\STAB{\rotatebox[origin=c]{90}{AR1 RMSE}}} & GDP & $7\cdot 10^{-3}$ & $9\cdot 10^{-3}$ & $10\cdot 10^{-3}$ & $10\cdot 10^{-3}$ & $10\cdot 10^{-3}$ & $11\cdot 10^{-3}$ \\
 & Inflation & $8\cdot 10^{-3}$ & $10\cdot 10^{-3}$ & $12\cdot 10^{-3}$ & $12\cdot 10^{-3}$ & $12\cdot 10^{-3}$ & $12\cdot 10^{-3}$ \\
 & Household Cons. & $7\cdot 10^{-3}$ & $10\cdot 10^{-3}$ & $10\cdot 10^{-3}$ & $11\cdot 10^{-3}$ & $11\cdot 10^{-3}$ & $12\cdot 10^{-3}$ \\
 & Government Cons. & $8\cdot 10^{-3}$ & $10\cdot 10^{-3}$ & $11\cdot 10^{-3}$ & $12\cdot 10^{-3}$ & $12\cdot 10^{-3}$ & $13\cdot 10^{-3}$ \\
 & Investment & $28\cdot 10^{-3}$ & $39\cdot 10^{-3}$ & $41\cdot 10^{-3}$ & $43\cdot 10^{-3}$ & $45\cdot 10^{-3}$ & $49\cdot 10^{-3}$ \\
\cmidrule(l){1-8}
\multirow{5}{*}{\centering\STAB{\rotatebox[origin=c]{90}{IIASA Model}}} & GDP & -3\% & -5\% & -8\% & -9\% & -12\% & -10\% \\
 & Inflation & -2\% & 0\% & 0\% & 2\% & 4\% & 2\% \\
 & Household Cons. & -12\% & -15\% & -16\% & -19\% & -19\% & -22\% \\
 & Government Cons. & 1\% & 0\% & -1\% & -1\% & -2\% & 0\% \\
 & Investment & 30\% & 30\% & 28\% & 28\% & 31\% & 31\% \\
\cmidrule(l){1-8}
\multirow{5}{*}{\centering\STAB{\rotatebox[origin=c]{90}{\npe}}} & GDP & 41\% & 43\% & 38\% & 43\% & 63\% & 62\% \\
 & Inflation & 44\% & 55\% & 49\% & 50\% & 59\% & 63\% \\
 & Household Cons. & 60\% & 55\% & 35\% & 38\% & 47\% & 44\% \\
 & Government Cons. & 73\% & 71\% & 67\% & 59\% & 48\% & 45\% \\
 & Investment & 62\% & 53\% & 51\% & 57\% & 61\% & 58\% \\
\cmidrule(l){1-8}
\multirow{5}{*}{\centering\STAB{\rotatebox[origin=c]{90}{\nre}}} & GDP & 49\% & 48\% & 38\% & 44\% & 50\% & 50\% \\
 & Inflation & 44\% & 55\% & 52\% & 53\% & 61\% & 66\% \\
 & Household Cons. & 63\% & 62\% & 47\% & 43\% & 44\% & 40\% \\
 & Government Cons. & 66\% & 75\% & 69\% & 59\% & 48\% & 37\% \\
 & Investment & 56\% & 58\% & 53\% & 59\% & 62\% & 56\% \\
\bottomrule
\end{tabularx}
\vspace{0.5cm}
\caption{The first super-column (``AR1 RMSE'') shows the median RMSE of an AR(1) model predicting five different economic aggregates for six different forecasting horizons. The median is taken over the full set of 38 countries, 20 times of initialisation, and 1000 trajectories. The remaining rows of Table \ref{table:forecasting_gdp} compare the performance of the agent-based models (the IIASA model and our calibrated models) to the AR(1) by showing the relative improvement in forecasting error (measured as a median across countries and initialisation times) compared to the AR(1).}
\label{table:forecasting_gdp}
\end{table}

\section{Conclusion}
\label{sec:conclusion}

The work we have presented introduces a data-driven agent-based model that is designed to simulate the world economy. It is built using a flexible modelling platform\footnote{We call this a modelling platform in the sense that the different modules -- that is, the agents, markets and the rules that drive their behaviour -- can be easily changed by the modeller. In this sense, this is not a single model because it allows the running and calibrating of a multitude of models.} That allows the modeller to change configurations and behavioural rules easily. Here, we use the model as a macromodel by aggregating, but it can also potentially be used for finer-grained predictions and analysis.

We show the capabilities of our model by running it for 38 different countries and at 20 different initialisation times, with each run simulating the entire world economy represented by one explicitly modelled country trading with an aggregated Rest of the World. We calibrate these runs using approximate Bayesian estimation methods, and we show that our calibrated model outperforms both a benchmark time series model and the model developed at IIASA~\citep{polednaEconomicForecastingAgentbased2020} by obtaining lower forecasting errors for all considered macroeconomic aggregates in 33 out of 38 countries. In combination with Bayes factors, which allow to compare the relative performance of a model over another, we show that the evidence in favor of our model is statistically significant. 


The improvements highlighted here are but a small exploration of the capabilities of our model, which is built in a modular way and uses modern programming principles. In future work, we intend to explore our model's capabilities further, making the individual behavioural rules more realistic and using more granular and extensive datasets. In the long term, we will run all 38 countries \emph{in parallel}, capturing all the relevant feedback loops of international trade. 

\clearpage

\bibliographystyle{elsarticle-num-names} 
\bibliography{cas-refs}

\appendix
\section{Model Description}
\label{sec:model_description}
This section provides an in-depth description of the behavioural rules of agents, details on market clearing, and model parameters. The model can be calibrated to a set of countries, where each country comprises firms and individuals grouped into households, a government, banks, and a central bank. The model includes four markets: a market for goods matching buyers (firms, households, government entities, the rest of the world) with sellers (firms, the rest of the world), a market for labour matching firms with individuals, a market for credit matching firms and households with banks, and a market for housing matching households with properties. The firm sector comprises $18$ industry sectors according to the NACE-2 classification, as shown in Table \ref{table:nace2}.

\tablefontsize
\begin{xltabular}{\textwidth}{Xl}
\toprule
Description & NACE Rev. 2 \\
\cmidrule(l){1-2}
Agriculture, forestry and fishing & A \\
Mining and quarrying & B \\
Manufacturing & C \\
Electricity, gas, steam and air conditioning supply & D \\
Water supply; sewerage; waste management, and remediation activities & E \\
Construction & F \\
Wholesale and retail trade; repair of motor vehicles and motorcycles & G \\
Transporting and storage & H \\
Accommodation and food service activities & I \\
Information and communication & J \\
Financial and insurance activities & K \\
Real estate activities & L \\
Professional, scientific and technical activities & M \\
Administrative and support service activities & N \\
Public administration and defence; compulsory social security & O \\
Education & P \\
Human health and social work activities & Q \\
Arts, entertainment and recreation + Other services activities & R + S \\
\bottomrule
\\
\caption{Classification of economic activities NACE Rev. 2.}
\label{table:nace2}
\end{xltabular}
\normalsize

The model is calibrated using OECD data\footnote{See \url{https://data-explorer.oecd.org}.}, IMF data\footnote{See \url{https://www.imf.org/en/Data}.}, World Bank data\footnote{See \url{https://data.worldbank.org/}.}, data provided by the Bank for International Settlemnts \footnote{See \url{https://data.bis.org/}}, Household Finance and Consumption Survey data \footnote{See \url{https://www.ecb.europa.eu/stats/ecb_surveys/hfcs/html/index.en.html}.}, and Compustat data\footnote{See \url{https://www.lseg.com/en/data-analytics/financial-data/company-data/fundamentals-data/standardized-fundamentals/sp-compustat-database}.}.

\paragraph{Notation}
By overlines $\overline{x}$, we denote values predicted by agents and with hats $\hat{x}$, we denote target values. If not otherwise specified, values are in nominal terms. We write $[x]^+ = \max(0,x)$, $[x]^-=-\min(0,x)$, and $\Delta x(t)=x(t) - x(t-1)$.

\subsection{Economic Aggregates}
\label{sec:economic_aggregates}
In this section, we briefly discuss economic aggregates in the model. These variables are defined in Table \ref{table:aggregate_vars_and_sec_weights} below.

\tablefontsize
\begin{xltabular}{\textwidth}{cXl}
\toprule
Category & Description & Notation \\
\cmidrule(l){1-3}
Sets & Sectors & $\mathcal{S}$ \\
\cmidrule(l){1-3}
\multirow{4}{*}{\STAB{\rotatebox[origin=c]{0}{Production}}} & Total real production & $Y(t)$ \\
 & Predicted total real production & $\overline{Y}(t)$ \\
 & Sectoral real production & $Y_s(t)$ \\
 & Predicted sectoral real production & $\overline{Y}_s(t)$ \\
\cmidrule(l){1-3}
 \multirow{18}{*}{Price} & Sectoral price index & $P_s(t)$ \\
 & Producer price index & $P^\text{PPI}(t)$ \\
 & Predicted producer price index & $\overline{P}^\text{PPI}(t)$ \\
 & Consumer price index & $P^\text{CPI}(t)$ \\
 & Predicted consumer price & $\overline{P}^\text{CPI}(t)$ \\
 & House price index & $P^\text{HPI}(t)$ \\
 & Predicted house price index & $\overline{P}^\text{HPI}(t)$ \\
 & Rental price index & $P^\text{RPI}(t)$ \\
 & Predicted rental price index & $\overline{P}^\text{RPI}(t)$ \\
 & Producer price index inflation & $\pi^\text{PPI}(t)$ \\
 & Predicted producer price index inflation & $\overline{\pi}^\text{PPI}(t)$ \\
 & Consumer price index inflation & $\pi^\text{CPI}(t)$ \\
 & Predicted consumer price index inflation & $\overline{\pi}^\text{CPI}(t)$ \\
 & House price index inflation & $\pi^\text{HPI}(t)$ \\
 & Predicted house price index inflation & $\overline{\pi}^\text{HPI}(t)$ \\
 & Rental price index inflation & $\pi^\text{RPI}(t)$ \\
 & Predicted rental price index inflation & $\overline{\pi}^\text{RPI}(t)$ \\
\cmidrule(l){1-3}
\multirow{7}{*}{\STAB{\rotatebox[origin=c]{0}{Credit}}} & Total debt & $L(t)$ \\
 & Total debt of firms operating in sector $s$ & $L_s^\text{L}(t)$ \\
 & Total household debt in consumption loans & $L^\text{C}(t)$ \\
 & Total mortgage debt & $L^\text{M}(t)$ \\
 & NPL ratio of firm loans in sector $s$ & $\nu^\text{F}_s(t)$ \\
 & NPL ratio of household consumption loans & $\nu^\text{C}(t)$ \\
 & NPL ratio of mortgages & $\nu^\text{M}(t)$ \\
\bottomrule
\\
\caption{Aggregate variables and sectoral weights in the model.}
\label{table:aggregate_vars_and_sec_weights}
\end{xltabular}
\normalsize

\paragraph{Production}
The total/sectoral real gross output of domestically producing firms is set to
\begin{align}
    Y(t) ={}& \sum_{f\in \mathcal{F}} Y_f(t) \\
    Y_s(t) ={}& \sum_{f\in \mathcal{F}_s} Y_f(t)
\end{align}
where $\mathcal{F}$ denotes the set of firms, $\mathcal{F}_s$ the set of firms operating in sector $s$, and $Y_f(t)$ real production of firm $f$.

\paragraph{Price}
The \emph{producer price index} $P^\text{PPI}(t)$ and the \emph{consumer price index} $P^\text{CPI}(t)$ are updated as\footnote{From looking at Eqs. \eqref{eq:ppi}-\eqref{eq:cpi}, $P^\text{PPI}(t)$ and $P^\text{CPI}(t)$ look more like weighted price averages than price indexes. However, as explained in \ref{sec:ic_firms}, we set all initial prices to 1, so $P^\text{PPI}(t)$ and $P^\text{CPI}(t)$ can be correctly interpreted as price indexes.}
\begin{align}
    P^\text{PPI}(t) ={}& \frac{\sum_{f\in \mathcal{F}} P_f(t) (Y_f(t) + S_f(t-1)) + \text{IMP}^\text{N}(t)}{\sum_{f\in \mathcal{F}} \left(Y_f(t)+S_f(t-1)\right) + \text{IMP}^\text{R}(t)} \label{eq:ppi}\\
    P^\text{CPI}(t) ={}& \sum_s b_s^\text{CPI} P_s(t) \label{eq:cpi}
\end{align}
where $\mathcal{F}$ is the set of firms, $P_f(t)$ is the price set by firm $f$, $Y_f(t)$ is real output of firm $f$, $S_f(t-1)$ are real finished goods inventories of firm $f$, $\text{IMP}^\text{N}(t)$ / $\text{IMP}^\text{R}(t)$ are total imports in nominal / real terms, $b_s^\text{CPI}$ are aggregate household consumption weights, and $P_s(t)$ denotes the price index for goods produced in sector $s$,
\begin{equation}
    P_s(t) = \frac{\sum_{f\in \mathcal{F}_s} P_f(t) (Y_f(t) + S_f(t-1)) + \text{IMP}^\text{N}_s(t)}{\sum_{f\in \mathcal{F}_s} \left(Y_f(t)+S_f(t-1)\right) + \text{IMP}^\text{R}_s(t)}
\end{equation}
where $\text{IMP}^\text{R}_s(t)$ are imports of sector $s$ in real terms and $\text{IMP}^\text{N}_s(t)$ are imports of sector $s$ in nominal terms.

\paragraph{Housing}
The \emph{house price index} is updated according to
\begin{equation}
    P^\text{HPI}(t) = \frac{\sum_{p\in \mathcal{P}} V_p(t)}{\sum_{p\in \mathcal{P}} V_p(0)}
\end{equation}
where $\mathcal{P}$ is the set of properties and $V_p(t)$ is the value of property $p$ at time $t$. The \emph{rental price index} is updated according to
\begin{equation}
    P^\text{RPI}(t) = \frac{\sum_{p\in \mathcal{P}} r_p(t)}{\sum_{p\in \mathcal{P}} r_p(0)}
\end{equation}
where $\mathcal{P}$ is the set of properties and $r_p(t)$ is the (imputed) rent of property $p$ at time $t$.

\paragraph{Credit}
The total amount of loans granted is computed as
\begin{equation}
    L(t) = \sum_{b\in\mathcal{B}} L_b(t)
\end{equation}
where $\mathcal{B}$ is the set of banks, and $L_b(t)$ is the total amount of loans granted by bank $b$. The total amount of loans granted to firms of sector $s$ is given by
\begin{equation}
    L_s^\text{F}(t) = \sum_{f\in\mathcal{F}_s} L_f(t)
\end{equation}
where $\mathcal{F}_s$ is the set of firms operating in sector $s$, and $L_f(t)$ the debt of firm $f$. The total amount of household consumption loans is given by
\begin{equation}
    L^\text{C}(t) = \sum_{h\in\mathcal{H}} L^\text{C}_h(t)
\end{equation}
where $\mathcal{H}$ is the set of households and $L^\text{C}_h(t)$ the debt of household $h$ in consumption loans. The total amount of mortgage debt is given by
\begin{equation}
    L^\text{M}(t) = \sum_{h\in\mathcal{H}} L^\text{M}_h(t)
\end{equation}
where $L^\text{M}_h(t)$ the mortgage debt of household $h$. 

The ratio of Non-Performing Loans (NPL) to firms in sector $s$ is given by
\begin{equation}
    \nu^\text{F}_s(t) = \frac{\sum_{f\in\mathcal{F}'_s(t)} L_f(t)}{L_s^\text{F}(t)}
\end{equation}
where $\mathcal{F}'_s(t)$ is the set of firms operating in sector $s$ that become insolvent at time $t$. The ratio of non-performing household consumption loans is given by
\begin{equation}
    \nu^\text{C}(t) = \frac{\sum_{h\in\mathcal{H}'(t)} L^\text{C}_h(t)}{L^\text{C}(t)}
\end{equation}
where $\mathcal{H}'(t)$ is the set of insolvent households. Similarly, the ratio of non-performing mortgages is given by
\begin{equation}
    \nu^\text{M}(t) = \frac{\sum_{h\in\mathcal{H}'(t)} L^\text{M}_h(t)}{L^\text{M}(t)}.
\end{equation}

\paragraph{GDP Identity}
GDP in the model can be calculated via the output approach, the expenditure approach, and the income approach:
\begin{equation}
\begin{aligned}
    \text{GDP}(t) ={}& \underbrace{\sum_{s\in\mathcal{S}}\tau_s^\text{PROD}\sum_{f\in\mathcal{F}_s} P_f(t) Y_f(t) + \tau^\text{VAT} \sum_{h\in\mathcal{H}} C_h(t) + \tau^\text{CF} \sum_{h\in\mathcal{H}} K_h(t) + \tau^\text{EXP} \text{EXP}(t)}_\text{Taxes on Products} \\
    & +\underbrace{\sum_{s\in\mathcal{S}}\left(1 - \tau_s^\text{PROD}\right)\sum_{f\in\mathcal{F}_s} P_f(t) Y_f(t)}_\text{Production} - \underbrace{\sum_{s\in\mathcal{S}} \sum_{s'\in\mathcal{S}} \frac{1}{m_{s's}} \sum_{f\in\mathcal{F}_s} P_s(t) Y_f(t)}_\text{Intermediate inputs} \\
    ={}& \underbrace{\sum_{h\in\mathcal{H}} \left(1+\tau^\text{VAT}\right) C_h(t)}_\text{Household consumption} + \underbrace{\sum_{g\in\mathcal{G}} C_g(t)}_\text{Government consumption} + \underbrace{\left(1 + \tau^\text{EXP}\right) \text{EXP}(t)}_\text{Exports} \\
    & - \underbrace{\text{IMP}(t)}_\text{Imports} + \underbrace{\left(1+\tau^\text{CF}\right)\sum_{h\in\mathcal{H}} K_h(t) + \sum_{s\in\mathcal{S}} \sum_{f\in\mathcal{F}_s} P_s(t) K_f(t)}_\text{Gross fixed capital formation} \\
    & + \underbrace{\sum_{s\in\mathcal{S}} \left[\sum_{f\in\mathcal{F}_s} P_s(t) \Delta S_f(t) + \sum_{s'\in\mathcal{S}} P_{s'}(t) \left(\Delta M_{fs'}(t) - \frac{1}{m_{s's}} Y_f(t)\right)\right]}_\text{Changes in stocks and inventories} \\
    ={}& \underbrace{\sum_{s\in\mathcal{S}}\tau_s^\text{PROD}\sum_{f\in\mathcal{F}_s} P_f(t) Y_f(t) + \tau^\text{VAT} \sum_{h\in\mathcal{H}} C_h(t) + \tau^\text{CF} \sum_{h\in\mathcal{H}} K_h(t) + \tau^\text{EXP} \text{EXP}(t)}_\text{Taxes on Products} \\
    & + \underbrace{\sum_{s\in\mathcal{S}}\left(1 - \tau_s^\text{PROD}\right)\sum_{f\in\mathcal{F}_s} P_f(t) Y_f(t) - \sum_{i\in\mathcal{I}^\text{E}(t)} w_i(t) - \sum_{s\in\mathcal{S}}\sum_{f\in\mathcal{F}_s} \sum_{s'\in\mathcal{S}} P_{s'}(t) \frac{1}{m_{s's}} Y_f(t))}_\text{Gross operating surplus and mixed income} \\
    & + \underbrace{\sum_{i\in\mathcal{I}^\text{E}(t)} w_i(t)}_\text{Compensation of employees}
\end{aligned}
\end{equation}
where $\mathcal{F}_s$ is the set of firms operating in sector $s$, $\mathcal{H}$ is the set of households, $\mathcal{G}$ is the set of government entities, and $\mathcal{I}^\text{E}(t)$ is the set of employed individuals at time $t$. In addition, $\tau_s^\text{PROD}$ are tax rates on production in sector $s$, $\tau^\text{VAT}$ is the value-added tax rate on household consumption $C_h(t)$, $\tau^\text{CF}$ is the tax rate on capital formation $K_h(t)$ of households, and $\tau^\text{EXP}$ is the tax rate on exports $\text{EXP}(t)$. Finally, $P_f(t)$ is the price set by firm $f$, $Y_f(t)$ is the real production of firm $f$, $K_f(t)$ is the real capital investment of firm $f$, $\text{IMP}(t)$ are total imports, $P_s(t)$ are average sectoral prices, $\Delta S_f(t)$ changes in real inventories of firm $f$, $\Delta M_{fs}(t)$ changes in real intermediate inputs of sector $s$, $m_{s's}$ denotes the real amount of intermediate inputs of sector $s'$ necessary to produce one real unit of output of sector $s$, and $w_i(t)$ is the wage paid to individual $i$.

\subsection{Expectations}
\label{sec:expectations}
The section summarizes the setting of agent expectations on growth, inflation, and housing prices.

\paragraph{Modeling Expectations}
The standard approach to model expectations in DSGE models is to use rational, or model-consistent, expectations, where the expectation operator coincides with the realized future value of the variables, leading to a fixed point dynamical equation that in most cases must be log-linearized around the steady state to be solved. This approach is hard to consider in an ABM, as we cannot analytically aggregate the expectations of heterogeneous interacting agents.\footnote{Rational expectations are also challenging for DSGE models that incorporate heterogeneity, such as Heterogeneous Agent New Keynesian (HANK) models. See, for instance, \cite{molllecture}.}

When it comes to deciding which expectation rules we should consider, we are faced with a choice between simple and more complicated decision rules. Simple decision rules that are less likely to lead to overfitting are often optimal \citep{artinger2022satisficing}, and we employ a deterministic autoregressive process with lag one for expectation formation. Hence, output stochasticity for our model results from the random search-and-matching processes on the goods, labour, credit, and housing markets. This distinguishes our model from the IIASA model \citep{polednaEconomicForecastingAgentbased2020}, which assumes an autoregressive model with lag one (AR(1)) process for expectation formation.\footnote{Most ABMs in the literature employ linear forecasts; an exception is the model in \cite{seppecherFlexibilityWagesMacroeconomic2012}, where the dynamics of household expectations are given by evolving sentiments in the population. Other approaches are discussed in \cite{braytonRoleExpectationsFRB1997}.} The autoregressive coefficient is recalibrated at every timestep using a mix of real-world data (from 2000-Q1 up to the initialisation time) and simulation output (from the initialisation time until the previous timestep) for the relevant variable. Therefore, the length of the time series used for calibration increases with every timestep. The following details the data we use to inform expectation setting.

\paragraph{Expectations on Growth}
Forecasts for total real gross output $\bar{Y}(t)$ and total real sectoral gross output $\bar{Y}_s(t)$ are used by every agent in updating their decisions. Realised real gross output and total real sectoral gross output enter the autoregressive process in log levels.\footnote{OECD: Quarterly GDP and Components - Output Approach (Code: \emph{DSD\_NAMAIN1@DF\_QNA\_BY\_ACTIVITY\_OUTPUT}).} Predicted growth in real gross output and predicted sectoral real gross output growth are set as
\begin{align}
    \bar{\gamma}(t) ={}& \log\frac{\bar{Y}(t)}{Y(t - 1)} \\
    \bar{\gamma}_s(t) ={}& \log\frac{\bar{Y}_s(t)}{Y_s(t - 1)}
\end{align}
where $Y(t-1)$ is the previous realised total gross output and $Y_s(t-1)$ is the previous realised total gross output of sector $s$.

\paragraph{Expectations on Inflation}
Similarly to growth, every agent uses global PPI / CPI / HPI / RPI forecasts to update their decisions.\footnote{One reason for using global forecasts is that it is unclear whether forecasting inflation improves when considering individual components \citep{romaAggregateNotAggregate2004}.} These forecasts are again obtained using autoregressive models given previous PPI\footnote{IMF: International Financial Statistics (Code: \emph{PPPI\_IX}).}, CPI\footnote{IMF: International Financial Statistics (Code: \emph{PCPI\_IX}).}, HPI/RPI\footnote{OECD: Analytical House Prices Indicators (Code: \emph{DSD\_AN\_HOUSE\_PRICES@DF\_HOUSE\_PRICES})}. Then, predicted inflation is set as
\begin{align}
    \overline{\pi}^\text{PPI}(t) ={}& \log\frac{\overline{P}^\text{PPI}(t)}{P^\text{PPI}(t-1)} \\
    \overline{\pi}^\text{CPI}(t) ={}& \log\frac{\overline{P}^\text{CPI}(t)}{P^\text{CPI}(t-1)} \\
    \overline{\pi}^\text{HPI}(t) ={}& \log\frac{\overline{P}^\text{HPI}(t)}{P^\text{HPI}(t-1)} \\
    \overline{\pi}^\text{RPI}(t) ={}& \log\frac{\overline{P}^\text{RPI}(t)}{P^\text{RPI}(t-1)}
\end{align}
where $\overline{P}^\text{PPI}(t)$ is predicted PPI, $P^\text{PPI}(t-1)$ is previous PPI, $\overline{P}^\text{CPI}(t)$ is predicted CPI, $P^\text{CPI}(t-1)$ is previous CPI, $\overline{P}^\text{HPI}(t)$ is predicted HPI, $P^\text{HPI}(t-1)$ is previous HPI, $\overline{P}^\text{RPI}(t)$ is predicted RPI, $P^\text{RPI}(t-1)$ is previous RPI.

\subsection{Banks}
In our model, banks set interest rates, hold central bank reserves, household and firm deposits, and grant loans to firms (long-term and short-term) and households (consumption loans and mortgages).\footnote{Different models mainly vary in terms of lending conditions. In some models, all credit demands are satisfied \cite{seppecherFlexibilityWagesMacroeconomic2012, mandelAgentbasedDynamicsDisaggregated2010}, others introduce probabilities of loan approvals \cite{ashrafHowInflationAffects2016,assenzaEmergentDynamicsMacroeconomic2015}, or assume (regulatory) upper bounds on the volume of credit a bank can grant \cite{dawidEconomicConvergencePolicy2014,dosiSchumpeterMeetingKeynes2010,polednaEconomicForecastingAgentbased2020}.} When firms they lent to go bankrupt, they take losses. For simplicity, we do not explicitly model persons employed by banks. We also assume that banks do not have investment strategies in derivatives and real assets. Banks may fail and will be bailed in.\footnote{In \cite{assenzaEmergentDynamicsMacroeconomic2015,ashrafHowInflationAffects2016}, banks will always be bailed out by the central government, in \cite{dawidEconomicConvergencePolicy2014} by the central bank. In \cite{seppecherFlexibilityWagesMacroeconomic2012}, the single bank's bankruptcy ends the simulation.}

Table \ref{table:bank_variables} shows variables and parameters related to banks, and Table \ref{table:loan_variables} shows variables and parameters related to loans (long/short-term firm loans, consumption loans, mortgages) granted by banks.

\tablefontsize
\begin{xltabular}{\textwidth}{cXl}
\toprule
Category & Description & Notation \\
\cmidrule(l){1-3}
\multirow{5}{*}{\STAB{\rotatebox[origin=c]{0}{Sets}}} & Set of banks & $\mathcal{B}$ \\
 & Set of insolvent banks & $\mathcal{B}'(t)$ \\
 & Loans provided & $\mathcal{L}_b(t)$ \\
 & Firms with deposits at the bank & $\mathcal{F}_b(t)$ \\
 & Households with deposits at the bank & $\mathcal{H}_b(t)$ \\
\cmidrule(l){1-3}
\multirow{4}{*}{\STAB{\rotatebox[origin=c]{0}{Assets}}} & Reserves & $R_b(t)$ \\
 & Total loans granted to firms of sector $s$ & $V_{bs}^\text{F}(t)$ \\
 & Total consumption loans granted to households & $V_b^\text{C}(t)$ \\
 & Total mortgages granted to households & $V_b^\text{M}(t)$ \\
\cmidrule(l){1-3}
\multirow{2}{*}{\STAB{\rotatebox[origin=c]{0}{Liabilities}}} & Liability & $L_b(t)$ \\
 & Equity & $E_b(t)$ \\
\cmidrule(l){1-3}
\multirow{1}{*}{\STAB{\rotatebox[origin=c]{0}{P\&L Account}}} & Profits & $\Pi_b(t)$ \\
\cmidrule(l){1-3}
\multirow{5}{*}{\STAB{\rotatebox[origin=c]{0}{Credit Supply}}} & Maximum credit willing to grant & $V_b^\text{max}(t)$ \\
 & Supply of credit to firms of sector $s$ & $\hat{V}_{bs}^\text{F}(t)$ \\
 & Supply of credit to households for consumption & $\hat{V}_b^\text{C}(t)$ \\
 & Supply of mortgage credit to households & $\hat{V}_b^\text{M}(t)$ \\
 & Influence of the NPL ratio when allocating credit supply & $\phi^\text{CS}$ \\
\cmidrule(l){1-3}
\multirow{2}{*}{\STAB{\rotatebox[origin=c]{0}{Interest Rates}}} & Interest rate on firm deposit overdrafts & $r_b^\text{F-O}(t)$ \\
 & Interest rate on household deposit overdrafts & $r_b^\text{H-O}(t)$ \\
\cmidrule(l){1-3}
\multirow{10}{*}{\STAB{\rotatebox[origin=c]{0}{\makecell{Regulatory\\Requirements}}}} & Bank capital adequacy ratio & $\rho^\text{CAR}$ \\
 & Solvency ratio between equity and assets & $\rho^\text{SR}$ \\
 & Firm loan debt-to-equity ratio & $\rho^\text{DtE}$ \\
 & Firm loan return-on-equity ratio & $\rho^\text{RoE}$ \\
 & Firm loan return-on-assets ratio & $\rho^\text{RoA}$ \\
 & Household consumption loan-to-income ratio & $\rho^\text{LTI-C}$ \\
 & Mortgage loan-to-value ratio & $\rho^\text{LTV}$ \\
 & Mortgage loan-to-income ratio & $\rho^\text{LTI-M}$ \\
 & Mortgage debt-service-to-income ratio & $\rho^\text{DSTI}$ \\
 & Equity injections for other insolvent banks & $I_b(t)$ \\
\bottomrule
\\
\caption{Variables and parameters in the model related to banks.}
\label{table:bank_variables}
\end{xltabular}
\normalsize

\tablefontsize
\begin{xltabular}{\textwidth}{ccXl}
\toprule
Agent & Category & Description & Notation \\
\cmidrule(l){1-4}
\multirow{3}{*}{\STAB{\rotatebox[origin=c]{0}{Loan $l$}}} & \multirow{3}{*}{\STAB{\rotatebox[origin=c]{0}{Attributes}}} & Amount & $V_l(t)$ \\
 & & Interest rate & $r_l$ \\
 & & Maturity & $m_l$ \\
\bottomrule
\\
\caption{Variables and parameters in the model related to loans.}
\label{table:loan_variables}
\end{xltabular}
\normalsize

\subsubsection{Initial Conditions}
In this section, we briefly summarize the initial conditions for banks.

\paragraph{Drawing Banks from Compustat Data}
Banks are sampled with replacement from Compustat data\footnote{CRSP/Compustat Merged Database - Bank Quarterly (Code: \emph{crsp\_a\_ccm}).} so that the total number of banks matches IMF aggregates\footnote{IMF: Financial Access Survey (Code: \emph{FAS}).}. This includes total deposits (Code: \emph{dptcq}) and total liabilities (Code: \emph{ltq}).

\paragraph{Initial Profits}
Initial profits of bank $b$ (see Eq. \eqref{eq:bank_profits}) are set according to
\begin{equation}
\begin{aligned}
    \Pi_b(0) ={}& \underbrace{\sum_{l\in \mathcal{L}_b(0)} r_l V_l(0)}_\text{Interest received on loans} + \underbrace{r_b^\text{F-O}(0) \sum_{f\in\mathcal{F}_b(0)} \left[D_f(0)\right]^- + r_b^\text{H-O}(0) \sum_{h\in\mathcal{H}_b(0)} \left[D_h(0)\right]^-}_\text{Interest received on overdrafts} + \underbrace{r(0) \left[R_b(0)\right]^+}_\text{Interest received on reserves} \\
    & - \underbrace{r(0) \left(\sum_{f\in \mathcal{F}_b} \left[D_f(0)\right]^+ + \sum_{h\in \mathcal{H}_b} \left[D_h(0)\right]^+\right)}_\text{Interest paid on deposits} - \underbrace{r(0) \left[R_b(0)\right]^-}_\text{Interest paid on reserves}
\end{aligned}
\end{equation}
which is the interest on granted loans to firms or households plus the interest received on overdrafts plus the interest received on reserves (if positive) minus the interest paid on deposits or reserves (if negative). In the equation above, $r_l$ is the interest rate on loan $l$, $V_l$ its amount, $r_b^\text{F-0}(0)$ / $r_b^\text{H-0}(0)$ are the initial overdraft rates on firm/household deposits, and $r(0)$ is the initial central bank policy rate.

\paragraph{Initial Reserves}
Initial central bank reserves (see Eq. \eqref{eq:bank_reserves}) of bank $b$ are set according to
\begin{equation}
    R_b(0) = \underbrace{\sum_{f\in \mathcal{F}_b(0)} D_f(0)}_\text{Firm deposits} + \underbrace{\sum_{h\in \mathcal{H}_b(0)} D_h(0)}_\text{Household deposits} + \underbrace{E_b(0)}_\text{Bank equity} - \underbrace{\sum_{l\in \mathcal{L}_b(0)} V_l(0)}_\text{Loans granted}
\end{equation}
which is total deposits and equity at the bank minus total loans granted to firms or households. Here $\mathcal{F}_b(0)$/$\mathcal{H}_b(0)$ denote the initial set of firms/households with deposits at bank $b$, and $\mathcal{L}_b(0)$ the total initial set of loans granted by the bank.

\paragraph{Initial Equity}
Initial bank equity (See Eq. \ref{eq:bank_equity}) is set proportionally to the total amount of loans granted and rescaled so that the sum of equity of all banks matches OECD aggregates\footnote{OECD: Quarterly Financial Balance Sheets (Code: DSD\_NASEC20@DF\_T720R\_Q), non-consolidated equity (Code: \emph{F5}) and financial net worth (Code: \emph{BF90}) of monetary financial institutions other than the central bank (Code: \emph{S12T}).}.

\paragraph{Initial Matching with Firms}
Banks are initially matched with firms by solving a linear sum assignment problem so that the sum of the differences in total firm deposits plus debt to total bank deposits to firms and debt in short-term and long-term firm loans is minimal. In other words, at each model initialisation, we create a bipartite graph between banks and firms that matches their respective initial states the closest.

\paragraph{Initial Matching with Households}
Banks are initially matched with households by solving a linear sum assignment problem so that the sum of the differences in total household deposits plus debt to total bank deposits to households and debt in household consumption loans and mortgages is minimal.

\subsubsection{Parameters}
Bank parameters are regulatory requirements and loan maturities.

\paragraph{Regulatory Requirements}
We assume a bank's capital adequacy ratio to be $\rho^\text{CAR}=0.08$, corresponding to Basel III, and a solvency ratio between equity and assets to be $\rho^\text{SR}=0.1$. For firm loans, we assume the firm loan debt-to-equity ratio to be $\rho^\text{DtE}=1.0$, the firm loan return-on-equity ratio to be $\rho^\text{RoE}=0.15$, and the firm loan return-on-assets ratio to be $\rho^\text{RoA}=0.05$. For household consumption loans, we assume the household consumption loan to income ratio to be $\rho^\text{LTI-C}=0.36$. For mortgages ($\rho^\text{LTV}$, $\rho^\text{LTI-M}$, $\rho^\text{DSTI}$), we follow the \emph{Overview of national macroprudential measures}\footnote{\url{https://www.esrb.europa.eu/national_policy/html/index.en.html}} compiled by the \emph{European Systemic Risk Board}.


\paragraph{Loan Maturities}
Firm short-term loans are assumed to mature in one quarter, long-term loans in two years, household consumption loans in one quarter, and mortgages in 25 years. This follows standard maturity limits set by central banks and compiled by the European Systemic Risk Board (ESRB) in their \emph{Overview of national macroprudential measures}.

\paragraph{Credit Supply}
The influence of the NPL ratio when allocating credit supply $\phi^\text{CS}$ is set to be $2.0$. The investigation of the effect of this parameter is out of the scope of this paper.

\subsubsection{Rules}
In this section, we discuss the behavioural rules for each bank, most notably the supply of credit, setting interest rates, and lending requirements.

\paragraph{Interest Rates Setting} 
Bank interest rates are set using a single equation error correction model derived from an Autoregressive Distributed Lags (ARDL) model.\footnote{For more extensive models, see \cite{leroyStructural2016}.}$^{,}$\footnote{The models in \cite{polednaEconomicForecastingAgentbased2020} and \cite{hommesCANVAS2022} assume fixed markups on the policy rate. In \cite{assenzaEmergentDynamicsMacroeconomic2015}, the interest rate increases with the borrower's leverage; in \cite{seppecherFlexibilityWagesMacroeconomic2012}, the interest rate is fixed.} We estimate a time-series model specified as
\begin{equation}
\label{eq:bank_ir}
\begin{aligned}
    \Delta r_l(t) ={}& \underbrace{\phi^\text{EC}\left(r_l(t-1) - \phi^\text{LR} r(t)\right)}_\text{Error correction mechanism} + \underbrace{\sum_{j=1}^{p-1} \alpha_j \Delta r_l(t-j)}_\text{Lagged interest rate} + \underbrace{\sum_{j=0}^{q-1} \beta_j \Delta r(t-j)}_\text{Lagged policy rate} \\
    & + \underbrace{\sum_{j=0}^{r-1} \gamma_j \Delta \pi^\text{PPI}(t-j)}_\text{Lagged inflation} + \underbrace{\sum_{j=0}^{s-1} \delta_j \Delta \nu(t-j)}_\text{Lagged NPL ratio} + \mu + \varepsilon_t
\end{aligned}
\end{equation}
for each type of loan $l$. The first term is the error correction mechanism, $\phi^\text{LR}$ is the long-run pass-through between the central bank policy rate $r(t)$\footnote{Obtained from the BIS central bank policy rates data set.} and the interest rate $r_l(t)$\footnote{ECB: Corporates' total loans (Code: \emph{MIR.M.U2.B.A2A.A.R.A.2240.EUR.N}), Household loans for consumption (Code: \emph{MIR.M.U2.B.A2B.A.R.A.2250.EUR.N}), Household loans for house purchase (Code: \emph{MIR.M.U2.B.A2C.A.R.A.2250.EUR.N}).}, $\phi^\text{EC}$ is the error-correction term. We additionally consider PPI inflation\footnote{IMF: International Financial Statistics (Code: \emph{PPPI\_IX}).} and the ratio of non-performing loans\footnote{World Bank Data: bank non-performing loans to gross loans (Code: \emph{GFDD.SI.02}).} ($\nu_s(t)$ for loans to firms in sector $s$, $\nu^\text{C}(t)$ for household consumption loans, $\nu^\text{M}(t)$ for mortgages), $\alpha_j$, $\beta_j$, $\gamma_j$, $\delta_j$ are coefficients, and the choice of lag structure $p$, $q$, $r$, $s$ is based on Akaike’s Information Criterion. These parameters are estimated using historic real data at model initialisation.

We further assume that the interest rate on firm or household deposits is equal to the policy rate, that the household deposits overdraft rate $r_b^\text{H-O}(t)$ is equal to the interest rate on household consumption loans, and that the firm deposits overdraft rate $r_b^\text{F-O}(t)$ is equal to the interest rate on short-term firm loans.

\paragraph{Firm Loans Lending Requirements}
For short- or long-term firm loans, banks consider three ratios limiting the maximum amount of a new loan:
\begin{enumerate}
\item \textbf{Debt to Equity}: The credit provided to a firm $f$ cannot exceed
\begin{equation}
    V_l(t) \leq \rho^\text{DtE} \underbrace{\sum_{s\in \mathcal{S}} P_s(t) K_{fs}(t)}_\text{Capital stock value} - \underbrace{L_f(t-1)}_\text{Loans} + \underbrace{\left[D_f(t-1)\right]^-}_\text{Overdrafts} + \underbrace{r_b^\text{F-O}(t) \left[D_f(t-1)\right]^- - \sum_{l\in\mathcal{L}_f(t)} r_l V_l(t)}_\text{Interest payments}
\end{equation}
where $\rho^\text{DtE}$ is a parameter, $\sum_{s\in \mathcal{S}} P_s(t) K_{fs}(t)$ is the current value of the capital stock of firm $f$, $L_f(t)$ is the current debt of the firm, $D_f(t)$ are firm deposits at its bank, $r_b^\text{F-O}(t)$ are firm overdraft rates, $\mathcal{L}_f(t)$ is the set of loans of firm $f$, $r_l$ is the interest rate of loan $l$, and $V_l(t)$ is the amount of loan $l$.
\item \textbf{Return on Equity}: The credit provided to a firm $f$ cannot exceed
\begin{equation}
    V_l(t) \leq \underbrace{\sum_{s\in \mathcal{S}} P_s(t) K_{fs}(t)}_\text{Capital stock value} + \underbrace{D_f(t-1)}_\text{Firm deposits} - \underbrace{L_f(t-1)}_\text{Firm debt} - \frac{\overline{\Pi}_f(t)}{\rho^\text{RoE}}
\end{equation}
where $\rho^\text{RoE}$ is a parameter and $\overline{\Pi}_f(t)$ are predicted profits of firm $f$.
\item \textbf{Return on Assets}: Credit is only provided to a firm $f$ if
\begin{equation}
    \frac{\overline{\Pi}_f(t)}{L_f(t-1) + E_f(t-1)} \geq \rho^\text{RoA}
\end{equation}
where $\rho^\text{RoA}$ is a parameter.
\end{enumerate}

\paragraph{Household Consumption Loans Lending Requirement}
Banks perform a household risk assessment based on their average income over the last six months (corresponding to the previous two timesteps). The credit provided to a household $h$ cannot exceed
\begin{equation}
    V_l(t) \leq \rho^\text{LTI-C} \underbrace{\frac{1}{2}\left(Y_h(t-2) + Y_h(t-1) \right)}_\text{Average income over the last 6 months} - \underbrace{L_h(t-1)}_\text{Household debt}
\end{equation}
where $\rho^\text{LTI-C}$ is a parameter.

\paragraph{Mortgage Lending Requirements}
Each bank has its own policies for extending mortgages to households. Generally, banks consider three ratios:
\begin{enumerate}
    \item \textbf{Loan-to-Value}: Bank $b$ sets a maximum value for the principal the household can borrow as a function of the down payment to be made. Since the down payment is the household's full financial wealth, the maximum loan value is restricted by
    \begin{equation}
        V_l(t) \leq \frac{\rho^\text{LTV}}{1 - \rho^\text{LTV}} \underbrace{W_h^\text{FA}(t)}_\text{Wealth in financial assets}
    \end{equation}
    where $\rho^\text{LTV}$ is a parameter.
    \item \textbf{Loan-to-Income}: The maximum loan value is restricted by
    \begin{equation}
        V_l(t) \leq \rho^\text{LTI-M} \underbrace{\frac{1}{2}\left(Y_h(t-2) + Y_h(t-1) \right)}_\text{Average income over the last 6 months} - \underbrace{L_h(t-1)}_\text{Household debt}
    \end{equation}
    where $\rho^\text{LTI-M}$ is a parameter.
    \item \textbf{Debt-Service-to-Income}: The maximum loan value is restricted by
    \begin{equation}
        V_l(t) \leq \rho^\text{DSTI} \underbrace{\frac{1}{2}\left(Y_h(t-2) + Y_h(t-1) \right)}_\text{Average income over the last 6 months} \frac{1 - (1 + r_l)^{-m_l}}{r_l}
    \end{equation}
    where $\rho^\text{DSTI}$ is a parameter, $r_l$ is the mortgage rate offered by the bank, and $m_l$ is the maturity of the mortgage.
\end{enumerate}

\paragraph{Supply-Side Credit Constraints}
Each bank $b$ also has supply-side constraints. The bank cannot grant loans above
\begin{equation}
    V_b^\text{max}(t) = \frac{E_b(t-1)}{\rho^\text{CAR}} - \underbrace{\sum_{l\in \mathcal{L}_b(t)} V_l(t)}_\text{Loans granted}
\end{equation}
where $E_b(t-1)$ is previous bank equity, and $\rho^\text{CAR}$ is a capital adequacy ratio. Each bank distributes the total amount of loans they're willing to grant among firm loans, household consumption loans, and mortgages based on the corresponding ratios of previous non-performing loans, specifically so that
\begin{align}
    \hat{V}_{bs}^\text{F}(t) \propto{}& V_{bs}^\text{F}(0) \exp\left(-\phi^\text{CS} \nu_s^\text{F}(t-1) \right) \\
    \hat{V}_b^\text{C}(t) \propto{}& V_b^\text{C}(0) \exp\left(-\phi^\text{CS} \nu^\text{C}(t-1) \right) \\
    \hat{V}_b^\text{M}(t) \propto{}& V_b^\text{M}(0) \exp\left(-\phi^\text{CS} \nu^\text{M}(t-1) \right) \\
    V_b^\text{max}(t) ={}& \sum_{s\in \mathcal{S}}\hat{V}_{bs}^\text{F}(t) + \hat{V}_b^\text{C}(t) + \hat{V}_b^\text{M}(t)
\end{align}
where $V_{bs}^\text{F}(0)$, $V_b^\text{C}(0)$, $V_b^\text{M}(0)$ are the initial amounts of credit granted by bank $b$ to firms of sector $s$,
\begin{equation}
    V_{bs}^\text{F}(0) = \sum_{f\in \mathcal{F}_b(0)\cap \mathcal{F}_s} L_{fb}(0),
\end{equation}
where $L_{fb}(0)$ are initial loans from bank $b$ to firm $f$; households for consumption,
\begin{equation}
    V_b^\text{C}(0) = \sum_{h\in \mathcal{H}_b(0)} L_{hb}^\text{C}(0),
\end{equation}
where $L_{hb}^\text{C}(0)$ are initial consumption loans from bank $b$ to household $h$; and mortgages
\begin{equation}
    V_b^\text{M}(0) = \sum_{h\in \mathcal{H}_b(0)} L_{hb}^\text{M}(0).
\end{equation}
where $L_{hb}^\text{M}(0)$ are initial mortgage loans from bank $b$ to household $h$. The parameter $\phi^\text{CS}$ governs the influence of the ratio of non-performing ratios in allocating the total supply of credit.

\paragraph{Profits}
Profits of bank $b$ are updated according to
\begin{equation}
\label{eq:bank_profits}
\begin{aligned}
\Pi_b(t) ={}& \underbrace{\sum_{l\in \mathcal{L}_b(t)} r_l V_l}_\text{Interest received on loans} + \underbrace{r_b^\text{F-O}(t) \sum_{f\in\mathcal{F}_b(t)} \left[D_f(t)\right]^- + r_b^\text{H-O}(t) \sum_{h\in\mathcal{H}_b(t)} \left[D_h(t)\right]^-}_\text{Interest received on overdrafts} + \underbrace{r(t) \left[R_b(t)\right]^+}_\text{Interest received on reserves} \\
& - \underbrace{r(t) \left(\sum_{f\in \mathcal{F}_b} \left[D_f(t)\right]^+ + \sum_{h\in \mathcal{H}_b} \left[D_h(t)\right]^+\right)}_\text{Interest paid on deposits} - \underbrace{r(t) \left[R_b(t)\right]^-}_\text{Interest paid on reserves} - \underbrace{I_b(t)}_\text{Equity injection}
\end{aligned}
\end{equation}
which is the interest on granted loans to firms or households plus the interest received on overdrafts plus the interest received on reserves (if positive) minus the interest paid on deposits or reserves (if negative). In the equation above, $r_l$ is the interest rate on loan $l$, $V_l$ its amount, $r_b^\text{F-0}(t)$/$r_b^\text{H-0}(t)$ are the overdraft rates on firm/household deposits, and $r(t)$ is the central bank policy rate. The equity injection $I_b(t)$ is for bailing in other insolvent banks.

\paragraph{Equity}
Bank equity is updated according to
\begin{equation}
\label{eq:bank_equity}
\begin{aligned}
E_b(t) ={}& \underbrace{E_b(t-1)}_\text{Previous bank equity} + \underbrace{\Pi_b(t)}_\text{Bank profits} - \underbrace{\tau^\text{CORP}\left[\Pi_b(t)\right]^+}_\text{Corporate Income Taxes} \\
& + \underbrace{\sum_{f\in \mathcal{F}_b(t)\cap\mathcal{F}'(t)} D_f(t) - \sum_{f\in \mathcal{F}'(t)} L_{fb}(t) + \sum_{h\in \mathcal{H}_b(t)\cap\mathcal{H}'(t)} W^\text{D}_h(t) - \sum_{h\in \mathcal{H}'(t)} L_{hb}(t)}_\text{Write-off of bad debt}
\end{aligned}
\end{equation}
where $\mathcal{F}'(t)$ / $\mathcal{H}'(t)$ is the set of insolvent firms / households, $L_{fb}(t)$ / $L_{hb}(t)$ are loans granted by bank $b$ to firm $f$ / household $h$, $D_f(t)$ are deposits of firm $f$, and $W^\text{D}_h(t)$ are current household deposits. Whenever a firm or a household goes bankrupt, its bank appropriates what is left as a deposit but loses all the loan (i.e. cannot resell the firm assets).

\paragraph{Liabilities} 
The liabilities of bank $b$ are updated as
\begin{equation}
\label{eq:bank_liabilities}
    L_b(t) = \underbrace{E_b(t)}_\text{Bank equity} + \underbrace{\sum_{f\in \mathcal{F}_b(t)} \left[D_f(t)\right]^+ + \sum_{h\in \mathcal{H}_b(t)} \left[ D_h(t)\right]^+}_\text{Deposits} - \underbrace{\left[R_b(t-1)\right]^-}_\text{Reserves}
\end{equation}
where $\mathcal{F}_b(t)$/$\mathcal{H}_b(t)$ is the set of firms/households that have their deposits at bank $b$.

\paragraph{Reserves}
The reserves of bank $b$ at the central bank are obtained by
\begin{equation}
\label{eq:bank_reserves}
    R_b(t) = \underbrace{\sum_{f\in \mathcal{F}_b(t)} D_f(t) + \sum_{h\in \mathcal{H}_b(t)} D_h(t)}_\text{Deposits} +  \underbrace{E_b(t)}_\text{Bank equity} - \underbrace{\sum_{l\in \mathcal{L}_b(t)} V_l(t)}_\text{Loans granted}
\end{equation}
where $\mathcal{F}_b(t)$/$\mathcal{H}_b(t)$ is the set of firms/households that have their deposits at bank $b$.

\paragraph{Insolvency}
If a bank has a low solvency ratio, it goes bankrupt. Specifically, if its ratio of equity to assets falls below a certain threshold
\begin{equation}
    \frac{E_b(t)}{L_b(t) + \left[R_b(t)\right]^+} < \rho^\text{SR}
\end{equation}
then the bank is \emph{bailed-in} by all other non-insolvent banks cancelling a fixed fraction of their debt until the equity of the insolvent bank is equal to the average equity of non-insolvent banks.

\subsection{Central Bank}
The central bank provides banks with liquidity and sets the central bank interest rate\footnote{Usually this is done using different variations of the Taylor rule, see the original paper \cite{taylorDiscretionPolicyRules1993}. In the models in \cite{mandelAgentbasedDynamicsDisaggregated2010} and \cite{ashrafHowInflationAffects2016}, the central bank sets the interest rate using a Taylor rule based on inflation and unemployment gaps. In the model in \cite{polednaEconomicForecastingAgentbased2020}, the central bank uses a version of the Taylor rule following \cite{blattnerRobustMonetaryPolicy2010}, which does not include an output gap.}.\footnote{In \cite{dawidEconomicConvergencePolicy2014} the central bank may perform quantitative easing.} Table \ref{table:central_bank_variables} shows variables and parameters related to the central bank.

\tablefontsize
\begin{xltabular}{\textwidth}{cXl}
\toprule
Category & Description & Notation \\
\cmidrule(l){1-3}
 \multirow{6}{*}{Policy Rate} & Policy Rate & $r(t)$ \\
 & Target CPI inflation & $\pi^\star$ \\
 & Autoregressive parameter & $\rho$ \\
 & Real equilibrium interest rate & $r^\star$ \\
 & Weight on inflation targeting & $\xi^\pi$ \\
 & Weight on economic growth & $\xi^\gamma$ \\ 
\bottomrule
\\
\caption{Variables and parameters in the model related to the central bank.}
\label{table:central_bank_variables}
\end{xltabular}
\normalsize

\subsubsection{Parameters}
We assume target CPI inflation to be $\pi^\star=0.02$.

\subsubsection{Rules}

\paragraph{Policy Rate}
Following \cite{poledna2023economic}, a Taylor rule models the policy rate according to
\begin{equation}
\label{eq:cb_policy_rate}
\begin{aligned}
    \bar{r}(t) = \left[\rho \bar{r}(t-1) + (1-\rho) \left(r^\star+\pi^\star+\xi^\pi(\pi^\text{CPI}(t)-\pi^\star)+\xi^\gamma \gamma(t)\right)\right]^+
\end{aligned}
\end{equation}
where $\rho$ is a measure for gradual adjustment of the policy rate, $r^\star$ is the real equilibrium interest rate, $\pi^\star$ is the inflation target by the central bank, $\xi^\pi$ is the weight the central bank puts on inflation targeting, and $\xi^\gamma$ the weight placed on economic growth. The parameters used in the Taylor rule are estimated at model initialisation using the central bank policy rate $r(t)$\footnote{Obtained from the BIS central bank policy rates data set.}, CPI inflation $\pi^\text{CPI}(t)$\footnote{IMF: International Financial Statistics (Code: \emph{PCPI\_IX}).}, and total growth rates $\gamma(t)$\footnote{IMF: International Financial Statistics (Code: \emph{NGDP\_SA\_XDC}).}. The OLS regression is given by
\begin{equation}
    r(t) = \alpha + \rho r(t-1) + \beta_{\pi} \left(\pi^\text{CPI}(t) - \pi^\star\right) + \beta_{\gamma}\gamma(t) + \varepsilon_t
\end{equation}
where $\alpha$ is the intercept, $\rho$, $\beta_{\pi}$ and $\beta_{\gamma}$ are coefficients to be estimated and $\varepsilon_t$ is an error term. From the Taylor rule equation, we obtain
\begin{align}
    r^\star ={}& \frac{\alpha}{1-\rho} - \pi^\star \\
    \xi^\pi ={}& \frac{\beta_\pi}{1-\rho} \\
    \xi^\gamma ={}& \frac{\beta_\gamma}{1-\rho}
\end{align}

\subsection{Firms}
Macroeconomic agent-based models typically distinguish between consumption good firms, which use labour and capital and sell to households, and capital good firms, which use labour and sell to consumption good firms.\footnote{Exceptions are the models in \cite{ashrafHowInflationAffects2016,polednaEconomicForecastingAgentbased2020}, where firms produce different types of goods, but there is no differentiation between consumption goods and capital, and the models in \cite{seppecherFlexibilityWagesMacroeconomic2012,lengnickAgentbasedMacroeconomicsBaseline2013}, where there is only one type of (consumption good) firm that uses labour only.} In our model, we make no such formal distinction. Firms set the quantity and the price of the goods they produce using labour, intermediate inputs, and capital inputs. Quantities are set by standard inventory planning, with prices as a markup on production costs. Firms also set the wages they pay to employees. Wages are based on whether a firm can fill all their vacancies or not\footnote{In \cite{lengnickAgentbasedMacroeconomicsBaseline2013} wages follow a stochastic process, in several models \cite{polednaEconomicForecastingAgentbased2020, dawidEconomicConvergencePolicy2014,dosiSchumpeterMeetingKeynes2010, mandelAgentbasedDynamicsDisaggregated2010} wages are also adjusted according to the evolution of labour productivity.}. Firms may have a financing gap and will ask for short-term or long-term loans from the bank to fill this gap. Firms may become bankrupt and leave the market; for simplicity, a new firm will immediately replace the old firm\footnote{The model in \cite{ashrafHowInflationAffects2016} explicitly describes market entry and the associated decisions a firm needs to make.}.

Table \ref{table:firm_variables} shows variables and parameters related to firms.

\tablefontsize
\begin{xltabular}{\textwidth}{cXl}
\toprule
Category & Description & Notation \\
\cmidrule(l){1-3}
\multirow{6}{*}{\STAB{\rotatebox[origin=c]{0}{Sets}}} & Set of firms & $\mathcal{F}$ \\
 & Set of insolvent firms & $\mathcal{F}'(t)$ \\
 & Set of firms operating in sector $s$ & $\mathcal{F}_s$ \\
 & Set of insolvent firms operating in sector $s$ & $\mathcal{F}'_s(t)$ \\
 & Individuals employed by the firm & $\mathcal{I}_f(t)$ \\
 & Loans taken out by the firm & $\mathcal{L}_f(t)$ \\
\cmidrule(l){1-3}
\multirow{10}{*}{\STAB{\rotatebox[origin=c]{0}{Production}}} & Real production & $Y_f(t)$ \\
 & Target real production & $\hat{Y}_f(t)$ \\
 & Real production limited by labour inputs & $H_f(t)$ \\
 & Real production limited by intermediate inputs & $M_f(t)$ \\
 & Real production limited by capital inputs & $K_f(t)$ \\
 & Predicted growth & $\overline{\gamma}_f(t)$ \\
 & Target inventory to production fraction & $\phi^\text{StY}$ \\
 & Influence of labour inputs in limiting target production & $\chi^\text{H}$ \\
 & Influence of intermediate inputs in limiting target production & $\chi^\text{M}$ \\
 & Influence of capital inputs in limiting target production & $\chi^\text{K}$ \\
\cmidrule(l){1-3}
\multirow{4}{*}{\STAB{\rotatebox[origin=c]{0}{Demand}}} & Real demand & $Q_f(t)$ \\
 & Real demand & $\overline{Q}_f(t)$ \\
 & Realised real demand & $\tilde{Q}_f(t)$ \\
 & Demand adjustment speed on firm growth & $\phi^Q_F$ \\
\cmidrule(l){1-3}
\multirow{17}{*}{\STAB{\rotatebox[origin=c]{0}{\makecell{Inventory\\and Stocks}}}} & Real inventory & $S_f(t)$ \\
 & Sectoral depreciation rates of inventory & $\delta_s$ \\
 & Real stock of intermediate inputs & $M_{fs}(t)$ \\
 & Intermediate inputs utilisation rate & $\omega^\text{M}$ \\
 & Target real intermediate inputs before financial frictions & $\hat{M}_{fs}^\text{FF}(t)$ \\
 & Influence of the financial situation on target intermediate inputs & $\phi^\text{FM}$ \\
 & Target real intermediate inputs & $ \hat{M}_{fs}(t)$ \\
 & Realised real intermediate inputs & $ \tilde{M}_{fs}(t)$ \\
 & Influence of existing stock in setting target intermediate inputs & $\phi^M$ \\
 & Real stock of capital inputs & $K_{fs}(t)$ \\
 & Capital inputs utilisation rate & $\omega^\text{K}$ \\
 & Target real capital inputs before financial frictions & $\hat{K}_{fs}^\text{FF}(t)$ \\
 & Influence of the financial situation on target capital inputs & $\phi^\text{FK}$ \\
 & Target real capital inputs & $\hat{K}_{fs}(t)$ \\
 & Realised real capital inputs & $\tilde{K}_{fs}(t)$ \\
 & Influence of existing stock in setting target capital inputs & $\phi^K$ \\
 & Delay in capital acquisition & $T^\text{KD}_s$ \\
\cmidrule(l){1-3}
 \multirow{8}{*}{\STAB{\rotatebox[origin=c]{0}{Labour}}} & Labour inputs & $H_f(t)$ \\
 & Target labour inputs & $\hat{H}_f(t)$ \\
 & Labour productivity factor & $h_f(t)$ \\
 & Maximum increase in work effort & $h^\text{max}$ \\
 & Total wages paid & $w_f(t)$ \\
 & Markup factor on new employee wages & $\mu^\text{WN}_f$ \\
 & Adjustment of the average wage for new employees & $\phi^\text{WN}$ \\
 & Time window for wage markups & $T^\text{WN}$ \\
\cmidrule(l){1-3}
\multirow{6}{*}{\STAB{\rotatebox[origin=c]{0}{Price}}} & Price & $P_f(t)$ \\
 & Unit costs & $U_f(t)$ \\
 & Demand-pull inflation & $\overline{\pi}_f^\text{DP}(t)$ \\
 & Cost-push inflation & $\overline{\pi}_f^\text{CP}(t)$ \\
 & Influence of demand-pull inflation when setting prices & $\phi^\text{DP}$ \\
 & Influence of cost-push inflation when setting prices & $\phi^\text{CP}$ \\
\cmidrule(l){1-3}
\multirow{12}{*}{\STAB{\rotatebox[origin=c]{0}{Financials}}} & Deposits & $D_f(t)$ \\
 & Predicted change in deposits & $\overline{D}^\Delta_f(t)$ \\
 & Profits & $\Pi_f(t)$ \\
 & Predicted profits & $\overline{\Pi}_f(t)$ \\
 & Production costs & $C_f(t)$ \\
 & Equity & $E_f(t)$ \\
 & Debt & $L_f(t)$ \\
 & Target short-term loans & $\hat{L}^\text{S}_f(t)$ \\
 & Target long-term loans & $\hat{L}^\text{L}_f(t)$ \\
 & Acquired short-term loans & $L^\text{S}_f(t)$ \\
 & Acquired long-term loans & $L^\text{L}_f(t)$  \\
 & Total loans from bank $b$ & $L_{fb}(t)$ \\
\cmidrule(l){1-3}
\multirow{4}{*}{\STAB{\rotatebox[origin=c]{0}{Weights}}} & Intermediate inputs of sector $s'$ required to produce one unit of sector $s$ & $m_{s's}$ \\
 & Capital inputs of sector $s'$ required to produce one unit of sector $s$ & $k_{s's}$ \\
 & Depreciation of capital goods of sector $s'$ for production of goods of sector $s$ & $d_{s's}$ \\
 & Labour productivity individuals employed in sector $s$ & $h_s$ \\
\bottomrule
\\
\caption{Variables and parameters in the model related to firms.}
\label{table:firm_variables}
\end{xltabular}
\normalsize

\subsubsection{Parameters}
This section summarizes firm parameters.

\paragraph{Sector-specific Weights}
We are assuming that initial sectoral prices are set to $1$. The matrix $m_{ss'}$ denotes the real amount of intermediate inputs of sector $s$ required to produce one real unit of sector $s'$. It is computed by dividing each column of the intermediate-use matrix of the input-output table\footnote{OECD: Inter-country input-output tables (Code: \emph{ICIO}).} by the corresponding sectoral gross output.

To obtain a matrix for capital inputs depreciation, we assume that at initialisation, only depreciated capital inputs are replaced. The matrix $d_{ss'}$, which denotes the real amount of capital inputs of sector $s$ that depreciate for one real unit of produced output of sector $s'$, is calculated by dividing each column of a matrix for sectoral capital compensation by corresponding sectoral gross output. The matrix for sectoral capital compensation is obtained from the capital compensation row of the input-output table so that the column-sums of the matrix are equal to the total capital compensation row and so that the row-sums are proportional to the gross fixed capital formation column attributed to firms.

We make a similar argument to find a matrix $k_{ss'}$, which denotes the necessary real capital inputs of sector $s$ to produce one real unit of sector $s'$. We create a matrix of sectoral net fixed assets from aggregate sectoral net fixed assets\footnote{OECD: Annual fixed assets by economic activity and by asset (Code: \emph{DSD\_NAMAIN10@DF\_TABLE9A}).} so that the column-sums of the matrix are equal to the total amount of net fixed assets and so that the row sums are proportional to the gross fixed capital formation column attributed to firms. Then, $k_{ss'}$ is obtained by dividing the columns of the matrix of net fixed assets by the corresponding sectoral gross output.

The average sectoral labour productivity $h_s$ is calculated by dividing total sectoral gross output by the number of people employed in that sector.

\paragraph{Inventories (Finished Goods)}
In our modelling framework, we assume that firms decide on production also considering a target level of finished goods inventories (see \ref{eq:firm_exp_demand}). Since in this paper, we are running the model over short time horizons, we assume inventory depreciation $\delta_s=0$ for all sectors $s$ (see Eq. \eqref{eq:firm_inventory}).

\paragraph{Stocks (Materials and Supplies)}
After a sensitivity analysis concluded that the initial intermediate inputs and capital goods utilisation rates do not affect simulation output in a range of $\omega^\text{M},\omega^\text{K}\in [0.5,0.9]$, we follow \cite{poledna2023economic} and set $\omega^\text{M}=\omega^\text{K}=0.85$. Since in this paper, we are running the model over short time horizons, we set the sector-specific delay between the purchase of capital and being able to use that capital in production to be $T^\text{KD}_s=1$ for all sectors $s$. We also assume that firms attempt to keep their fraction of intermediate inputs and capital goods to production constant, assuming $\phi^M=\phi^K=1$ (see Eq. \ref{eq:firm_target_intermediate_inputs} and Eq. \ref{eq:firm_target_capital_inputs}).

\paragraph{Wage Adjustments}
The discussion of heterogeneity in individual labour productivity is out of the scope of this paper. Therefore, we assume no wage mark-ups when firms fail to meet the labour targets, $\phi^\text{WN}=0$ (see Eq. \ref{eq:wn_adjustment}).

\paragraph{Maximum Increase in Work Effort}
Following \cite{poledna2023economic}, the maximum increase in work effort is assumed to be $h^\text{max}=1.5$, see Eq. \ref{eq:we}.

\paragraph{Target Setting}
The study of the effect of firm liquidity shortages on target production is left for future research $\phi^\text{FM}=\phi^\text{FK}=0$, see Eq. \ref{eq:ii_demand} and Eq. \ref{eq:cap_demand}).

\subsubsection{Initial Conditions}
\label{sec:ic_firms}
This section discusses setting the initial conditions for firms.

\paragraph{Drawing Firms from Compustat Data}
Firms are sampled with replacement from Compustat data\footnote{Compustat Global - Fundamentals Annual/Quarterly (Code: \emph{comp\_global\_daily}).} so that the number of firms by sector (Code: \emph{gsector}) matches OECD aggregates\footnote{OECD: Structural Business Statistics by Size Class and Economic Activity (Code: \emph{DSD\_SDBSBSC\_ISIC4@DF\_SDBS\_ISIC4}).}. This includes the total number of employees (Code: \emph{emp}) which is rescaled to match OECD aggregates\footnote{OECD: Quarterly Employment by Economic activity (Code: \emph{DSD\_NAMAIN1@DF\_QNA\_BY\_ACTIVITY\_EMPDC}).}, total debt (Code: \emph{dlttq}) which is rescaled to match OECD aggregates\footnote{OECD: Quarterly Financial Balance Sheets (Code: \emph{DSD\_NASEC20@DF\_T710R\_Q}), loans (Code: \emph{F4}) of non-financial corporations (Code: \emph{S11}).}, total deposits (Code: \emph{dptbq}) which is rescaled to match OECD aggregates\footnote{OECD: Quarterly Financial Balance Sheets (Code: \emph{DSD\_NASEC20@DF\_T710R\_Q}), non-consolidated deposits (Code: \emph{F2}) held by non-financial corporations (Code: \emph{S11}).}.

\paragraph{Initial Total Wages}
The initial total wages paid by a firm $f$ are proportional to its number of employees $|\mathcal{I}_f(0)|$,
\begin{equation}
    w_f(0) = \frac{|\mathcal{I}_f(0)|}{|\mathcal{I}_s(0)|} w_s(0),
\end{equation}
where $w_s(0)$ is the initial total labour compensation of sector $s$ obtained from socio-economic accounts and $\mathcal{I}_s(0)$ is the set of individuals employed in sector $s$.

\paragraph{Initial Production}
For each sector $s$, the initial production of the firms operating in the sector is set proportional to the initial number of employees of each firm. For each sector $s$ and each firm $f\in\mathcal{F}_s$,
\begin{equation}
    Y_f(0) = \frac{|\mathcal{I}_f(0)|}{|\mathcal{I}_s(0)|} Y_s(0)
\end{equation}
where $Y_s(0)$ is the total sectoral output and $\mathcal{I}_s(0)$ is the set of individuals employed in sector $s$.

\paragraph{Initial Prices}
Initial prices are set at $1$. For each firm $f$,
\begin{equation}
    P_f(0) = 1.
\end{equation}

\paragraph{Initial Demand}
Initial firm demand is set to be equal to initial production. For each firm $f$,
\begin{equation}
    Q_f(0) = Y_f(0).
\end{equation}

\paragraph{Initial Inventory}
The initial inventory of firm $f$ is set according to
\begin{equation}
    S_f(0) = \phi^\text{StY} Y_f(0).
\end{equation}
where $Y_f(0)$ is initial production and $\phi^\text{StY}$ is the target inventory to production fraction.

\paragraph{Initial Stock of Intermediate Inputs}
The initial stock of intermediate inputs of firm $f$ operating in industry $s$ is given by
\begin{equation}
    M_{fs'}(0) = \frac{1}{\omega^\text{M}} \frac{1}{m_{s's}} Y_f(0).
\end{equation}
where $m_{s's}$ denotes the real amount of intermediate inputs of sector $s'$ required to produce one real unit of sector $s$ and $\omega^\text{M}$ is the initial utilisation rate of intermediate inputs.

\paragraph{Initial Stock of Capital Inputs}
The initial stock of capital inputs of firm $f$ operating in industry $s$ is given by
\begin{equation}
    K_{fs'}(0) = \frac{1}{\omega^\text{K}} \frac{1}{k_{s's}} Y_f(0).
\end{equation}
where $k_{s's}$ denotes the real amount of capital inputs of sector $s'$ required to produce one real unit of sector $s$ and $\omega^\text{K}$ is the initial utilisation rate of capital inputs.

\paragraph{Initial Matching with Employees}
Firms are initially matched with employees by solving a linear sum assignment problem so that the sum of the differences between total firm wages and the sum of received wages before taxes by employees is minimal.

\paragraph{Initial Matching with Banks}
Firms are initially matched with banks by solving a linear sum assignment problem so that the sum of the differences between total firm deposits plus debt to total bank deposits and debt is minimal.

\paragraph{Initial Profits}
Total initial production costs of firm $f$ operating in sector $s$ with deposits at bank $b$ are computed as
\begin{equation}
\begin{aligned}
    C_f(0) ={}& \underbrace{w_f(t)}_\text{Labour costs} + \underbrace{\sum_{s'\in \mathcal{S}} m_{s's} P_{s'}(0) Y_f(0)}_\text{Intermediate inputs bought} + \underbrace{\sum_{s'\in \mathcal{S}} d_{s's} P_{s'}(0) Y_f(0)}_\text{Capital inputs bought} + \underbrace{\tau_s^\text{PROD} P_f(0) Y_f(0)}_\text{Taxes on production} \\
    & + \underbrace{r_b^\text{F-O}(0)\left[D_f(0)\right]^- - r(0) \left[D_f(0)\right]^+}_\text{Interest paid on deposits} + \underbrace{\sum_{l\in \mathcal{L}_f(0)} r_l V_l(0),}_\text{Interest on loans}
\end{aligned}
\end{equation}
where $w_f(0)$ are total initial wages paid by firm $f$, $P_s(0)=1$ are average initial sectoral prices, $r_b^\text{F-O}(0)$ is the initial overdraft rate on firm deposits, $r(0)$ is the initial central bank policy rate, and $\mathcal{L}_f(0)$ is the initial set of loans that firm took out. The matrix $d_{s's}$ denotes depreciation rates of capital goods of sector $s'$ for firms producing goods of sector $s$. We assume here that at initialisation, the firm buys as many real units of intermediate inputs as used for production and that the depreciated capital is exactly replaced.

Initial firm profits are then computed as
\begin{equation}
    \Pi_f(0) = \underbrace{P_f(0) Y_f(0)}_\text{Output} - \underbrace{C_f(0).}_\text{Production costs}
\end{equation}

\subsubsection{Rules}
\label{sec:firm_rules}
Firms estimate demand and set production targets. They also decide on the wages they pay to employees, target credit when they face liquidity shortages, and may go bankrupt.

\paragraph{Predicted Idiosyncratic Growth}
Firms also make idiosyncratic estimates for future growth. Predicted growth of firm $f$ is set as
\begin{equation}
    \label{eq:firm_exp_growth}
    \overline{\gamma}_f(t) = \frac{Q_f(t-1)}{Y_f(t-1) + S_f(t-2)}
\end{equation}
if there is either (real) excess demand ($Y_f(t-1) + S_f(t-2)\leq Q_f(t-1)$) and the firms' price is above the market average ($P_f(t-1)\geq P_s(t-1)$) or if there is (real) excess supply ($Y_f(t-1) + S_f(t-2)\geq Q_f(t-1)$) and the firms' price is below the market average ($P_f(t-1)\geq P_s(t-1)$); otherwise $\overline{\gamma}_f = 0$. In the equation, $Q_f(t-1)$ is previous real demand, $Y_f(t-1)$ is previous real production, and $S_f(t-2)$ is real inventory previously offered on the goods market alongside newly produced goods. 

\paragraph{Predicted Demand}
The predicted real demand faced by firm $f$ operating in sector $s$ is set based on predicted sectoral growth and predicted idiosyncratic firm growth,
\begin{equation}
    \label{eq:firm_exp_demand}
    \overline{Q}_f(t) = \underbrace{\left(1 + \overline{\gamma}_s(t)\right)}_\text{Predicted sectoral growth} \times \underbrace{\left(1 + \phi^Q_F \overline{\gamma}_f(t)\right)}_\text{Predicted idiosyncratic growth} \times \underbrace{Q_f(t-1)}_\text{Previous demand}
\end{equation}
where $\phi^Q_F$ is a parameter that controls the extent to which predicted firm-specific growth determines predicted demand.

\paragraph{Predicted Profits}
Predicted profits of firm $f$ operating in sector $s$ are set according to
\begin{equation}
    \label{eq:firm_exp_profits}
    \overline{\Pi}_f(t) = \underbrace{\left(1+ \overline{\pi}^\text{PPI}(t) \right)}_\text{Predicted PPI inflation} \times \underbrace{\left(1 + \overline{\gamma}_f(t)\right)}_\text{Predicted idiosyncratic growth} \times \underbrace{\Pi_f(t-1)}_\text{Previous profits}
\end{equation}
where $\Pi_f(t-1)$ are realised previous profits.

\paragraph{Target Production}
A firm's real target production is set based on predicted real demand $\overline{Q}_f(t)$, current real inventory, and the current financial situation of the firm. It may also be limited by the firm's current workforce, stock of intermediate inputs, and stock of capital inputs. Specifically,
\begin{equation}
\label{eq:firm_target_production}
\begin{aligned}
    \hat{Y}_f(t) = \min\Bigg(&\underbrace{\overline{Q}_f(t) + \phi^\text{StY} Y_f(t-1) - S_f(t-1)}_\text{Predicted demand given current inventory},\, \underbrace{\overline{Q}_f(t) + \chi^H \left(H_f(t) - \overline{Q}_f(t)\right)}_\text{Labour inputs},\\
    & \underbrace{\overline{Q}_f(t) + \chi^M \left(M_f(t) - \overline{Q}_f(t)\right)}_\text{Intermediate inputs},\,\underbrace{\overline{Q}_f(t) + \chi^K \left(K_f(t) - \overline{Q}_f(t)\right)}_\text{Capital inputs}
    \Bigg)
\end{aligned}
\end{equation}
where $\phi^\text{StY}$ is the target inventory to production fraction and $S_f(t-1)$ are real inventories kept from the previous timestep. Target production (as in production, see \eqref{eq:firm_production}) is additionally constrained by the firm's labour inputs $H_f(t)$, intermediate inputs $M_f(t)$ and capital inputs $K_f(t)$. Parameters $\chi^\text{H},\chi^\text{M},\chi^\text{K}\in [0, 1]$ determine the influence of these limiting factors on target production; for instance, if $\chi^\text{H}=0$, firms would ignore labour inputs in determining target production and if $\chi^\text{H}=1$, they would fully consider labour input constraints. Target production determines the firm's demand for labour, intermediate inputs, and capital.

The maximum production allowed given the stock of intermediate inputs $M_f(t)$ and the maximum production allowed given the stock of capital inputs $K_f(t)$ is set as
\begin{align}
    M_f(t) ={}& \min_{s'\in\mathcal{S}}\left(\frac{M_{fs'}(t-1)}{m_{s's}}\right) \\
    K_f(t) ={}& \min_{s'\in\mathcal{S}}\left(\frac{K_{fs'}(t-1)}{k_{s's}}\right).
\end{align}
These functional forms are substantially different compared to \cite{poledna2023economic}. Indeed, \cite{poledna2023economic} aggregates all intermediate inputs into a single intermediate composite and all capital inputs into a single capital composite, essentially assuming a linear production function in each intermediate and capital input. Here, we are stricter about substitution possibilities in the economy and assume a Leontief production function for each input instead. Our framework also accommodates other less stringent production functions. One example is the Partially Binding Leontief production function \citep{pichler2022forecasting}, which is a Leontief production function that, for each industry, only considers certain inputs to be critical for production, as evaluated in a survey of industry analysts.

\paragraph{Labour Inputs}
The labour inputs of firm $f$ are given by
\begin{equation}
    H_f(t) = h_f(t) \underbrace{\sum_{i\in\mathcal{I}_f(t)} H_i(t),}_\text{Labour supply from employees}
\end{equation}
where $\mathcal{I}_f(t)$ is the set of individuals employed by firm $f$, $H_i(t)$ are labour inputs from individual $i$, and $h_f(t)$ is a firm-specific factor denoting work effort, set as
\begin{align}
    \label{eq:we}
    \phi^\text{WE} ={}& \min\left(\underbrace{h^\text{max}}_\text{Maximum factor}, \underbrace{\frac{\min\left(M_f(t),K_f(t)\right)}{h_f(0) \sum_{i\in\mathcal{I}_f(t)} H_i(t)}}_\text{Additional production factor}\right) \\
    h_f(t) ={}& \phi^\text{WE} h_f(0).
\end{align}
This follows \cite{poledna2023economic}, where $\phi^\text{WE}>1.0$ corresponds to overtime work allowed by the availability of intermediate inputs $M_f(t)$ and capital inputs $K_f(t)$. The parameter $h^\text{max}$ is the maximum increase in work effort. Similarly, $\phi^\text{WE}<1.0$ corresponds to part-time employment.

The target labour inputs of firms are set based on target production,
\begin{equation}
\label{eq:firm_target_labour_inputs}
    \hat{H}_f(t) = \hat{Y}_f(t).
\end{equation}

\paragraph{Wages}
The wages paid by firm $f$ to its currently employed individuals $i\in\mathcal{I}_f(t)$ are set according to
\begin{equation}
\label{eq:firm_wages}
    w_i(t) = \underbrace{\left(1+ \overline{\pi}^\text{PPI}(t) \right)}_\text{Predicted PPI inflation} \times \underbrace{\left(1 + \mu^\text{WN}_f\right)}_\text{Labour market tightness} \times \underbrace{\phi^\text{WE}}_\text{Work effort} \times \underbrace{w_i(t-1)}_\text{Previous wage}
\end{equation}
where $\phi^\text{WE}$ denotes the level of work effort defined in equation \eqref{eq:we} above, $w_i(t-1)$ is the previous wage paid to employee $i$, and $\mu_f^\text{WN}$ is a firm-specific markup on the average salary paid to improve the chances of keeping the current or hiring new employees. It is set to be
\begin{equation}
\label{eq:wn_adjustment}
    \mu_f^\text{WN} = \phi^\text{WN} \frac{1}{T^\text{WN}} \sum_{t'=1}^{T^\text{WN}} \underbrace{\left[\frac{\hat{H}_f(t-t') - H_f(t-t')}{\hat{H}_f(t-t')} \right]^+}_\text{Relative failure in hiring labour}
\end{equation}
where $\phi^\text{WN}$ is a parameter, and $T^\text{WN}$ is the time frame firms consider when raising prices after failing to meet their labour targets $\hat{H}_f(t-t') - H_f(t-t')$. The offered wage to a prospective employee $i$ that contributes labour inputs $H_i(t)$ is given by
\begin{equation}
    w_i(t) = \underbrace{\left(1 + \mu^\text{WN}_f\right)}_\text{Labour market tightness} \times \underbrace{\phi^\text{WE}}_\text{Work effort} \times \underbrace{\frac{\sum_{i'\in \mathcal{I}_f(t-1)} w_{i'}(t-1)}{\sum_{i'\in \mathcal{I}_f(t-1)} H_{i'}(t)}}_\text{Average wage by labour inputs} \underbrace{H_i(t)}_\text{Prospective labour inputs}.
\end{equation}

\paragraph{Production}
Firms produce goods of the sector they operate in according to a Leontief production function using labour inputs $H_f(t)$, intermediate inputs $M_f(t)$, and capital inputs $K_f(t)$.\footnote{In most models, there are two types of firms: consumption goods firms that require capital and labour, and capital goods firms that only require labour. There is a vast variety of used production technologies in different models, for instance, a Leontief production function in \cite{assenzaEmergentDynamicsMacroeconomic2015,hommesCANVAS2022}, Cobb-Douglas in \cite{dawidAgentbasedMacroeconomics2018}, or a CRS technology in \cite{gaffeoAdaptiveMicrofoundationsEmergent2008}.} Firm $f$'s production is given by
\begin{equation}
\label{eq:firm_production}
    Y_f(t) = \min\left(\underbrace{\hat{Y}_f(t)}_\text{Target production}, \underbrace{H_f(t)}_\text{Labour supply},\underbrace{M_f(t-1)}_\text{Intermediate inputs},\underbrace{K_f(t-1)}_\text{Capital inputs}\right).
\end{equation}
Production by firm $f$ may not be equal to target production $\hat{Y}_f(t)$, and it has to scale down activity if the available labour supply, intermediate inputs, or capital inputs limit it. This approach is consistent with the data and in line with similar large models \citep{polednaEconomicForecastingAgentbased2020,hommesCANVAS2022}.

\paragraph{Prices}
Firms set prices based on their expectations for inflation. The price set by firm $f$ operating in sector $s$ is given by
\begin{equation}
    \label{eq:firm_prices}
    P_f(t) = \underbrace{\left(1+  \overline{\pi}^\text{PPI}(t)\right)}_\text{Predicted PPI inflation} \times  \underbrace{\left(1+\phi^\text{DP} \overline{\pi}_f^\text{DP}(t)\right)}_\text{Demand-pull inflation} \times \underbrace{\left(1 + \phi^\text{CP} \overline{\pi}_f^\text{CP}(t)\right)}_\text{Cost-push inflation} \times \underbrace{P_f(t-1)}_\text{Previous price}
\end{equation}
where $\phi^\text{DP}$ and $\phi^\text{CP}$ are parameters, demand-pull inflation is defined as
\begin{equation}
\label{eq:dp_inflation}
    \overline{\pi}^\text{DP}_f(t) = \frac{Q_f(t-1)}{Y_f(t-1) + S_f(t-2)} - 1
\end{equation}
if the total good real supply offered by the firm $Y_f(t-1) + S_f(t-2)$ exceeds previous real demand $Q_f(t-1)$ and the firm had a price $P_f(t-1)$ that was higher than the average sector price $P_s(t-1)$, or if total good supply is lower than previous demand, and firm prices are lower than the average sector price, otherwise
\begin{equation}
    \overline{\pi_f}^\text{DP}(t)=0.
\end{equation}
On the other hand, cost-push inflation is defined as
\begin{equation}
\label{eq:cp_inflation}
    \overline{\pi}_f^\text{CP}(t) = \frac{U_f(t-1)}{P_f(t-1)} - 1.
\end{equation}
where $U_f(t-1)$ are previous unit costs, computed as
\begin{equation}
    U_f(t) = \underbrace{\frac{w_f(t)}{Y_f(t)}}_\text{Labour costs} + \underbrace{\sum_{s'\in \mathcal{S}} m_{s's} P_{s'}(t-1)}_\text{Intermediate inputs used} + \underbrace{\sum_{s'\in \mathcal{S}} d_{s's} P_{s'}(t-1)}_\text{Depreciation} + \underbrace{\tau_s^\text{PROD} P_f(t-1)}_\text{Taxes on production}
\end{equation}
where the first term $\frac{w_f(t)}{Y_f(t)}$ denotes total wages paid by real units of output.

\paragraph{Demand for Goods before Financial Frictions}
Each firm needs intermediate inputs for production. Each firm holds a real stock of intermediate inputs $M_{fs}(t)$ that is taken out from when producing. Target intermediate inputs before financial frictions of sector $s'$ of firm $f$ operating in sector $s$ are based on the chosen target production $\hat{Y}_f(t)$ and current stocks and given by
\begin{equation}
\label{eq:firm_target_intermediate_inputs}
    \hat{M}_{fs'}^\text{FF}(t) = \left[\underbrace{m_{s's} \hat{Y}_f(t)}_\text{Targetted amount} - \phi^M \underbrace{\left(M_{fs'}(t-1) - M_{fs'}(0)\frac{Y_f(t)}{Y_f(0)}\right)}_\text{Existing stock}\right]^+
\end{equation}
where the parameter $\phi^M$ governs how much the current stock of intermediate inputs is considered.

Similarly, each firm needs capital inputs for production. Each firm holds a real stock of capital inputs $K_{fs}(t)$ that depreciates when used for production. Target capital inputs before financial frictions are set as
\begin{equation}
\label{eq:firm_target_capital_inputs}
    \hat{K}_{fs'}^\text{FF}(t) = \left[\underbrace{d_{s's} \hat{Y}_f(t)}_\text{Targetted amount} - \phi^K \underbrace{\left(K_{fs'}(t-1) - K_{fs'}(0)\frac{Y_f(t)}{Y_f(0)}\right)}_\text{Existing stock}\right]^+
\end{equation}
where the parameter $\phi^K$ governs how much the current stock of capital inputs is considered.

\paragraph{Demand for Loans}
The predicted change in deposits without new loans and without taking the purchase of new inputs into account is given by
\begin{equation}
\begin{aligned}
    \overline{D}^\Delta_f(t) ={}& \underbrace{P_f(t) Y_f(t)}_\text{Production} - \underbrace{w_f(t)}_\text{Total wages paid} - \underbrace{\tau^\text{CORP} \left[\overline{\Pi}_f(t)\right]^+}_\text{Predicted corporate taxes} \\
    & - \underbrace{r_b^\text{F-O}(t)\left[D_f(t-1)\right]^- - r(t) \left[D_f(t-1)\right]^+}_\text{Interest paid on deposits} - \underbrace{\sum_{l\in \mathcal{L}_f(t)} r_l V_l(t)}_\text{Interest on loans} \\
    & - \underbrace{\tau_s^\text{PROD} P_f(t) Y_f(t)}_\text{Taxes on production} - \underbrace{\sum_{l\in \mathcal{L}_f(t)} \frac{V_l(t)}{m_l}}_\text{Debt installment}
\end{aligned}
\end{equation}
where $w_f(t)$ are total wages paid by firm $f$, $\overline{\Pi}_f(t)$ are predicted firm profits, $r_b^\text{F-O}(t)$ is the overdraft rate on firm deposits, $r(t)$ is the central bank policy rate, and $\mathcal{L}_f(t)$ is the set of loans that firm took out. 

The firms' financing needs determine the demand for credit. If the operating costs exceed the internal funds, the firm has that difference as a financing gap and will apply for a bank loan. Firms apply for short-term and long-term loans. Each firm applies for short-term loans to cover any financing gap due to wages or intermediate inputs purchasing costs by applying for loans to the value of
\begin{equation}
    \hat{L}_f^\text{S}(t) = \left[\underbrace{D_f(t-1) + \overline{D}^\Delta_f(t)}_\text{Predicted deposits} - \underbrace{\sum_{s'\in\mathcal{S}} P_{s'}(t-1) \hat{M}^\text{FF}_{fs'}(t),}_\text{Costs of intermediate inputs}\right]^-
\end{equation}
where $P_{s'}(t-1)$ is the previous average price of goods from sector $s$. Each firm may also apply for long-term loans to finance the purchase of capital inputs,
\begin{equation}
    \hat{L}_f^\text{L}(t) = \left[\underbrace{D_f(t-1) + \Delta \overline{D}_f(t)}_\text{Predicted deposits} - \underbrace{\sum_{s' \in \mathcal{S}} P_{s'}(t-1) \hat{K}^\text{FF}_{fs'}(t).}_\text{Costs of capital inputs}\right]^-
\end{equation}

\paragraph{Constrained Demand for Goods}
Firms purchase fewer intermediate inputs and capital goods if they fail to acquire the loans they applied for. Specifically, target intermediate inputs are set as
\begin{equation}
\label{eq:ii_demand}
    \hat{M}_{fs'}(t) = \underbrace{\hat{M}^\text{FF}_{fs'}(t)}_\text{Demand before frictions} - \phi^\text{FM} \underbrace{\left(\frac{\hat{L}_f^\text{S}(t) - L_f^\text{S}(t)}{\left(1 + \overline{\pi}^\text{PPI}(t)\right) P_s(t-1)}\right),}_\text{Credit gap}
\end{equation}
where $\hat{L}_f^\text{S}(t) - L_f^\text{S}(t)$ is the difference between requested and granted short-term loans, and $\phi^\text{FM}$ is a parameter. Target capital inputs are set as
\begin{equation}
\label{eq:cap_demand}
    \hat{K}_{fs'}(t) = \underbrace{\hat{K}^\text{FF}_{fs'}(t)}_\text{Demand before frictions} - \phi^\text{FK} \underbrace{\left(\frac{\hat{L}_f^\text{L}(t) - L_f^\text{L}(t)}{\left(1 + \overline{\pi}^\text{PPI}(t)\right) P_s(t-1)}\right),}_\text{Credit gap}
\end{equation}
where $\hat{L}_f^\text{L}(t) - L_f^\text{L}(t)$ is the difference between requested and granted long-term loans, and $\phi^\text{FK}$ is a parameter.

\paragraph{Inventory and Stocks}
The inventory of finalised goods of firm $f$ operating in sector $s$ is updated according to
\begin{equation}
\label{eq:firm_inventory}
    S_f(t) = \max\left(0,\,\underbrace{\left(1-\delta_s\right) S_f(t-1)}_\text{Previous inventory} + \underbrace{Y_f(t)}_\text{Production} - \underbrace{\tilde{Q}_f(t)}_\text{Units sold}\right)
\end{equation}
where $\delta_s$ is sector-specific depreciation of finalised goods $S_f(t)$ in the inventory of firm $f$. There is no direct distinction between goods bought as intermediate inputs and goods bought as capital inputs in the goods market. Firms prioritize bought goods to satisfy their demand for intermediate inputs first; the remainder goes into their capital stock. The updated stock of intermediate inputs is
\begin{equation}
\label{eq:firm_intermediate_inputs}
    M_{fs'}(t) = \left[\underbrace{M_{fs'}(t-1)}_\text{Previous stock} - \underbrace{m_{s's} Y_f(t)}_\text{Used stock} + \underbrace{\tilde{M}_{fs'}(t)}_\text{Newly bought stock}\right]^+
\end{equation}
and the new stock of capital inputs is
\begin{equation}
\label{eq:firm_capital_inputs}
    K_{fs'}(t) = \left[\underbrace{K_{fs'}(t-1)}_\text{Previous stock} - \underbrace{d_{s's} Y_f(t)}_\text{Depreciated stock} + \underbrace{\tilde{K}_{fs'}(t - T^\text{D}_{s'})}_\text{Newly bought stock}\right]^+
\end{equation}
where $T^\text{D}_{s'}$ is a sector-specific delay between purchasing capital and using that capital in production.

\paragraph{Demand}
Firms face demand for goods from other firms, households, government entities, and the rest of the world. The real demand a firm faces is how much that firm managed to sell to buyers plus what they could have sold additionally. Specifically, firm $f$'s real demand is given by
\begin{equation}
    \label{eq:firm_demand}
    Q_f(t) = \underbrace{\tilde{Q}_f(t)}_\text{Amount sold} + \underbrace{\tilde{Y}_f^\text{E}(t)}_\text{Excess demand}
\end{equation}
where $\tilde{Q}_f(t)$ is how much firm $f$ sold on the goods market and $\tilde{Y}_f^\text{E}(t)$ is how much firm $f$ could have sold additionally if it had more supply. To calculate excess demand, firms get allocated a share of the total real amount potential buyers would have purchased at the offered price if the random search-and-matching process of the goods market had continued after firms ran out of supply. The share the firm receives is based on the firms' characteristics; see \ref{sec:goods_market}.

\paragraph{Profits}
Total production costs of firm $f$ operating in sector $s$ with deposits at bank $b$ are computed as
\begin{equation}
\label{eq:firm_profits}
\begin{aligned}
    C_f(t) ={}& \underbrace{w_f(t)}_\text{Total wages paid} + \underbrace{\sum_{s'\in \mathcal{S}} P_{s'}(t) \left(M_{fs'}(t) - M_{fs'}(t-1) + m_{s's} Y_f(t)\right)}_\text{Intermediate inputs bought} \\
    & + \underbrace{\sum_{s'\in \mathcal{S}} P_{s'}(t) \left(K_{fs'}(t) - K_{fs'}(t-1) + d_{s's} Y_f(t)\right)}_\text{Capital inputs bought} \\
    & + \underbrace{r_b^\text{F-O}(t)\left[D_f(t-1)\right]^- - r(t) \left[D_f(t-1)\right]^+}_\text{Interest paid on deposits} + \underbrace{\sum_{l\in \mathcal{L}_f(t)} r_l V_l(t)}_\text{Interest on loans} + \underbrace{\tau_s^\text{PROD} P_f(t) Y_f(t),}_\text{Taxes on production}
\end{aligned}
\end{equation}
where $w_f(t)$ are total wages paid by firm $f$, $r_b^\text{F-O}(t)$ is the overdraft rate on firm deposits, $r(t)$ is the central bank policy rate, and $\mathcal{L}_f(t)$ is the set of loans that firm took out. Firm profits are then computed as
\begin{equation}
    \Pi_f(t) = \underbrace{P_f(t) \tilde{Q}_f(t)}_\text{Sales} + \underbrace{P_f(t) \Delta S_f(t)}_\text{Inventory change} - \underbrace{C_f(t).}_\text{Production costs}
\end{equation}

\paragraph{Deposits}
Deposits of firm $f$ that has its deposits with bank $b$ and is operating in sector $s$ are updated as
\begin{equation}
\label{eq:firm_deposits}
\begin{aligned}
    D_f(t) ={}& \underbrace{D_f(t-1)}_\text{Previous deposits} + \underbrace{P_f(t) \tilde{Q}_f(t)}_\text{Units sold} - \underbrace{C_f(t)}_\text{Production costs} - \underbrace{\tau^\text{CORP} \left[\Pi_f(t)\right]^+}_\text{Corporate taxes} - \underbrace{\sum_{l\in \mathcal{L}_f(t)} \frac{V_l(t)}{m_l}}_\text{Debt installment} \\
& + \underbrace{\Delta L_f(t).}_\text{New credit}
\end{aligned}
\end{equation}

\paragraph{Debt}
The debt of firm $f$ is updated as
\begin{equation}
\label{eq:firm_debt}
    L_f(t) = \underbrace{L_f(t-1)}_\text{Previous debt} - \underbrace{\sum_{l\in \mathcal{L}_f(t)} \frac{V_l(t)}{m_l}}_\text{Debt installment} + \underbrace{\Delta L_f(t).}_\text{New credit}
\end{equation}

\paragraph{Equity}
Equity of firm $f$ evolves according to
\begin{equation}
\label{eq:firm_equity}
    E_f(t) = \underbrace{D_f(t)}_\text{Deposits} + \underbrace{P_f(t) S_f(t)}_\text{Inventory} + \underbrace{\sum_{s'\in \mathcal{S}} P_{s'}(t) \left(M_{fs'}(t) + K_{fs'}(t)\right)}_\text{Stock} - \underbrace{L_f(t).}_\text{Debt}
\end{equation}

\paragraph{Bankruptcy}
If a firm $f$ is cash-flow insolvent ($D_f(t)<0$) and balance-sheet insolvent ($E_f(t)<0$), it goes bankrupt. A bankrupt firm is replaced by a new firm that enters the same sector. That new firm keeps the same stock and inventory as the bankrupt firm and is initialised with
\begin{equation}
    \underbrace{D_f(t+1)}_\text{Deposits} = \underbrace{L_f(t+1)}_\text{Debt} = 0.
\end{equation}
As described in Eq. \eqref{eq:bank_equity}, the losses are born by the bank $b$ that gave credit to firm $f$.

\subsection{Government}
\label{sec:government}
The government consumes according to fixed fractions, collects taxes, and pays social benefits to individuals and households. Variables and parameters related to the government are summarised in Table \ref{table:central_government_parameters}.

\tablefontsize
\begin{xltabular}{\textwidth}{cXl}
\toprule
Category & Description & Notation \\
\cmidrule(l){1-3}
\multirow{1}{*}{Sets} & Set of government entities & $\mathcal{G}$ \\
\cmidrule(l){1-3}
\multirow{9}{*}{Revenue} & Revenue & $Y^\text{CG}(t)$ \\
 & Sectoral taxes on production & $\tau^\text{PROD}_s$ \\
 & Income tax rate & $\tau^\text{INC}$ \\
 & Corporate tax rate & $\tau^\text{CORP}$ \\
 & Value-added tax rate & $\tau^\text{VAT}$ \\
 & Capital formation tax rate & $\tau^\text{CF}$ \\
 & Social insurance tax rate (employers) & $\tau^\text{SIF}$ \\
 & Social insurance tax rate (employees) & $\tau^\text{SIW}$ \\
 & Exports tax rate & $\tau^\text{EXP}$ \\
\cmidrule(l){1-3}
\multirow{5}{*}{Expenditures} & Consumption & $C^\text{CG}(t)$ \\
 & Target consumption & $\hat{C}^\text{CG}(t)$ \\
 & Government consumption weights & $c_s^\text{CG}$ \\
 & Per-capita real unemployment benefits & $w^\text{U}(t)$ \\
 & Total real other social benefits & $\text{sb}^\text{O}(t)$ \\
\cmidrule(l){1-3}
\multirow{1}{*}{Deficit} 
 & Deficit & $\Pi^\text{CG}(t)$ \\
\cmidrule(l){1-3}
\multirow{1}{*}{Debt} & Debt & $L^\text{CG}(t)$ \\
\bottomrule
\\
\caption{Variables and parameters in the model related to the government.}
\label{table:central_government_parameters}
\end{xltabular}
\normalsize

\subsubsection{Initial Conditions}
This section summarizes the government's initial conditions on consumption, social benefits, and initial debt.

\paragraph{Number of Entities}
Following, \cite{poledna2023economic}, the number of government entities in each country is set to be 25\% of the number of domestically producing firms.

\paragraph{Initial Consumption}
Initial total consumption $C^\text{CG}(0)=\hat{C}^\text{CG}(0)$ and government consumption weights $c_s^\text{CG}$ are matched to input-output tables\footnote{OECD: Inter-country input-output tables (Code: \emph{ICIO}).}, see Eq. \eqref{eq:gov_cons}. Consumption is distributed evenly among all government entities.

\paragraph{Initial Social Benefits}
Initial social benefits are matched with OECD aggregates\footnote{OECD: Social Expenditure Aggregates (Code: \emph{DSD\_SOCX\_AGG@DF\_SOCX\_AGG}).} and distributed according to \emph{Household Finance and Consumption Survey} microdata, see Eq. \eqref{eq:gov_ub} and Eq. \eqref{eq:gov_ob}.

\paragraph{Initial Debt}
Initial government debt $L^\text{CG}(0)$ is initialised using World Bank data\footnote{World Bank Data: Government debt (Code: \emph{GC.DOD.TOTL.GD.ZS})}, see Eq. \eqref{eq:gov_debt}.

\subsubsection{Parameters}
Sectoral taxes on production $\tau^\text{PROD}_s$ are taken from input-output tables\footnote{OECD: Inter-country input-output tables (Code: \emph{ICIO}).}. The income tax rate, corporate tax rate, export taxes, value-added tax rate, and social insurance rates are taken directly from the OECD database.\footnote{OECD: Annual Government Taxes and Social Contributions Receipts (Code: \emph{DSD\_NASEC10@DF\_TABLE10}) with income taxes (Code: \emph{D51A}), corporate taxes (Code: \emph{D51B}), export taxes (Code: \emph{D214K}), value-added taxes (Code: \emph{D211}), employees' contributions to social insurance (Code: \emph{D613CE}), employers' contribution to social insurance (Code: \emph{D611}).}

\subsubsection{Rules}
The government consumes and pays social benefits, based on which we update revenue, deficits, and debt.

\paragraph{Consumption}
Government entities update their total real target consumption $\hat{C}^\text{CG}(t)$ based on an AR(1) model on historical data. Nominal sectoral target consumption is then given by
\begin{equation}
\label{eq:gov_cons}
    \hat{C}_s^\text{CG}(t) = \underbrace{c_s^\text{CG}}_\text{Weights} \times \underbrace{\left(1 + \overline{\pi}^\text{PPI}(t)\right)}_\text{Predicted PPI inflation} \times \underbrace{P_s(t-1)}_\text{Previous average price} \times \underbrace{\hat{C}^\text{CG}(t)}_\text{Total target consumption}
\end{equation}
which is then distributed evenly among each government entity $g\in\mathcal{G}$.

\paragraph{Social Benefits}
The government pays unemployment benefits to unemployed individuals and other social benefits to households based on household composition and individual characteristics (income, wealth, and debt). Per-capita real unemployment benefits are increased when the economy enters a recession,
\begin{equation}
\label{eq:gov_ub}
  w^\text{U}(t) = \underbrace{\max\left(1,\,\frac{1}{1 + \overline{\gamma}(t)}\right)}_\text{Inverse predicted growth} w^\text{U}(t-1)
\end{equation}
Total real other benefits grow according to
\begin{equation}
\label{eq:gov_ob}
    \text{sb}^\text{O}(t) = \underbrace{\left(1 + \overline{\gamma}(t)\right)}_\text{Predicted growth} \text{sb}^\text{O}(t-1).
\end{equation}

\paragraph{Revenue}
The government collects social contributions ($\tau^\text{SIF}$ employers', $\tau^\text{SIW}$ employees') and taxes according to fixed rates on labour income ($\tau^\text{INC}$), corporate income ($\tau^\text{CORP}$), value-added ($\tau^\text{VAT}$), household capital formation ($\tau^\text{CF}$), net taxes on production ($\tau^\text{PROD}_s$), and exports ($\tau^\text{EXP}$):
\begin{equation}
\label{eq:gov_revenue}
\begin{aligned}
    Y^\text{G}(t) ={}& \underbrace{P^\text{CPI}(t) \left(\tau^\text{SIW}+\tau^\text{INC}\left(1-\tau^\text{SIW}\right)\right) \sum_{i\in \mathcal{I}^\text{E}(t)} w_i(t)}_\text{Social security contributions and labour income taxes} + \underbrace{\tau^\text{INC} \sum_{h\in \mathcal{H}} \sum_{p\in \mathcal{P}^\text{R}_h(t)} r_p(t)}_\text{Rental income taxes} \\
    & + \underbrace{\tau^\text{CORP}\left(\sum_{f\in \mathcal{F}} \left[\Pi_f(t)\right]^+ + \sum_{b\in \mathcal{B}} \left[\Pi_b(t)\right]^+\right)}_\text{Corporate income taxes} + \underbrace{\tau^\text{VAT} \sum_{h\in\mathcal{H}} C_h(t)}_\text{Value added taxes} \\
    & + \underbrace{\tau^\text{CF} \sum_{h\in\mathcal{H}} K_h(t)}_\text{Taxes on capital formation} + \underbrace{\sum_{s\in \mathcal{S}} \tau^\text{PROD}_s \sum_{f\in \mathcal{F}_s} P_f(t) Y_f(t)}_\text{Taxes on Production} + \underbrace{\tau^\text{EXP} \text{EXP}(t)}_\text{Export taxes}
\end{aligned}
\end{equation}
where $P^\text{CPI}(t)$ is the consumer price index, $\mathcal{I}^\text{E}(t)$ is the set of employed individuals, $w_i(t)$ is the wage paid by the employer of individual $i$, $\mathcal{P}^\text{R}_h(t)$ is the set of properties owned by household $h$ that are rented-out to other households, $r_p(t)$ is the rent paid for living in property $p$, $\Pi_f(t)$ / $\Pi_b(t)$ are profits of firm $f$ / bank $b$, $C_h(t)$ is consumption of household $h$, $K_h(t)$ is investment of household $h$, and $\text{EXP}(t)$ are total nominal exports.

\paragraph{Deficit/Surplus}
Government deficits $\Pi^\text{CG}(t)$ are updated according to
\begin{equation}
\label{eq:gov_deficit}
    \Pi^\text{CG}(t) = \underbrace{P^\text{CPI}(t) \left(\text{sb}^\text{O}(t) + \left|\mathcal{I}^\text{U}(t)\right| w^\text{U}(t)\right)}_\text{Social benefits} + \underbrace{C^\text{CG}(t)}_\text{Spending} + \underbrace{r(t) L^\text{CG}(t-1)}_\text{Interest payments} - \underbrace{Y^\text{CG}(t).}_\text{Revenue}
\end{equation}
where $\text{sb}^\text{O}(t)$ are total real other social benefits and $\left|\mathcal{I}^\text{U}(t)\right| w^\text{U}(t)$ are total real unemployment benefits. We are assuming that social housing revenue is equal to costs.

\paragraph{Debt}
Government debt is updated according to
\begin{equation}
\label{eq:gov_debt}
    L^\text{CG}(t) = \underbrace{L^\text{CG}(t-1)}_\text{Previous debt} + \underbrace{\Pi^\text{CG}(t).}_\text{Deficit}
\end{equation}

\subsection{Households}
Households are groups of individuals. Households, as such, are the relevant units when it comes to the goods, credit and housing market. They may own housing, live in rented properties or social housing, and apply for consumption loans or mortgages. They receive income as the sum of income of their individuals plus additional other social benefits, income from financial assets, and rental income. They hold wealth as real and financial assets. Table \ref{table:hh_variables} shows variables and parameters related to households.

\tablefontsize
\begin{xltabular}{\textwidth}{cXl}
\toprule
Category & Description & Notation \\
\cmidrule(l){1-3}
\multirow{6}{*}{\STAB{\rotatebox[origin=c]{0}{Sets}}} & Set of households & $\mathcal{H}$ \\
 & Set of insolvent households & $\mathcal{H}'(t)$ \\
 & Set of corresponding individuals & $\mathcal{I}_h(t)$ \\
 & Set of loans & $\mathcal{L}_h(t)$ \\
 & Set of owned properties & $\mathcal{P}_h(t)$ \\
 & Set of rented-out properties & $\mathcal{P}^\text{R}_h(t)$ \\
\cmidrule(l){1-3}
\multirow{6}{*}{\STAB{\rotatebox[origin=c]{0}{Income}}} & Income & $Y_h(t)$ \\
 & Predicted income & $\overline{Y}_h(t)$ \\
 & Rent & $r_h(t)$ \\
 & Other social benefits & $\text{sb}^\text{O}_h(t)$ \\
 & Income from financial assets coefficient & $\phi^\text{FA}$ \\
 & Variance of the Gaussian noise impacting financial assets income & $\sigma^\text{FA}$ \\
\cmidrule(l){1-3}
\multirow{6}{*}{\STAB{\rotatebox[origin=c]{0}{Consumption}}} & Consumption & $C_h(t)$ \\
 & Target consumption & $\hat{C}_{hs}(t)$ \\
 & Aggregate consumption weights & $c_s^\text{CPI}(t)$ \\
 & Saving rate & $\phi^\text{SR}_h(t)$ \\
 & Consumption smoothing fraction & $\phi^\text{CO}$ \\
 & Time frame for consumption smoothing & $T^\text{CO}$ \\
\cmidrule(l){1-3}
\multirow{4}{*}{\STAB{\rotatebox[origin=c]{0}{Investment}}} & Investment & $K_h(t)$ \\
 & Target investment & $\hat{K}_{hs}(t)$ \\
 & Investment weights & $k_s$ \\
 & Investment rate & $\phi^\text{IR}$ \\
\cmidrule(l){1-3}
\multirow{9}{*}{\STAB{\rotatebox[origin=c]{0}{Wealth}}} & Wealth & $W_h(t)$ \\
 & Wealth in properties & $W^\text{P}_h(t)$ \\
 & Wealth in other real assets & $W^\text{ORA}_h(t)$ \\
 & Wealth in real assets & $W^\text{RA}_h(t)$ \\
 & Wealth in deposits & $W^\text{D}_h(t)$ \\
 & Wealth in other financial assets & $W^\text{OFA}_h(t)$ \\
 & Wealth in financial assets & $W^\text{FA}_h(t)$ \\
 & Net Wealth & $W^\text{N}_h(t)$ \\
 & Rate of depreciation of real assets & $d^\text{RA}$ \\
\cmidrule(l){1-3}
\multirow{10}{*}{\STAB{\rotatebox[origin=c]{0}{Debt}}} & Total debt & $L_h(t)$ \\
 & Debt to bank $b$ & $L_{hb}(t)$\\
 & Debt in consumptions loans & $L^\text{C}_h(t)$ \\
 & Debt in consumptions loans to bank $b$ & $L^\text{C}_{hb}(t)$ \\
 & Debt in mortgages & $L^\text{M}_h(t)$ \\
 & Debt in mortgages to bank $b$ & $L^\text{M}_{hb}(t)$ \\
 & Demand for consumption loans & $\hat{L}^\text{C}_h(t)$ \\
 & Demand for mortgages & $\hat{L}^\text{M}_h(t)$ \\
 & Newly granted consumption loans & $\hat{L}^{\Delta\text{C}}_h(t)$ \\
 & Newly granted mortgages & $\hat{L}^{\Delta\text{M}}_h(t)$ \\
\cmidrule(l){1-3}
\multirow{11}{*}{\STAB{\rotatebox[origin=c]{0}{\makecell{Housing as\\a Tenant}}}} & Probability of not changing housing when renting & $p^\text{RS}$ \\
 & Maximum price for buying a property & $P_h(t)$ \\
 & Factor for setting the maximum price & $\phi^\text{HP}$ \\
 & Exponent for setting the maximum price & $\beta^\text{HP}$ \\
 & Mean of the Gaussian noise impacting the maximum price & $\mu^\text{HP}$ \\
 & Variance of the Gaussian noise impacting the maximum price & $\sigma^\text{HP}$ \\
 & Psychological pressure of renting & $\mu^\text{PS}$ \\
 & Influence of costs in deciding whether to buy or rent & $\phi^\text{B}$ \\
 & Factor for setting the maximum rent & $\phi^\text{HR}$ \\
 & Exponent for setting the maximum rent & $\beta^\text{HR}$ \\
 & Probability of buying over renting & $p_h^\text{B}(t)$ \\
\cmidrule(l){1-3}
\multirow{4}{*}{\STAB{\rotatebox[origin=c]{0}{\makecell{Housing as an\\Owner-Occupier}}}} & Probability of not changing housing when owning & $p^\text{OS}$ \\
 & Probability of price reduction & $p^\text{PM}$ \\
 & Mean of the Gaussian noise reducing the price & $\mu^\text{PM}$ \\
 & Variance of the Gaussian noise reducing the price & $\sigma^\text{PM}$ \\
\cmidrule(l){1-3}
\multirow{5}{*}{\STAB{\rotatebox[origin=c]{0}{\makecell{Housing\\as an Investor}}}} & Probability of rent reduction & $p^\text{RM}$ \\
 & Mean of the Gaussian noise reducing the rent & $\mu^\text{RM}$ \\
 & Variance of the Gaussian noise reducing the rent & $\sigma^\text{RM}$ \\
 & Partial indexation for setting rent & $\phi^\text{PIR}$ \\
 & Lag for setting rent & $T^\text{PIR}$ \\
\bottomrule
\\
\caption{Variables and parameters in the model related to households.}
\label{table:hh_variables}
\end{xltabular}
\normalsize

\subsubsection{Initial Conditions}
\label{sec:ic_hhs}
This section describes setting the initial conditions of households.

\paragraph{Drawing Households from HFCS Data}
Households are sampled from \emph{Household Finance and Consumption Survey} data to match OECD aggregates\footnote{OECD: Infra-annual Labour Statistics (Code: \emph{DSD\_NASEC20@DF\_T720R\_Q}).} according to the given weights (Code: \emph{HW0010}). The data includes microdata for the fields described in Table \ref{table:hfcs_codes_households}.

\begin{xltabular}{\textwidth}{cXl}
\toprule
Category & Description & Code \\
\cmidrule(l){1-3}
\multirow{2}{*}{\STAB{\rotatebox[origin=c]{0}{Attributes}}} & Type & \emph{DHHTYPE} \\
 & Country & \emph{SA0100} \\
\cmidrule(l){1-3}
\multirow{4}{*}{\STAB{\rotatebox[origin=c]{0}{Income}}} & Rental income from real estate & \emph{DI1300} \\
 & Income from financial assets & \emph{DI1400} \\
 & Income from regular social transfers & \emph{DI1620} \\
 & Income & \emph{DI2000} \\
\cmidrule(l){1-3}
\multirow{13}{*}{\STAB{\rotatebox[origin=c]{0}{Assets}}} & Value of main residence & \emph{DA1110} \\
 & Value of other properties & \emph{DA1120} \\
 & Value of vehicles & \emph{DA1130} \\
 & Value of valuables & \emph{DA1131} \\
 & Wealth in deposits & \emph{DA2101} \\
 & Mutual funds & \emph{DA2102} \\
 & Bonds & \emph{DA2103} \\
 & Value of private businesses & \emph{DA2104} \\
 & Shares & \emph{DA2105} \\
 & Managed accounts & \emph{DA2106} \\
 & Money owed & \emph{DA2107} \\
 & Other assets & \emph{DA2108} \\
 & Voluntary Pension & \emph{DA2109} \\
\cmidrule(l){1-3}
\multirow{5}{*}{\STAB{\rotatebox[origin=c]{0}{Liabilities}}} & Outstanding balance of mortgages on the main residence & \emph{DL1110} \\
 & Outstanding balance of mortgages on other properties & \emph{DL1120} \\
 & Outstanding balance of credit line & \emph{DL1210} \\
 & Outstanding balance of credit card debt & \emph{DL1220} \\
 & Outstanding balance of other non-mortgage loans & \emph{DL1230} \\
\cmidrule(l){1-3}
\multirow{3}{*}{\STAB{\rotatebox[origin=c]{0}{Housing}}} & Tenure status of the main residence & \emph{HB0300} \\
 & Rent paid & \emph{HB2300} \\
 & Number of properties other than the main residence & \emph{HB2410} \\
\cmidrule(l){1-3}
Consumption & Household income share spent on consumer goods and services & DOCOGOODP \\
\bottomrule
\\
\caption{Household Finance and Consumption Survey codes related to households.}
\label{table:hfcs_codes_households}
\end{xltabular}
\normalsize

\paragraph{Initial Income}
Initial household income (see Eq. \eqref{eq:hh_income}) from real estate and initial income from financial assets is matched to HFCS data. Other social benefits paid to households are set proportional to regular social transfers and rescaled to match OECD aggregates\footnote{OECD: Social Expenditure Aggregates (Code: \emph{DSD\_SOCX\_AGG@DF\_SOCX\_AGG}).}. Income after rent $Y^\text{-r}_h(0)$ is set as the difference between total income (Code: \emph{DI2000}) and rent paid (Code: \emph{HB2300}). 

\paragraph{Initial Wealth}
Total household wealth (see Eq. \eqref{eq:hh_wealth}) $W_h(0)$, wealth in properties $W^\text{P}_h(0)$, and wealth in deposits $W_h^\text{D}(0)$ is matched to HFCS data. The initial wealth in other real assets $W_h^\text{ORA}(0)$ is set to be the sum of the values of vehicles (Code: \emph{DA1130}) and of other valuables (Code: \emph{DA1131}). Wealth in other financial assets $W_h^\text{OFA}$, since not explicitly modelled, are assumed to be the sum of mutual funds (Code: \emph{DA2102}), bonds (Code: \emph{2103}), shares (Code: \emph{2105}), managed accounts (Code: \emph{2106}), money owed (Code: \emph{DA2107}), other assets (Code: \emph{2108}), and voluntary pensions (Code: \emph{DA2109}).

\paragraph{Initial Debt}
Total household mortgage debt $L_h^\text{M}(0)$ is set to be the sum of outstanding balances of mortgages on household main residences (Code: \emph{DL1110}) and outstanding balances of mortgages on other properties (Code: \emph{DL1120}). Total household debt in consumption loans $L_h^\text{C}(0)$ is set to be the sum of outstanding balances of lines of credit (Code: \emph{DL1210}), credit card debt (Code: \emph{DL1220}), and other non-mortgage loans (Code: \emph{DL1230}).

\paragraph{Initial Consumption}
The initial household saving rate $\phi^\text{SR}_h(0)$ is set to be proportional to 1 $-$ the household income share spent on consumer goods and services (Code: \emph{DOCOGOODP}), and rescaled so that initial household consumption given by
\begin{equation}
    C_{hs}(0) = \frac{1}{\tau^\text{VAT}}\left(1 - \phi^\text{SR}_h(0)\right) c_s^\text{CPI} Y_h(t)
\end{equation}
matches the aggregate input-output tables\footnote{OECD: Inter-country input-output tables (Code: \emph{ICIO}).} household final consumption expenditure column. The value-added tax rate is denoted by $\tau^\text{VAT}$. 

\paragraph{Initial Investment}
The investment rate $\phi^\text{IR}$ is assumed to be homogenous among households and set so that initial household investment
\begin{equation}
    K_{hs}(0) = \frac{1}{\tau^\text{CF}} \phi^\text{IR} k_s Y_h(t)
\end{equation}
matches the input-output table. The investment weights $k_s$ are set to match input-output tables, and $\tau^\text{CF}$ are taxes on capital formation.

\paragraph{Initial Matching with Banks}
Households are initially matched with banks by solving a linear sum assignment problem so that the sum of the differences between total household deposits plus debt to total bank deposits and debt from households is minimal.

\paragraph{Initial Matching with Properties}
Households are initially matched with properties by solving a linear sum assignment problem so that the sum of the differences between property values to the households' total wealth in real assets is minimal. Properties themselves are generated based on the tenure status of the main residence (Codes: \emph{HB0300}, \emph{DA1110}) of each household and their additional number of properties owned (Code: \emph{HB2410}, \emph{DA1120}). 

\subsubsection{Parameters}
This section summarizes setting household initial conditions.

\paragraph{Coefficient for Income from Financial Assets}
The coefficient for household income from financial assets $\phi^\text{FA}$ is obtained as the slope of the linear regression of household income from financial assets on household wealth in other financial assets. We assume $\sigma^\text{FA}=0$ for simplicity; see Eq. \eqref{eq:hh_income}.

\paragraph{Consumption Smoothing Time Frame}
The time frame for consumption smoothing is assumed to be $T^\text{CO}=12$, see Eq. \eqref{eq:hh_consumption_full}.

\paragraph{Depreciation of Other Real Assets}
We are assuming a depreciation rate of $d^\text{RA}=5\%$, see Eq. \eqref{eq:hh_ora_upt}.

\paragraph{Housing as a Tenant}
We are taking the estimated values for household parameters related to housing as a tenant from \cite{carro2023heterogeneous}. Specifically, when setting the maximum purchasing price in \eqref{eq:mpp}, we assume $\phi^\text{HP}=42.9036$, $\beta^\text{HP}=0.7892$, $\mu^\text{HP}=-0.0177$, and $\sigma^\text{HP}=0.1684$. The psychological pressure of renting is set as $\mu^\text{PS}=0.4$, and the influence of costs in deciding whether to buy or rent is $\phi^\text{B}=0.001$. When setting the desired rental price in \eqref{eq:target_rental_price}, the factor is $\phi^\text{HR}=17.2166$ and the exponent is $\beta^\text{HR}=0.3464$. We are assuming that the average period for which tenants hold their property is $2$ years so that the quarterly probability of not changing housing is given by $p^\text{RS}=\sfrac{7}{8}$.

\paragraph{Housing as an Owner-Occupier}
We are taking the values for household parameters related to housing as an owner-occupier from \cite{carro2023heterogeneous}. Specifically, the probability of quarterly price reduction for properties not yet sold is given by $p^\text{PM}=0.1964$, so that $\mu^\text{PM}=1.4531$ and $\sigma^\text{PM}=0.4889$ in \eqref{eq:price_red}. We are assuming that the average period for which owner-occupiers hold their houses is $20$ years so that the quarterly probability of not changing housing is given by $p^\text{OS}=\sfrac{79}{80}$.

\paragraph{Housing as a Buy-to-Let Investor}
We are taking the values for household parameters related to housing as a buy-to-let investor from \cite{carro2023heterogeneous}. Specifically, the probability of quarterly rent reduction for properties not yet rented out is given by $p^\text{RM}=0.2848$, so that $\mu^\text{RM}=1.6559$ and $\sigma^\text{RM}=0.7855$ in \eqref{eq:rent_red}. For simplicity, we assume that rents are fully indexed with CPI, $\phi^\text{PIR}=1$ with a lag of one quarter, $T^\text{PIR}=1$.

\subsubsection{Rules}
\label{sec:hh_rules}
Households receive income, and based on their income, they choose their target consumption. They operate on the housing market, apply for consumption loans and mortgages, and update their wealth. Households can go bankrupt.

\paragraph{Predicted Income}
The predicted income of household $h$ is the sum of the predicted incomes of its individuals, predicted other social transfers to the household that depend on household properties (household type and wealth), rental income, and predicted income from financial assets,
\begin{equation}
\label{eq:hh_exp_income}
    \overline{Y}_h(t) = \underbrace{\sum_{i\in \mathcal{I}_h(t)} \overline{Y}_i(t)}_\text{Predicted individual incomes} + \underbrace{\overline{P}^\text{CPI}(t)\text{sb}^\text{O}_h(t)}_\text{Predicted social transfers} + \underbrace{\sum_{p\in \mathcal{P}^\text{R}_h(t)} r_p(t)}_\text{Rental income} + \underbrace{\phi^\text{FA} W^\text{OFA}_h(t-1)}_\text{Predicted income from financial assets}
\end{equation}
where $\mathcal{I}_h(t)$ is the set of individuals corresponding to household $h$, $\overline{P}^\text{CPI}(t)$ is the predicted consumer price index, $\text{sb}^\text{O}_h(t)$ are real other social transfers to the household, $\mathcal{P}^\text{R}_h(t)$ is the set of properties the household has rented out, $\phi^\text{FA}$ is a parameter, and $W_h^\text{OFA}(t)$ are other financial assets of the household.

\paragraph{Income}
The income of household $h$ is the sum of the incomes of its individuals, other social transfers to the household that depend on household properties (household type and wealth), rental income, and income from financial assets,
\begin{equation}
\label{eq:hh_income}
    Y_h(t) = \underbrace{\sum_{i\in \mathcal{I}_h(t)} Y_i(t)}_\text{Individual incomes} + \underbrace{P^\text{CPI}(t)\text{sb}^\text{O}_h(t)}_\text{Social transfers} + \underbrace{\sum_{p\in \mathcal{P}^\text{R}_h(t)} r_p(t)}_\text{Rental income} + \underbrace{\left(1+\varepsilon\right) \phi^\text{FA} W^\text{OFA}_h(t-1)}_\text{Income from financial assets}
\end{equation}
where $\mathcal{I}_h(t)$ is the set of individuals corresponding to household $h$, $P^\text{CPI}(t)$ is the consumer price index, $\text{sb}^\text{O}_h(t)$ are other social transfers to the household, $\mathcal{P}^\text{R}_h(t)$ is the set of properties the household has rented out, $\varepsilon\sim N(0,\sigma^\text{FA})$ is noise, $\phi^\text{FA}$ is a parameter, and $W_h^\text{ORA}(t)$ are other financial assets of the household.

\paragraph{Consumption}
Households participate in the goods market. Target consumption of household $h$ is determined given a minimum consumption amount, an average over previous consumption levels, and a fraction of predicted income\footnote{Most generally, the consumption budget is a linear combination of human and financial wealth, e.g. \cite{dawidAgentbasedMacroeconomics2018}. In \cite{dosiSchumpeterMeetingKeynes2010}, the household's consumption budget is precisely its income; in \cite{mandelAgentbasedDynamicsDisaggregated2010}, the consumption budget is a linear function of real money balances, in \cite{lengnickAgentbasedMacroeconomicsBaseline2013} it is an increasing concave function of real money balances, and in \cite{assenzaEmergentDynamicsMacroeconomic2015,ashrafHowInflationAffects2016}, it is a function of past incomes and financial wealth. A more complicated implementation in \cite{seppecherFlexibilityWagesMacroeconomic2012} incorporates consumer sentiment and opinion dynamics.},
\begin{equation}
\begin{aligned}
    \label{eq:hh_consumption_full}
    \hat{C}_{hs}(t) = \frac{c_s^\text{CPI}(t)}{1+\tau^\text{VAT}}\max\Bigg(&\underbrace{\left(1-\phi_h^\text{SR}(t)\right) \overline{P}^\text{CPI}(t) w^\text{U}(t)}_\text{Minimal consumption},\,\underbrace{\left(1-\phi_h^\text{SR}(t)\right) \overline{Y}_h(t)}_\text{Fraction of predicted income},\, \underbrace{\frac{\phi^\text{CO}}{T^\text{CO}}\sum_{t'=1}^{T^\text{CO}} C_h(t-t')}_\text{Consumption smoothing},\Bigg)
\end{aligned}
\end{equation}
where $c_s^\text{CPI}(t)$ are consumption weights, $\phi_h^\text{SR}(t)$ are saving rates of household $h$, $\overline{P}^\text{CPI}(t)$ is the predicted consumer price index, $w^\text{U}(t)$ are unemployment benefits, $\phi^\text{CO}$ and $T^\text{CO}$ determine consumption smoothing, $C_h(t)$ is total consumption, and $\overline{Y}_h(t)$ is predicted income. Saving rates $\phi_h^\text{SR}(t)$ are updated using a linear model estimated at model initialisation using the Household Finance and Consumption Survey (HFCS) data on household income, wealth, and debt.

\paragraph{Investment}
Households also purchase investment goods in the goods market. The target investment of household $h$ is set as
\begin{equation}
    \hat{K}_{hs}(t) = \frac{k_s}{1 + \tau^\text{CF}} \underbrace{\phi_h^\text{IR} \overline{Y}_h(t)}_\text{Fraction of predicted income}
\end{equation}
where $\tau^\text{CF}$ are taxes on capital formation, $\phi_h^\text{IR}$ is the investment rate, and $\overline{Y}_h(t)$ is predicted income of household $h$. 

\paragraph{Housing}
The household may live in social housing, rent, or own its main residence. It may also own additional properties rented out to other households. The household's decisions depend on its current financial situation.

\begin{itemize}
\item \emph{Households in Social Housing}:
Households in social housing first decide on a desired purchase price based on their predicted income, then predict the maximum housing value they can afford with the purchase price, and then compare the costs of buying with the cost of renting a house of the same value. The household is willing to pay up to
\begin{equation}
    \label{eq:mpp}
    P_h(t) = \phi^\text{HP} \overline{Y}_h(t)^{\beta^\text{HP}} \exp\left(\varepsilon\right)
\end{equation}
where $\phi^\text{HP}$ and $\beta^\text{HP}$ are parameters, and $\varepsilon\sim N(\mu^\text{HP},\sigma^\text{HP})$. Households then estimate the value $V^\star$ of housing they can afford by regressing the value of previously sold properties on corresponding prices. They then compare the annual costs of renting a property of value $V^\star$ to the cost of purchasing a property of that value. The annual cost of renting is given by
\begin{equation}
    C^\text{R}_{V^\star}(t) = 4 \left(1 + \mu^\text{PS}\right) \overline{r}_{V^\star}(t)
\end{equation}
where $\mu^\text{PS}$ represents the psychological pressure of having to rent and $\overline{r}_{V^\star}(t)$ is the predicted rent of a property of value $V^\star$, also regressed on previously previous sales. The annual cost of purchasing the house is given by
\begin{equation}
    C^\text{B}_{V^\star}(t) = \underbrace{4\frac{P_h(t) - W_h^\text{FA}(t)}{m_l} +  4\frac{r^\star\left(P_h(t) - W_h^\text{FA}(t)\right)}{1 - \left(1 + r^\star\right) ^{m_l}}}_\text{Annual mortgage repayment} - \underbrace{\left(\left(1+\overline{\gamma}^\text{HPI}(t)\right)^4 - 1\right) V^\star}_\text{Predicted house appreciation/depreciation}
\end{equation}
where $W_h^\text{FA}(t)$ is the financial wealth of household $h$, $m_l$ is the maturity of the potential new mortgage, $r^\star$ is the average mortgage interest rate observed in the previous timestep, and $\overline{\gamma}^\text{HPI}(t)$ is predicted house price index growth. The first term of the annual mortgage repayment above corresponds to paying back the principal, and the second term corresponds to interest. We assume that the mortgage is fixed-rate and that there is no remortgaging.

Then, the probability of buying over renting is given by
\begin{equation}
    p^\text{B}_h(t) = \frac{1}{1 + \exp\left(\phi^\text{B}\left(C^\text{R}_{V^\star}(t) - C^\text{B}_{V^\star}(t)\right)\right)}
\end{equation}
where $\phi^\text{B}$ determines the influence of the difference in predicted costs. If a household buys the property and its financial wealth $W_h^\text{FA}(t)$ is greater than the price of the property $P_h(t)$, it will pay for the property using its existing financial wealth (other financial assets first, then deposits). Otherwise, it will apply for a mortgage. If deciding to rent, the rent the household is willing to pay depends on its income and is given by
\begin{equation}
    \label{eq:target_rental_price}
    r_h(t) \leq \phi^\text{HR} Y_h(t)^{\beta^\text{HR}}
\end{equation}
where $\phi^\text{HR}$ and $\beta^\text{HR}$ are parameters.

\item\emph{Households Currently Renting}:
Households renting attempt to move with a probability $1-p^\text{RS}$. Otherwise, they follow the same decision process as households in social housing above.

\item\emph{Households Currently Owning}:
Households currently owning do not attempt to move with a probability $p^\text{OS}$. Otherwise, based on their income, households currently decide whether to move into a rented property or buy a different property in the same way as presently renting households. If a household finds a property they would like to rent, they will put their current property $p$ on the market at a price of
\begin{equation}
    P_p(t) = \underbrace{\left(1 + \overline{\pi}^\text{HPI}(t) \right)}_\text{Markup} \underbrace{V_p(t),}_\text{Property value}
\end{equation}
where $V_p(t)$ is the value of the property and $\overline{\pi}^\text{HPI}(t)$ is the predicted house price index inflation. Each timestep they do not manage to sell the property, the household will reduce the price with a probability of $p^\text{PM}$ by some random fraction
\begin{equation}
    \label{eq:price_red}
    P_p(t) = \underbrace{\left(1-\exp\varepsilon\right)}_\text{Random reduction} P_p(t-1)
\end{equation}
where $\varepsilon\sim N(\mu^\text{PM},\sigma^\text{PM})$, until they manage to sell it.

If the household decides to buy a new home for themselves, they will put their current home on the market and take out a mortgage to buy a new home with the price as before. They will choose their full wealth in financial assets as a down payment.

\item \emph{Households as Buy-to-Let Investors}: Buy-to-let investors rent-out their vacant properties. A property $p$ of value $V^\star$ is put on the rental market at a rent
\begin{equation}
    r_p(t) = \underbrace{\left(1 + \overline{\pi}^\text{RPI}(t)\right)}_\text{Markup} \underbrace{r_{V^\star}(t)}_\text{Average property rent}
\end{equation}
where $\overline{\pi}^\text{RPI}(t)$ is the predicted RPI inflation, $r_{V^\star}(t)$ is the average rent of a property of value $V^\star$, obtained by regressing the value of previously newly rented-out properties on corresponding rents.

Each quarter, a property remains on the rental market, the offered rent is reduced according to
\begin{equation}
    \label{eq:rent_red}
    r_p(t) = \underbrace{\left(1-\exp \varepsilon\right)}_\text{Random reduction} r_p(t-1)
\end{equation}
with probability $p^\text{RM}$, where $\varepsilon\sim N(\mu^\text{RM},\sigma^\text{RM})$.

If a property $p$ is rented out, the rent set by the owner of the house is partially indexed with lagged CPI inflation,
\begin{equation}
    r_p(t) = \underbrace{\left(1+\phi^\text{PIR} \pi^\text{CPI}(t-T^\text{PIR})\right)}_\text{Lagged partially indexed inflation} r_p(t-1)
\end{equation}
where $\phi^\text{PIR}$ is the partial indexation and $T^\text{PIR}$ is a lag.
\end{itemize}

\paragraph{Demand for Credit}
Households apply for consumption loans to cover liquidity shortages and mortgages for purchasing properties.

If target consumption $\hat{C}_h(t)$ is above disposable income $Y^\text{-r}_h(t)$, the household needs to use a fraction of its wealth to make up for the difference. Households use their financial assets (first other financial assets, then deposits) to accommodate additional consumption. If that is not sufficient, the household applies for a consumption loan to cover the gap 
\begin{equation}
    \hat{L}_h^\text{C}(t) = \left[\underbrace{\hat{C}_h(t)}_\text{Target consumption} - \underbrace{Y^\text{-r}_h(t)}_\text{Disposable income} - \underbrace{W_h^\text{FA}(t-1)}_\text{Financial wealth}\right]^+.
\end{equation}
Households may additionally have decided to buy a property for themselves or for renting it out. If the price of the property $P_p(t)$ is above the household financial wealth minus additional necessary spending for consumption, the household will apply for a mortgage of
\begin{equation}
    \hat{L}_h^\text{M}(t) = \left[\underbrace{P_p(t)}_\text{Property price} - \underbrace{\left[W_h^\text{FA}(t-1) - \left(\hat{C}_h(t) - \overline{Y}^\text{-r}_h(t)\right)\right]^+}_\text{Down-payment}\right]^+
\end{equation}
where we assume that the household's desired down payment is their full wealth and financial assets.

\paragraph{Wealth}
Households\footnote{In \cite{dosiSchumpeterMeetingKeynes2010,assenzaEmergentDynamicsMacroeconomic2015,seppecherFlexibilityWagesMacroeconomic2012,mandelAgentbasedDynamicsDisaggregated2010} households' wealth consists only of deposits, whereas in \cite{lengnickAgentbasedMacroeconomicsBaseline2013}, households' wealth consists of only cash. In \cite{dawidEconomicConvergencePolicy2014}, households hold deposits at banks, a portfolio of shares, and government bonds.} hold wealth in real assets,
\begin{equation}
    W_h^\text{RA}(t) = \underbrace{W_h^\text{P}(t)}_\text{Wealth in properties} + \underbrace{W_h^\text{ORA}(t)}_\text{Wealth in other real assets}
\end{equation}
and in financial assets,
\begin{equation}
    W_h^\text{FA}(t) = \underbrace{W_h^\text{D}(t)}_\text{Wealth in deposits} + \underbrace{W_h^\text{OFA}(t)}_\text{Wealth in other financial assets}
\end{equation}
If the total growth of wealth in the current period $Y_h^\text{-r}(t) - C_h(t)$ is positive, that difference is shared among deposits and other financial assets in fixed fractions depending on current household income, wealth, and debt. If $Y_h^\text{-r}(t) - C_h(t)$ is negative, a portion of financial household wealth is used up, first other financial assets $W^\text{OFA}_h(t)$ and then deposits $W^\text{D}_h(t)$. If that is not sufficient, the household applies for a consumption loan to cover the gap $\hat{L}_h^\text{C}(t)$.

The wealth in properties of the household is updated as
\begin{equation}
    W_h^\text{P}(t) = \sum_{p\in\mathcal{P}_h(t)} \underbrace{V_p(t)}_\text{Value of the property}
\end{equation}
where $\mathcal{P}_h(t)$ is the set of properties owned by the household and $V_p(t)$ is the value of property $p$.
The value of other real assets depreciates
\begin{equation}
\label{eq:hh_ora_upt}
    W_h^\text{ORA}(t) = \underbrace{\left(1-d^\text{RA}\right) W_h^\text{ORA}(t-1)}_\text{Depreciated current wealth} + \underbrace{K_h(t)}_\text{New other real assets}
\end{equation}
where $d^\text{RA}$ is the rate of depreciation.
The deposits of the household are updated as
\begin{equation}
\begin{aligned}
    W_h^\text{D}(t) ={}& \underbrace{W_h^\text{D}(t-1)}_\text{Previous deposits} + \underbrace{\Delta W_h^\text{D}(t)}_\text{New deposits} - \underbrace{\sum_{l\in \mathcal{L}_h(t)} \frac{V_l(t)}{m_l}}_\text{Debt installment} + \underbrace{ L^{\Delta\text{C}}_h(t) + L^{\Delta\text{M}}_h(t)}_\text{Granted loans} \\
    & + \underbrace{r(t) \left[W_h^\text{D}(t-1)\right]^+ - r_b^\text{H-O}(t) \left[W_h^\text{D}(t-1)\right]^-}_\text{Interest on deposits} - \underbrace{\sum_{l\in \mathcal{L}_h(t)} r_l V_l(t)}_\text{Interest paid on debt} \\
    & - \underbrace{\tau^\text{CF} \sum_{h\in\mathcal{H}} K_h(t)}_\text{Taxes on capital formation}
\end{aligned}
\end{equation}
where $\mathcal{L}_h(t)$ is the set of household loans, $L_h^{\Delta\text{C}}(t)$ are newly granted consumption loans, $L_h^{\Delta\text{M}}(t)$ are newly granted mortgages, $r(t)$ is the policy rate, $r_b^\text{H-O}(t)$ is the overdraft rate on deposits, $V_l(t)$ is the amount of loan $l$, $m_l$ is the maturity of loan $l$, $r_l$ is the interest rate of loan $l$, $\tau^\text{CF}$ is the tax rate on capital formation, and $K_h(t)$ is new household investment.

The value of other financial assets is updated as
\begin{equation}
    W_h^\text{OFA}(t) = \underbrace{W_h^\text{OFA}(t-1)}_\text{Wealth in other financial assets} + \underbrace{\Delta W^\text{OFA}_h(t)}_\text{Change in other financial assets}
\end{equation}
and total wealth as
\begin{equation}
\label{eq:hh_wealth}
    W_h(t) = \underbrace{W_h^\text{P}(t) + W_h^\text{ORA}(t)}_\text{Real assets} + \underbrace{W_h^\text{D}(t) + W_h^\text{OFA}(t)}_\text{Financial assets}.
\end{equation}

\paragraph{Debt}
Household debt is set according to
\begin{equation}
\label{eq:hh_debt}
    L_h(t) = \underbrace{L_h(t-1)}_\text{Previous debt} - \underbrace{\sum_{l\in \mathcal{L}_h(t)} \frac{V_l(t)}{m_l}}_\text{Debt installment} + \underbrace{L^{\Delta\text{C}}_h(t) + L^{\Delta\text{M}}_h(t)}_\text{Granted loans}
\end{equation}
where $\mathcal{L}_h(t)$ is the set of current loans the household pays instalments on, $L^{\Delta\text{C}}_h(t)$ are newly granted household consumption loans, and $L^{\Delta\text{M}}_h(t)$ are newly granted mortgages.

\paragraph{Net Wealth}
Household net wealth is set to be
\begin{equation}
    W_h^\text{N}(t) = \underbrace{W_h(t)}_\text{Wealth} - \underbrace{L_h(t)}_\text{Debt}.
\end{equation}

\paragraph{Bankruptcy}
If a household $h$ is insolvent, $W_h^\text{N}(t)<0$ (negative net wealth) and $W^\text{D}_h(t)<0$ (negative deposits), it goes bankrupt. The banks receive all of the households' financial wealth, as well as its other owned properties $\mathcal{P}^\text{R}_h(t)$. The remaining debt and deposit overdrafts are written off. If the household is in the process of paying off a mortgage for its current residence, the bank takes the residence, and the household will seek to rent or buy in the next iteration.

\subsection{Individuals}
Individuals are part of households. They may be employed, unemployed, or not economically active. If employed, they supply firms with labour and receive a wage. If they are unemployed, they receive social benefits from the government.

Table \ref{table:ind_variables} shows variables and parameters related to individuals.

\tablefontsize
\begin{xltabular}{\textwidth}{cXl}
\toprule
Category & Description & Notation \\
\cmidrule(l){1-3}
\multirow{4}{*}{\STAB{\rotatebox[origin=c]{0}{Sets}}} & Set of individuals & $\mathcal{I}$ \\
 & Set of employed individuals & $\mathcal{I}^\text{E}(t)$ \\
 & Set of unemployed individuals & $\mathcal{I}^\text{U}(t)$ \\
 & Set of not-economically-active individuals & $\mathcal{I}^\text{N}$ \\
\cmidrule(l){1-3}
\multirow{5}{*}{\STAB{\rotatebox[origin=c]{0}{Income}}} & Income & $Y_i(t)$ \\
 & Predicted income & $\overline{Y}_i(t)$ \\
 & Wage & $w_i(t)$ \\
 & Reservation wage & $w^\text{RW}_i(t)$ \\
 & Time period for reservation wages & $T^\text{RW}$ \\
\cmidrule(l){1-3}
\multirow{3}{*}{\STAB{\rotatebox[origin=c]{0}{\makecell{Labour\\inputs}}}} & Labour inputs contributed per timestep & $H_i(t)$ \\
 & Increase in the quality of labour for employed individuals & $h^\text{E}$ \\
 & Decrease in the quality of labour for unemployed individuals & $h^\text{U}$ \\
\bottomrule
\\
\caption{Variables and parameters in the model related to individuals.}
\label{table:ind_variables}
\end{xltabular}
\normalsize

\subsubsection{Parameters}
We are running the model on relatively short time scales so that we will assume no change in the quality of labour provided by individuals, $h^\text{U}=h^\text{E}=0$. The time period for reservation wages is set to be $T^\text{RW}=8$, see Eq. \eqref{eq:ind_reservation_wages}.

\subsubsection{Initial Conditions}
This section describes setting initial conditions for individuals.

\paragraph{Drawing Individuals from HFCS Data}
Each of the randomly drawn households in \ref{sec:ic_hhs} is linked to corresponding individuals (Code: \emph{iid}). The data on individuals includes microdata for the fields described in Table \ref{table:hfcs_codes_individuals}.

\begin{xltabular}{\textwidth}{cXl}
\cmidrule(l){1-3}
Category & Description & Code \\
\cmidrule(l){1-3}
\multirow{5}{*}{\STAB{\rotatebox[origin=c]{0}{Attributes}}} & Gender & \emph{RA0200} \\
 & Age & \emph{RA0300} \\
 & Labour status & \emph{PE0100} \\
 & Education level & \emph{PA0200} \\
 & Employment industry & \emph{PE0400} \\
\cmidrule(l){1-3}
\multirow{3}{*}{\STAB{\rotatebox[origin=c]{0}{Income}}} & Employment income & \emph{PG0110} \\
 & Self-employment income & \emph{PG0210} \\
 & Income from unemployment benefits & \emph{PG0510} \\
\cmidrule(l){1-3}
\caption{Household Finance and Consumption Survey codes related to individuals.}
\label{table:hfcs_codes_individuals}
\end{xltabular}
\normalsize

\paragraph{Initial Labour Status}
The initial labour status of an individual is matched to HFCS data. It is adjusted to match the unemployment rate\footnote{World Bank Data: Unemployment rate (Code: \emph{API\_SL.UEM.TOTL.ZS\_DS2}).} and the vacancy rate\footnote{OECD: Infra-annual Registered Unemployment and Job Vacancies (Code: \emph{DSD\_OLAB@DF\_OIALAB\_INDIC}).}.

\paragraph{Initial Employment Industry}
The initial employment industry of an individual is matched to HFCS data. It is adjusted to match industry aggregates\footnote{OECD: Quarterly Employment by Economic Activity (Code: \emph{DSD\_NAMAIN1@DF\_QNA\_BY\_ACTIVITY\_EMPDC}).} if necessary, by changing the employment industry of randomly chosen individuals.

\paragraph{Initial Labour Inputs}
Each individual's initial labour inputs are assumed to be one: for each $i\in\mathcal{H}$,
\begin{equation}
    H_i(0) = 1.
\end{equation}

\paragraph{Initial Wages}
Initial wages $w_i(0)$ of employed individuals are assumed to be the sum of HFCS employment income (Code: \emph{PG0110}) plus HFCS self-employment income (Code: \emph{PG0210}), since self-employment is not specifically modelled. Initial wages of unemployed- or non-economically-active- individuals are assumed to be zero.

\paragraph{Initial Income}
The initial labour income of employed individuals is assumed to be just wages. Initial income of unemployed individuals is proportional to their income from unemployment benefits taken from HFCS data (Code: \emph{PG0510}) and rescaled to match aggregate social benefits\footnote{OECD: Social Expenditure Aggregates (Code: \emph{DSD\_SOCX\_AGG@DF\_SOCX\_AGG}).}. Not-economically active individuals do not directly receive income, but their corresponding household receives social benefits (e.g., for retirement).

\paragraph{Initial Matching with Firms}
Employed individuals are initially matched with firms by solving a linear sum assignment problem so that the sum of the differences between total firm wages and the sum of received wages before taxes by employees is minimal.

\subsubsection{Rules}
Individuals update their labour supply based on their employment status, choose their reservation wage, and update their predicted income and realized income.

\paragraph{Labour Supply}
Every individual supplies labour inputs $H_i(t)$ to the firm it employs. Not economically active individuals can not supply labour. The labour inputs of currently unemployed individuals decrease according to
\begin{equation}
    H_i(t) = \underbrace{\frac{1}{1 + h^\text{U}}}_\text{Markup} H_i(t-1)
\end{equation}
and the labour inputs of currently employed individuals increase according to
\begin{equation}
    H_i(t) = \underbrace{\left(1 + h^\text{E}\right)}_\text{Markup} H_i(t-1)
\end{equation}
where $h^\text{U}$ is a parameter that determines the fall in the quality of labour of unemployed individuals and $h^\text{E}$ is a parameter that determines the rise in the quality of labour of employed individuals.

\paragraph{Reservation Wages}
The reservation wage of an unemployed individual $i$ is set as the mean of previously received wages and at least what they get from unemployment benefits,
\begin{equation}
\label{eq:ind_reservation_wages}
    w_i^\text{R}(t) = \max\left(\underbrace{\overline{P}^\text{CPI}(t) w^\text{U}(t)}_\text{Unemployment benefits},\underbrace{\frac{1}{T^\text{RW}}\sum_{t'=1}^{T^\text{RW}} w_i(t-t')}_\text{Average previous wages}\right)
\end{equation}
where $\overline{P}^\text{CPI}(t)$ is the predicted consumer price index, $w^\text{U}(t)$ are real per-capita unemployment benefits, and $T^\text{RW}$ is the time-span over which an individual is considering its average wage.

\paragraph{Predicted Income}
The predicted income of an employed individual is 
\begin{equation}
\label{eq:exp_ind_income}
    \overline{Y}_i(t) = \overline{P}^\text{CPI}(t) w_i(t)\underbrace{\left(1-\tau^\text{SIW}-\tau^\text{INC}\left(1-\tau^\text{SIW}\right)\right)}_\text{Taxes}
\end{equation}
where $\overline{P}^\text{CPI}(t)$ is the predicted consumer price index, $\tau^\text{SIW}$ are employees' social contributions and $\tau^\text{INC}$ is the income tax rate. Each unemployed individual receives unemployment benefits, $\overline{Y}_i(t)=\overline{P}^\text{CPI}(t) w^\text{U}(t)$, not-economically-active individuals estimate their income as part of the other social benefits received by their household, which depends on the household type (see \ref{sec:hh_rules}).

\paragraph{Income}
The income of an employed individual is 
\begin{equation}
\label{eq:ind_income}
    Y_i(t) = P^\text{CPI}(t) w_i(t) \underbrace{\left(1-\tau^\text{SIW}-\tau^\text{INC}\left(1-\tau^\text{SIW}\right)\right)}_\text{Taxes}
\end{equation}
where $P^\text{CPI}(t)$ is the consumer price index, $\tau^\text{SIW}$ are employees' social contributions and $\tau^\text{INC}$ is the income tax rate. Each unemployed individual receives unemployment benefits, $Y_i(t)=P^\text{CPI}(t) w^\text{U}(t)$, not-economically-active individuals receive income as part of the other social benefits received by their household, which depends on the household type (see \ref{sec:hh_rules}).

\subsection{Rest of the World}
The \emph{rest-of-the-world} models every country we do not specifically simulate. It operates on the global goods market through imports and exports. Corresponding variables are summarised in Table \ref{table:row_variables}.

\tablefontsize
\begin{xltabular}{\textwidth}{cXl}
\toprule
Category & Description & Notation \\
\cmidrule(l){1-3}
\multirow{8}{*}{\STAB{\rotatebox[origin=c]{0}{Trade}}} & Total real exports & $Y^\text{ROW}(t)$ \\
 & Total real target exports & $\hat{Y}^\text{ROW}(t)$ \\
 & Export weights & $c^\text{ROW}_s$ \\
 & Total imports & $C^\text{ROW}(t)$ \\
 & Total target imports & $\hat{C}^\text{ROW}(t)$ \\
 & Import weights & $y^\text{ROW}_s$ \\
 & Adjustment speed\footnote{The economy of the rest of the world is not explicitly modelled and depends on the average growth and inflation of explicitly modelled countries. The parameter $\phi^\text{ROW}$ models the impact of growth and inflation of the modelled countries on the rest-of-the-world agent.} & $\phi^\text{ROW}$ \\
 & Net exports & $\text{NX}^\text{ROW}(t)$ \\
\cmidrule(l){1-3}
\multirow{1}{*}{Prices} & Sectoral price & $P_s^\text{ROW}(t)$  \\
\bottomrule
\\
\caption{Variables in the model related to the rest of the world.}
\label{table:row_variables}
\end{xltabular}
\normalsize

\subsubsection{Parameters}
The export weights $c_s^\text{ROW}$ and the import weights $y_s^\text{ROW}$ are matched to input-output tables\footnote{OECD: Inter-country input-output tables (Code: \emph{ICIO}).}.

Since in this paper, we are only simulating a single country interacting with the rest of the world, the adjustment speed is assumed to be $\phi^\text{ROW}=1.0$.

\subsubsection{Initial Conditions}
Initial conditions of the rest of the world are the initial exports $Y^\text{ROW}(0)$ and imports $C^\text{ROW}(0)$ of every country not explicitly modelled. These are obtained from input-output tables. Initial prices set by the rest of the world are assumed to be $P_s^\text{ROW}(0) = 1$ for all sectors $s$.

\subsubsection{Rules}
The rest of the world determines target imports and exports to the modelled countries and prices.

\paragraph{Prices}
The aggregate price index of all modelled countries is given by
\begin{equation}
    \tilde{P}(t) = \frac{\sum_{f\in \tilde{\mathcal{F}}} P_f(t) (Y_f(t) + S_f(t-1))}{\sum_{f\in \tilde{\mathcal{F}}} \left(Y_f(t)+S_f(t-1)\right)}
\end{equation}
where $\tilde{\mathcal{F}}$ is the set of all firms of all countries, $Y_f(t)$ is output of firm $f$, $S_f(t)$ is the level of inventories of firm $f$, and $P_f(t)$ is price of goods produced by firm $f$. The prices for goods bought from the rest of the world are indexed with the aggregate price index of each modelled country,
\begin{equation}
    P_s^\text{ROW}(t) = \left[\underbrace{\left(1 + \phi^\text{ROW} \left(\tilde{P}(t) - 1\right)\right)}_\text{Price index} \times \underbrace{P_s^\text{ROW}(0)}_\text{Initial prices}\right]^+
\end{equation}
where $\phi^\text{ROW}$ is a parameter.

\paragraph{Exports}
Target exports of the rest of the world are indexed by an aggregate production index
\begin{equation}
    \tilde{Y}(t) = \frac{\sum_{f\in \tilde{\mathcal{F}}} Y_f(t)}{\sum_{f\in \tilde{\mathcal{F}}} Y_f(0)}
\end{equation}
where $\tilde{\mathcal{F}}$ is the set of all firms of all countries. Total real target exports are then set as
\begin{equation}
    \hat{Y}^\text{ROW}(t) = \left[\underbrace{\left(1 + \phi^\text{ROW} \left(\tilde{Y}(t) - 1\right)\right)}_\text{Production index} \times \underbrace{\hat{Y}^\text{ROW}(0)}_\text{Initial exports}\right]^+
\end{equation}
with weights $y^\text{ROW}_s$.

\paragraph{Imports}
Target imports into the rest of the world are indexed with both the aggregate domestic price index and the aggregate production index,
\begin{equation}
    \hat{C}^\text{ROW}(t) = \left[\underbrace{\left(1 + \phi^\text{ROW} \left(\tilde{P}(t) - 1\right)\right)}_\text{Price index} \times \underbrace{\left(1 + \phi^\text{ROW} \left(\tilde{Y}(t) - 1\right)\right)}_\text{Production index} \times \underbrace{\hat{C}^\text{ROW}(0)}_\text{Initial imports}\right]^+
\end{equation}
with weights $c^\text{ROW}_s$.

\paragraph{Accounting}
The cumulative net exports of the rest of the world evolve according to
\begin{equation}
    \text{NX}^\text{ROW}(t) = \underbrace{\text{NX}^\text{ROW}(t-1)}_\text{Previous cumulative NX} + \underbrace{\sum_{s\in\mathcal{S}} P_s^\text{ROW}(t) Y_s^\text{ROW}(t)}_\text{Exports} - \underbrace{C^\text{ROW}(t)}_\text{Imports}.
\end{equation}
where $\mathcal{S}$ is the set of sectors.

\subsection{Goods Market}
\label{sec:goods_market}
Firms of all modelled countries sell goods and set prices. Firms also buy goods for use as intermediate and capital inputs. Households and government entities buy goods. The rest of the world sells goods, sets prices, and buys goods. Our random search-and-matching process\footnote{All models we reviewed choose buyers and sellers one at a time based on offered prices. In \cite{polednaEconomicForecastingAgentbased2020}, a household selects a firm to buy from additionally based on size.} preserves realistic export- and import ratios while also allowing us to prioritise agents, for instance, when they offer better prices.

\subsubsection{Parameters}
We assume $\phi^\text{GM}=2$, which controls the likelihood of choosing a seller based on the prices they offer, see \eqref{eq:gm}. This follows \cite{poledna2023economic}.

\subsubsection{Rules}
This section summarizes the modelling rules for clearing the goods market.

\paragraph{Exchange Rates}
Exchange rates between countries are assumed to be constant.

\paragraph{Clearing}
Supply and demand are allocated among firms, households, government entities, and the rest of the world proportional to their total supply or demand while keeping total flows among those agent types realistic. Individual agents are matched using a random search-and-matching process. Firms are always prioritised as buyers and as sellers if they are larger or offer a better price ranked by
\begin{equation}
\label{eq:gm}
    \underbrace{\frac{\exp(-\phi^\text{GM} P_f(t))}{\sum_{f'\in \mathcal{F}_s(t)} \exp(-\phi^\text{GM} P_{f'}(t))}}_\text{Relative prices} \times \underbrace{\frac{Y_f(t)}{\sum_{f'\in \mathcal{F}_s(t)} Y_{f'}(t)}}_\text{Relative production}
\end{equation}
where $\phi^\text{GM}$ is a parameter that governs the influence of the better price in prioritizing sellers. Equation \eqref{eq:gm} specifies the supply chain formation between agents and countries.

\paragraph{Excess Demand}
If demand is left after matching buyers and sellers, it is distributed among sellers as if the allocation process continues.

\subsection{Labour Market}
\label{sec:labour_market}
The labour market randomly matches firms with potential employees in each country separately. Each firm compares its target labour inputs $\hat{H}_f^\text{t}(t)$ to its current labour inputs $H_f(t-1)$. Firms with higher target production attempt to hire more individuals, and firms with current labour inputs higher than the target fire individuals. All firms fire first, and then the remaining firms hire.

Table \ref{table:lm_parameters} shows parameters related to the labour market.

\tablefontsize
\begin{xltabular}{\textwidth}{cXl}
\toprule
Category & Description & Notation \\
\cmidrule(l){1-3}
\multirow{4}{*}{\STAB{\rotatebox[origin=c]{0}{Clearing}}} & Number of fired individuals & $N^\text{F}_f(t)$ \\
 & Number of hired individuals & $N^\text{H}_f(t)$ \\
 & Firing speed of firms & $\gamma^\text{F}$ \\
 & Hiring speed of firms & $\gamma^\text{H}$ \\
\bottomrule
\\
\caption{Parameters in the model related to the labour market.}
\label{table:lm_parameters}
\end{xltabular}
\normalsize

\subsubsection{Parameters}
We assume $\gamma^\text{F}=\gamma^\text{H}=1$ to simplify the calibration exercise; see Eq. \eqref{eq:lm_firing} and Eq. \eqref{eq:lm_hiring}.

\subsubsection{Rules}
This section summarizes the rules for clearing the labour market.

\paragraph{Firing}
Each firm fires employees at random until firing any other employee brings its current labour inputs below its target labour inputs or until the labour inputs lost due to firing employees exceeds 
\begin{equation}
\label{eq:lm_firing}
    \underbrace{\gamma^\text{F}}_\text{Firing speed} \times \underbrace{\left(H_f(t-1) - \hat{H}_f(t)\right)}_\text{Excess labour supply}
\end{equation}
where $\gamma^\text{F}\in [0, 1]$ is the speed at which can be fired, and $H_f(t-1) - \hat{H}_f(t)$ is the difference between current and target labour inputs.

\paragraph{Hiring}
We iterate over each firm with hiring needs at random. Firms hire individuals seeking jobs in random order until their target labour inputs fall below their given labour inputs or until their labour inputs gained due to new hires exceeds
\begin{equation}
\label{eq:lm_hiring}
    \underbrace{\gamma^\text{H}}_\text{Hiring speed} \times \underbrace{\left(\hat{Y}_f(t) - H_f(t-1)\right)}_\text{Missing labour supply}
\end{equation}
where $\gamma^\text{H}\in [0, 1]$ is the speed at which can be hired and $\hat{H}_f(t) - H_f(t-1)$ is the difference target labour inputs and current labour inputs. Individuals only accept a job offer if the offered wage is at least at the level of their reservation wage.

\subsection{Credit Market}
\label{sec:credit_market}
The credit market randomly matches firms and households looking for loans to banks willing to provide loans. The market for short-term firm loans is cleared first, then long-term firm loans, then household consumption loans, and finally mortgages. In each case, firms/households are drawn at random, select a random subset of $n^\text{LF}$ / $n^\text{LH}$ banks, visit them in order of their offered interest rates, and succeed or fail to obtain loans based on the credit conditions of the corresponding bank.\footnote{There are various implementations of a credit market in the literature. See \cite{seppecherFlexibilityWagesMacroeconomic2012, mandelAgentbasedDynamicsDisaggregated2010} for models without any restrictions on credit demands, \cite{dawidEconomicConvergencePolicy2014,dosiSchumpeterMeetingKeynes2010,polednaEconomicForecastingAgentbased2020} for models with certain regulatory requirements, or \cite{ashrafHowInflationAffects2016} for a model that allows lenders to seize a collateral.}

Table \ref{table:cm_parameters} shows parameters related to the credit market.

\tablefontsize
\begin{xltabular}{\textwidth}{cXl}
\toprule
Category & Description & Notation \\
\cmidrule(l){1-3}
\multirow{2}{*}{\STAB{\rotatebox[origin=c]{0}{Clearing}}} & The maximum number of banks each firm visits & $n^\text{LF}$ \\
 & The maximum number of banks each household visits & $n^\text{LH}$ \\
\bottomrule
\\
\caption{Parameters in the model related to the credit market.}
\label{table:cm_parameters}
\end{xltabular}
\normalsize

\subsection{Housing Market}
\label{sec:housing_market}
The housing market module randomly matches buyers and sellers of properties and households looking to rent with households offering rental properties. There are separate markets for buying and renting properties. The market for buying properties is cleared first.

Table \ref{table:prop_variables} shows variables and parameters related to properties.

\tablefontsize
\begin{xltabular}{\textwidth}{cXl}
\toprule
Category & Description & Notation \\
\cmidrule(l){1-3}
\multirow{6}{*}{\STAB{\rotatebox[origin=c]{0}{Price}}} & Price & $P_p(t)$ \\
 & Predicted annual price of renting & $\overline{P}^\text{R}_p(t)$ \\
 & Predicted annual price of buying & $\overline{P}^\text{B}_p(t)$ \\
 & Property value & $V_p(t)$ \\
 & Rent & $r_p(t)$ \\
 & Predicted rental yield & $\overline{Y}_p(t)$ \\
\bottomrule
\\
\caption{Variables in the model related to properties.}
\label{table:prop_variables}
\end{xltabular}
\normalsize
 
\paragraph{Clearing}
Households looking to buy or rent are drawn at random. They visit the property whose price or rent is closest to what they hope to spend. The purchase of a new property does not go ahead if the household fails to obtain a mortgage if required.

\section{Details on Neural Network Architectures}
\label{sec:appendix_training}
This appendix provides details on the training procedures and the neural architectures for neural posterior estimation, neural ratio estimation, and Bayes factor estimation.

\subsection{Details on Neural Posterior Estimation}
\label{app:training_npe}
We use a Masked Autoregressive Flow (MAF, \citet{papamakarios2017masked}) as a density estimator. The MAF consists of 5 stacked Masked Autoencoders for Distribution Estimation (MADE, \citet{germain2015made}), each with 2 blocks and 50 hidden features. The density estimator is trained using Adam \citep{kingma2014adam} with a $5\times 10^{-4}$ learning rate. We use 90\% of the data for training and the remaining 10\% as a hold-out validation set and cease training after $20$ epochs have passed without improvement in performance on the validation set to avoid overfitting.

\subsection{Details on Neural Ratio Estimation}
\label{app:training_nre}
We use a residual neural network (ResNet, \citet{he2016deep}) with 2 layers of 50 hidden features as a classifier for learning the density ratio. We once again use 90\% of the data for training and reserve the remaining 10\% as a validation set. We optimise with Adam \citep{kingma2014adam} with a learning rate of $5\times 10^{-4}$. To avoid overfitting, we cease training after $20$ successive epochs with no improvement in the validation set performance.

\subsection{Details on Estimating the Bayes Factor}
\label{app:training_bf}
We estimate the Bayes factors using neural networks. We take this neural network to be a 
feedforward neural network with hidden sizes 32, 32, 32, 16, which produces as output a single real value estimating the log of the Bayes factor. Each hidden layer uses ReLU activations. 
As input features, we use the mean of real GDP, GDP deflator, real household consumption, real government consumption, real gross fixed capital formation, resulting in 5 input features. 
We train using AdamW \citep{loshchilov2018decoupled} and a learning rate of $10^{-3}$, using $80\%$ of the data for training and the remaining 20\% as validation. We train for a maximum of 500 epochs, with early stopping if the validation set performance does not improve for $50$ consecutive epochs, to avoid overfitting.

\section{Details on Running the IIASA-Model for all OECD Countries}
\label{sec:appendix_sp_reproduction}
We are running the IIASA model\footnote{See \url{https://github.com/iiasa/abm}.} for all 38 OECD member countries on a scale of 1/10000 for 18 industries (see Table \ref{table:nace2}). The model is calibrated using the OECD datasets\footnote{See \url{https://data-explorer.oecd.org}.} summarized in Table \ref{table:eurostat_oecd_codes} below. The table shows the Eurostat codes\footnote{See \url{https://ec.europa.eu/eurostat/data/database}.} from Table 1 in \cite{poledna2023economic} with the corresponding OECD codes.

\begin{table}[h]
\center
\small
\footnotesize
\begin{xltabular}{\textwidth}{Xll}
\toprule
Description & Eurostat Code & OECD Code \\
\cmidrule(l){1-3}
Population by current activity status & cens\_11an\_r2 & DSD\_LFS@DF\_IALFS\_INDIC \\
Business demography by legal form & bd\_9ac\_l\_form\_r2 & DSD\_SDBSBSC\_ISIC4@DF\_SDBS\_ISIC4 \\
Symmetric input-output table & naio\_10\_cp1700 & ICIO Tables \\
Cross-classification of fixed assets & nama\_10\_nfa\_st & DSD\_NAMAIN10@DF\_TABLE9A \\
Government revenue and main aggregates & gov\_10a\_main & DSD\_NASEC10@DF\_TABLE12\_REV \\
Government expenditure by function & gov\_10a\_exp & DSD\_NASEC10@DF\_TABLE11 \\
Government non-financial accounts & gov\_10q\_ggnfa & DSD\_NASEC1@DF\_QSA\_TRANSACTIONS\_C/D \\
Government debt & gov\_10q\_ggdebt & DSD\_NASEC20@DF\_T7PSD\_Q \\
Financial balance sheets & nasq\_10\_f\_bs & DSD\_NASEC20@DF\_T710R\_Q \\
Non-financial transactions & nasq\_10\_nf\_tr & DSD\_NASEC1@DF\_QSA\_TRANSACTIONS\_C \\
GDP and main components & namq\_10\_gdp & DSD\_NAMAIN1@DF\_QNA \\
Money market interest rates & irt\_st\_q & DSD\_STES@DF\_FINMARK \\
\bottomrule
\end{xltabular}
\vspace{0.25cm}
\caption{Datasets used for calibration of the IIASA model \citep{poledna2023economic} with corresponding Eurostat and OECD codes.}
\label{table:eurostat_oecd_codes}
\end{table}

In the IIASA model, the policy rate is set using a Taylor rule, which is calibrated to Euro area economic growth and inflation. For running non-Euro area countries, we instead calibrate the Taylor rule to the domestic level of economic growth and inflation.

\section{Further Results}
\label{sec:appendix_further_results}

\subsection{Bayes Factors}
\label{sec:appendix_bayes_factors}
Tables \ref{table:bayes_factors1} and \ref{table:bayes_factors2} show the 
classification of the 
Bayes factors are estimated by country and time of initialisation, using the classification scheme from \citet{jeffreys1998theory} (summarised in Table \ref{tab:bayes_factor_classes} below). Overall Bayes factors are computed for each initialisation period using the fact that each country is simulated separately and shares no parameters; thus, the overall Bayes factors are products of the country-wise Bayes factors within each initialisation period.

\begin{table}[H]
\centering
\renewcommand{\arraystretch}{0.93}
\begin{tabularx}{\textwidth}{X|cccc|cccc|cccc}
\toprule
\multirow{2}{*}{Country} & \multicolumn{4}{c}{2013} & \multicolumn{4}{c}{2014} & \multicolumn{4}{c}{2015} \\
 & Q1 & Q2 & Q3 & Q4 & Q1 & Q2 & Q3 & Q4 & Q1 & Q2 & Q3 & Q4 \\
\cmidrule(l){1-13}
AUS & \textcolor{blue}{D} & \textcolor{blue}{D} & \textcolor{blue}{D} & \textcolor{blue}{D} & \textcolor{blue}{D} & \textcolor{blue}{D} & VSt & VSt & \textcolor{blue}{D} & \textcolor{blue}{D} & Su & Su \\
AUT & \textcolor{red}{N} & \textcolor{red}{N} & Su & \textcolor{blue}{D} & VSt & \textcolor{blue}{D} & \textcolor{blue}{D} & \textcolor{blue}{D} & \textcolor{blue}{D} & \textcolor{blue}{D} & \textcolor{blue}{D} & \textcolor{blue}{D} \\
BEL & \textcolor{red}{N} & \textcolor{red}{N} & \textcolor{red}{N} & \textcolor{red}{N} & \textcolor{red}{N} & St & \textcolor{blue}{D} & \textcolor{red}{N} & Su & B & \textcolor{blue}{D} & \textcolor{blue}{D} \\
CAN & \textcolor{red}{N} & \textcolor{red}{N} & \textcolor{blue}{D} & St & \textcolor{red}{N} & \textcolor{red}{N} & \textcolor{blue}{D} & \textcolor{blue}{D} & B & VSt & St & St \\
CHL & \textcolor{red}{N} & \textcolor{red}{N} & \textcolor{red}{N} & \textcolor{red}{N} & \textcolor{red}{N} & \textcolor{red}{N} & \textcolor{red}{N} & \textcolor{red}{N} & \textcolor{red}{N} & \textcolor{red}{N} & \textcolor{red}{N} & \textcolor{red}{N} \\
COL & B & Su & \textcolor{red}{N} & \textcolor{red}{N} & \textcolor{red}{N} & Su & \textcolor{red}{N} & \textcolor{red}{N} & \textcolor{red}{N} & \textcolor{red}{N} & \textcolor{red}{N} & \textcolor{red}{N} \\
CRI & \textcolor{blue}{D} & \textcolor{blue}{D} & St & VSt & \textcolor{blue}{D} & \textcolor{blue}{D} & \textcolor{blue}{D} & \textcolor{blue}{D} & \textcolor{blue}{D} & \textcolor{blue}{D} & \textcolor{blue}{D} & \textcolor{blue}{D} \\
CZE & \textcolor{blue}{D} & St & \textcolor{red}{N} & St & \textcolor{red}{N} & \textcolor{red}{N} & \textcolor{red}{N} & Su & VSt & Su & \textcolor{blue}{D} & \textcolor{blue}{D} \\
DNK & \textcolor{blue}{D} & \textcolor{blue}{D} & \textcolor{red}{N} & \textcolor{blue}{D} & \textcolor{red}{N} & \textcolor{red}{N} & \textcolor{red}{N} & \textcolor{blue}{D} & \textcolor{blue}{D} & \textcolor{blue}{D} & \textcolor{red}{N} & \textcolor{red}{N} \\
EST & \textcolor{red}{N} & VSt & VSt & St & \textcolor{blue}{D} & \textcolor{blue}{D} & \textcolor{red}{N} & \textcolor{red}{N} & \textcolor{red}{N} & \textcolor{red}{N} & \textcolor{red}{N} & \textcolor{blue}{D} \\
FIN & B & \textcolor{red}{N} & \textcolor{blue}{D} & St & \textcolor{blue}{D} & VSt & \textcolor{blue}{D} & \textcolor{blue}{D} & \textcolor{blue}{D} & \textcolor{blue}{D} & \textcolor{blue}{D} & \textcolor{blue}{D} \\
FRA & VSt & VSt & Su & Su & \textcolor{red}{N} & Su & Su & \textcolor{blue}{D} & \textcolor{blue}{D} & \textcolor{blue}{D} & \textcolor{blue}{D} & \textcolor{blue}{D} \\
DEU & St & Su & \textcolor{blue}{D} & VSt & \textcolor{red}{N} & St & B & Su & VSt & \textcolor{blue}{D} & VSt & B \\
GRC & \textcolor{red}{N} & \textcolor{red}{N} & \textcolor{red}{N} & \textcolor{red}{N} & \textcolor{blue}{D} & \textcolor{blue}{D} & \textcolor{blue}{D} & \textcolor{blue}{D} & \textcolor{blue}{D} & \textcolor{blue}{D} & \textcolor{blue}{D} & \textcolor{red}{N} \\
HUN & \textcolor{red}{N} & \textcolor{red}{N} & St & \textcolor{red}{N} & \textcolor{red}{N} & \textcolor{red}{N} & \textcolor{red}{N} & \textcolor{red}{N} & Su & \textcolor{blue}{D} & \textcolor{red}{N} & \textcolor{red}{N} \\
ISL & \textcolor{red}{N} & \textcolor{red}{N} & \textcolor{blue}{D} & B & \textcolor{red}{N} & \textcolor{red}{N} & \textcolor{red}{N} & \textcolor{red}{N} & \textcolor{red}{N} & \textcolor{red}{N} & \textcolor{red}{N} & \textcolor{red}{N} \\
IRL & \textcolor{blue}{D} & \textcolor{blue}{D} & \textcolor{blue}{D} & \textcolor{blue}{D} & \textcolor{blue}{D} & \textcolor{blue}{D} & \textcolor{blue}{D} & \textcolor{blue}{D} & \textcolor{blue}{D} & \textcolor{blue}{D} & \textcolor{blue}{D} & \textcolor{blue}{D} \\
ISR & \textcolor{red}{N} & St & Su & \textcolor{red}{N} & B & \textcolor{red}{N} & St & VSt & \textcolor{blue}{D} & \textcolor{blue}{D} & \textcolor{blue}{D} & \textcolor{red}{N} \\
ITA & Su & \textcolor{red}{N} & St & \textcolor{blue}{D} & \textcolor{red}{N} & St & \textcolor{blue}{D} & \textcolor{blue}{D} & \textcolor{blue}{D} & VSt & Su & \textcolor{blue}{D} \\
JPN & \textcolor{red}{N} & \textcolor{red}{N} & \textcolor{blue}{D} & \textcolor{red}{N} & \textcolor{red}{N} & \textcolor{red}{N} & \textcolor{red}{N} & \textcolor{red}{N} & \textcolor{red}{N} & \textcolor{red}{N} & \textcolor{red}{N} & \textcolor{red}{N} \\
KOR & Su & \textcolor{blue}{D} & \textcolor{blue}{D} & Su & \textcolor{blue}{D} & \textcolor{blue}{D} & \textcolor{blue}{D} & \textcolor{blue}{D} & St & \textcolor{blue}{D} & St & St \\
LVA & \textcolor{red}{N} & \textcolor{blue}{D} & \textcolor{blue}{D} & VSt & \textcolor{red}{N} & \textcolor{blue}{D} & Su & \textcolor{blue}{D} & Su & \textcolor{red}{N} & \textcolor{blue}{D} & \textcolor{blue}{D} \\
LTU & \textcolor{blue}{D} & \textcolor{red}{N} & \textcolor{blue}{D} & \textcolor{red}{N} & \textcolor{red}{N} & \textcolor{blue}{D} & \textcolor{blue}{D} & \textcolor{blue}{D} & \textcolor{blue}{D} & \textcolor{blue}{D} & \textcolor{blue}{D} & \textcolor{blue}{D} \\
LUX & \textcolor{red}{N} & \textcolor{red}{N} & \textcolor{red}{N} & \textcolor{red}{N} & \textcolor{red}{N} & VSt & \textcolor{red}{N} & \textcolor{red}{N} & \textcolor{red}{N} & \textcolor{red}{N} & \textcolor{red}{N} & \textcolor{red}{N} \\
MEX & \textcolor{red}{N} & \textcolor{red}{N} & St & VSt & \textcolor{red}{N} & \textcolor{red}{N} & \textcolor{red}{N} & \textcolor{red}{N} & \textcolor{red}{N} & \textcolor{red}{N} & \textcolor{red}{N} & \textcolor{red}{N} \\
NLD & \textcolor{blue}{D} & \textcolor{blue}{D} & \textcolor{red}{N} & \textcolor{blue}{D} & \textcolor{blue}{D} & \textcolor{blue}{D} & \textcolor{blue}{D} & \textcolor{blue}{D} & \textcolor{blue}{D} & \textcolor{blue}{D} & \textcolor{blue}{D} & \textcolor{blue}{D} \\
NZL & \textcolor{blue}{D} & \textcolor{blue}{D} & St & \textcolor{blue}{D} & \textcolor{red}{N} & \textcolor{blue}{D} & \textcolor{red}{N} & \textcolor{blue}{D} & \textcolor{blue}{D} & \textcolor{blue}{D} & \textcolor{blue}{D} & \textcolor{blue}{D} \\
NOR & \textcolor{red}{N} & \textcolor{red}{N} & \textcolor{red}{N} & \textcolor{red}{N} & \textcolor{red}{N} & \textcolor{red}{N} & \textcolor{red}{N} & \textcolor{red}{N} & \textcolor{red}{N} & \textcolor{red}{N} & \textcolor{red}{N} & \textcolor{red}{N} \\
POL & \textcolor{red}{N} & \textcolor{blue}{D} & \textcolor{red}{N} & \textcolor{red}{N} & \textcolor{blue}{D} & \textcolor{blue}{D} & \textcolor{blue}{D} & \textcolor{blue}{D} & \textcolor{blue}{D} & \textcolor{blue}{D} & \textcolor{blue}{D} & \textcolor{blue}{D} \\
PRT & \textcolor{red}{N} & \textcolor{red}{N} & \textcolor{red}{N} & \textcolor{red}{N} & \textcolor{blue}{D} & \textcolor{blue}{D} & \textcolor{red}{N} & \textcolor{red}{N} & \textcolor{blue}{D} & \textcolor{blue}{D} & \textcolor{blue}{D} & \textcolor{blue}{D} \\
SVK & \textcolor{blue}{D} & \textcolor{blue}{D} & \textcolor{blue}{D} & \textcolor{blue}{D} & \textcolor{blue}{D} & \textcolor{blue}{D} & \textcolor{blue}{D} & \textcolor{blue}{D} & VSt & \textcolor{blue}{D} & \textcolor{red}{N} & \textcolor{red}{N} \\
SVN & \textcolor{red}{N} & \textcolor{red}{N} & \textcolor{red}{N} & \textcolor{red}{N} & \textcolor{blue}{D} & \textcolor{blue}{D} & \textcolor{blue}{D} & \textcolor{red}{N} & \textcolor{blue}{D} & \textcolor{blue}{D} & St & \textcolor{blue}{D} \\
ESP & Su & \textcolor{blue}{D} & \textcolor{blue}{D} & \textcolor{blue}{D} & \textcolor{blue}{D} & Su & \textcolor{blue}{D} & \textcolor{blue}{D} & St & VSt & St & VSt \\
SWE & \textcolor{blue}{D} & \textcolor{blue}{D} & \textcolor{blue}{D} & \textcolor{blue}{D} & \textcolor{blue}{D} & St & \textcolor{blue}{D} & \textcolor{blue}{D} & \textcolor{blue}{D} & \textcolor{blue}{D} & \textcolor{blue}{D} & B \\
CHE & \textcolor{blue}{D} & \textcolor{blue}{D} & \textcolor{blue}{D} & \textcolor{blue}{D} & B & \textcolor{blue}{D} & \textcolor{blue}{D} & \textcolor{blue}{D} & \textcolor{blue}{D} & \textcolor{blue}{D} & \textcolor{blue}{D} & Su \\
TUR & \textcolor{red}{N} & \textcolor{red}{N} & \textcolor{red}{N} & \textcolor{red}{N} & \textcolor{red}{N} & \textcolor{red}{N} & \textcolor{red}{N} & B & \textcolor{red}{N} & \textcolor{red}{N} & \textcolor{red}{N} & \textcolor{blue}{D} \\
GBR & \textcolor{blue}{D} & \textcolor{blue}{D} & \textcolor{blue}{D} & \textcolor{blue}{D} & \textcolor{blue}{D} & VSt & \textcolor{blue}{D} & \textcolor{blue}{D} & \textcolor{blue}{D} & B & Su & St \\
USA & \textcolor{red}{N} & B & \textcolor{blue}{D} & \textcolor{red}{N} & \textcolor{blue}{D} & \textcolor{blue}{D} & \textcolor{blue}{D} & \textcolor{blue}{D} & St & \textcolor{blue}{D} & \textcolor{blue}{D} & St \\
\cmidrule(l){1-13}
Overall & \textcolor{red}{N} & \textcolor{blue}{D} & \textcolor{red}{N} & \textcolor{red}{N} & \textcolor{blue}{D} & \textcolor{blue}{D} & \textcolor{blue}{D} & \textcolor{blue}{D} & \textcolor{blue}{D} & \textcolor{blue}{D} & \textcolor{blue}{D} & \textcolor{blue}{D}\\
\bottomrule
\end{tabularx}
\vspace{0.05cm}
\caption{Estimated Bayes factors by the time of initialisation and country, from 2013 to 2015.}
\label{table:bayes_factors1}
\end{table}

\begin{table}[H]
\centering
\renewcommand{\arraystretch}{0.93}
\begin{tabularx}{\textwidth}{X|cccc|cccc}
\toprule
\multirow{2}{*}{Country} & \multicolumn{4}{c}{2016} & \multicolumn{4}{c}{2017} \\
 & Q1 & Q2 & Q3 & Q4 & Q1 & Q2 & Q3 & Q4 \\
\cmidrule(l){1-9}
AUS & B & St & \textcolor{red}{N} & VSt & VSt & VSt & \textcolor{blue}{D} & \textcolor{blue}{D} \\
AUT & \textcolor{blue}{D} & \textcolor{blue}{D} & \textcolor{blue}{D} & \textcolor{blue}{D} & \textcolor{blue}{D} & \textcolor{red}{N} & \textcolor{red}{N} & \textcolor{red}{N} \\
BEL & \textcolor{blue}{D} & \textcolor{blue}{D} & \textcolor{red}{N} & \textcolor{blue}{D} & VSt & \textcolor{blue}{D} & \textcolor{blue}{D} & \textcolor{blue}{D} \\
CAN & St & St & VSt & Su & B & \textcolor{blue}{D} & \textcolor{red}{N} & \textcolor{blue}{D} \\
CHL & \textcolor{red}{N} & Su & \textcolor{blue}{D} & B & \textcolor{blue}{D} & VSt & \textcolor{blue}{D} & \textcolor{blue}{D} \\
COL & \textcolor{red}{N} & \textcolor{red}{N} & \textcolor{red}{N} & \textcolor{blue}{D} & \textcolor{red}{N} & \textcolor{red}{N} & \textcolor{blue}{D} & \textcolor{red}{N} \\
CRI & St & \textcolor{red}{N} & \textcolor{red}{N} & \textcolor{red}{N} & \textcolor{red}{N} & \textcolor{red}{N} & \textcolor{red}{N} & \textcolor{red}{N} \\
CZE & \textcolor{blue}{D} & \textcolor{blue}{D} & \textcolor{blue}{D} & \textcolor{blue}{D} & \textcolor{blue}{D} & \textcolor{blue}{D} & \textcolor{blue}{D} & \textcolor{blue}{D} \\
DNK & \textcolor{red}{N} & \textcolor{red}{N} & \textcolor{red}{N} & \textcolor{red}{N} & \textcolor{red}{N} & \textcolor{red}{N} & \textcolor{red}{N} & \textcolor{red}{N} \\
EST & \textcolor{red}{N} & \textcolor{blue}{D} & \textcolor{blue}{D} & \textcolor{red}{N} & St & \textcolor{red}{N} & \textcolor{red}{N} & \textcolor{blue}{D} \\
FIN & \textcolor{blue}{D} & St & \textcolor{red}{N} & \textcolor{red}{N} & \textcolor{blue}{D} & \textcolor{red}{N} & \textcolor{red}{N} & \textcolor{red}{N} \\
FRA & \textcolor{blue}{D} & \textcolor{blue}{D} & \textcolor{blue}{D} & \textcolor{blue}{D} & St & \textcolor{blue}{D} & \textcolor{blue}{D} & \textcolor{blue}{D} \\
DEU & \textcolor{red}{N} & \textcolor{red}{N} & \textcolor{red}{N} & \textcolor{red}{N} & \textcolor{red}{N} & \textcolor{blue}{D} & \textcolor{red}{N} & \textcolor{red}{N} \\
GRC & \textcolor{blue}{D} & \textcolor{blue}{D} & \textcolor{blue}{D} & \textcolor{blue}{D} & \textcolor{blue}{D} & \textcolor{blue}{D} & \textcolor{blue}{D} & Su \\
HUN & \textcolor{blue}{D} & \textcolor{blue}{D} & \textcolor{blue}{D} & \textcolor{blue}{D} & \textcolor{blue}{D} & \textcolor{red}{N} & \textcolor{red}{N} & B \\
ISL & \textcolor{red}{N} & \textcolor{red}{N} & \textcolor{red}{N} & \textcolor{red}{N} & \textcolor{red}{N} & \textcolor{red}{N} & \textcolor{red}{N} & \textcolor{red}{N} \\
IRL & \textcolor{blue}{D} & \textcolor{blue}{D} & \textcolor{blue}{D} & \textcolor{blue}{D} & \textcolor{blue}{D} & \textcolor{red}{N} & \textcolor{blue}{D} & \textcolor{blue}{D} \\
ISR & St & St & \textcolor{red}{N} & VSt & Su & VSt & \textcolor{red}{N} & \textcolor{blue}{D} \\
ITA & B & \textcolor{blue}{D} & \textcolor{red}{N} & B & \textcolor{red}{N} & B & \textcolor{red}{N} & \textcolor{blue}{D} \\
JPN & \textcolor{red}{N} & \textcolor{red}{N} & \textcolor{red}{N} & \textcolor{red}{N} & \textcolor{red}{N} & B & \textcolor{red}{N} & \textcolor{red}{N} \\
KOR & B & St & St & \textcolor{blue}{D} & \textcolor{blue}{D} & B & VSt & St \\
LVA & \textcolor{blue}{D} & \textcolor{blue}{D} & \textcolor{blue}{D} & \textcolor{blue}{D} & \textcolor{blue}{D} & \textcolor{red}{N} & \textcolor{red}{N} & \textcolor{red}{N} \\
LTU & \textcolor{blue}{D} & \textcolor{red}{N} & \textcolor{blue}{D} & \textcolor{blue}{D} & \textcolor{blue}{D} & St & \textcolor{blue}{D} & \textcolor{blue}{D} \\
LUX & \textcolor{red}{N} & \textcolor{red}{N} & VSt & \textcolor{red}{N} & \textcolor{red}{N} & \textcolor{red}{N} & \textcolor{blue}{D} & \textcolor{red}{N} \\
MEX & \textcolor{red}{N} & \textcolor{red}{N} & \textcolor{red}{N} & \textcolor{red}{N} & \textcolor{red}{N} & \textcolor{red}{N} & \textcolor{blue}{D} & \textcolor{blue}{D} \\
NLD & Su & St & \textcolor{red}{N} & VSt & \textcolor{blue}{D} & \textcolor{red}{N} & \textcolor{red}{N} & VSt \\
NZL & \textcolor{blue}{D} & \textcolor{blue}{D} & \textcolor{blue}{D} & \textcolor{blue}{D} & \textcolor{blue}{D} & \textcolor{blue}{D} & \textcolor{red}{N} & \textcolor{blue}{D} \\
NOR & \textcolor{red}{N} & \textcolor{red}{N} & \textcolor{red}{N} & \textcolor{red}{N} & \textcolor{red}{N} & \textcolor{red}{N} & \textcolor{red}{N} & \textcolor{red}{N} \\
POL & VSt & \textcolor{blue}{D} & \textcolor{blue}{D} & \textcolor{blue}{D} & \textcolor{blue}{D} & \textcolor{blue}{D} & \textcolor{blue}{D} & St \\
PRT & \textcolor{blue}{D} & \textcolor{blue}{D} & \textcolor{blue}{D} & \textcolor{blue}{D} & \textcolor{blue}{D} & \textcolor{blue}{D} & \textcolor{blue}{D} & \textcolor{red}{N} \\
SVK & VSt & VSt & \textcolor{blue}{D} & St & \textcolor{red}{N} & \textcolor{red}{N} & \textcolor{red}{N} & \textcolor{red}{N} \\
SVN & \textcolor{blue}{D} & \textcolor{blue}{D} & \textcolor{blue}{D} & \textcolor{blue}{D} & \textcolor{blue}{D} & \textcolor{red}{N} & \textcolor{blue}{D} & \textcolor{red}{N} \\
ESP & \textcolor{blue}{D} & \textcolor{blue}{D} & \textcolor{red}{N} & \textcolor{blue}{D} & \textcolor{red}{N} & \textcolor{blue}{D} & \textcolor{blue}{D} & \textcolor{blue}{D} \\
SWE & \textcolor{red}{N} & \textcolor{red}{N} & \textcolor{red}{N} & \textcolor{blue}{D} & \textcolor{red}{N} & \textcolor{red}{N} & \textcolor{red}{N} & \textcolor{red}{N} \\
CHE & \textcolor{blue}{D} & \textcolor{blue}{D} & \textcolor{blue}{D} & \textcolor{blue}{D} & VSt & \textcolor{red}{N} & VSt & \textcolor{red}{N} \\
TUR & St & \textcolor{red}{N} & VSt & \textcolor{red}{N} & \textcolor{red}{N} & \textcolor{blue}{D} & \textcolor{red}{N} & \textcolor{blue}{D} \\
GBR & \textcolor{blue}{D} & St & St & Su & B & \textcolor{blue}{D} & \textcolor{blue}{D} & \textcolor{blue}{D} \\
USA & \textcolor{red}{N} & \textcolor{blue}{D} & Su & \textcolor{red}{N} & \textcolor{blue}{D} & \textcolor{blue}{D} & \textcolor{blue}{D} & VSt \\
\cmidrule(l){1-9}
Overall & \textcolor{blue}{D} & \textcolor{blue}{D} & \textcolor{blue}{D} & \textcolor{blue}{D} & \textcolor{blue}{D} & \textcolor{blue}{D} & \textcolor{blue}{D} & \textcolor{blue}{D}\\
\bottomrule
\end{tabularx}
\vspace{0.05cm}
\caption{Estimated Bayes factors by the time of initialisation and country, from 2016 to 2017.}
\label{table:bayes_factors2}
\end{table}
\normalsize

\begin{table}[h]
    \centering
    \begin{tabular}{c|c|c}
    \toprule
       \textbf{Range} & \textbf{Interpretation} & \textbf{Count}\\\midrule
         $[0,1)$ & Negative (\textcolor{red}{N}) & 282\\
         $[1, 10^{1/2})$ & Barely Worth Mentioning (B) & 24\\
         $[10^{1/2}, 10)$ & Substantial (Su) & 33\\
         $[10, 10^{3/2})$ & Strong (St) & 47\\
         $[10^{3/2}, 10^2)$ & Very Strong (VSt) & 42\\
         $> 10^2$ & Decisive (\textcolor{blue}{D}) & 332\\
         \bottomrule
    \end{tabular}
    \vspace{0.05cm}
    \caption{Classification of Bayes factor estimates as reported in \citet{jeffreys1998theory}, and counts of each Bayes factor class in Tables \ref{table:bayes_factors1} and \ref{table:bayes_factors2}.}
    \label{tab:bayes_factor_classes}
\end{table}

\clearpage

\subsection{Country-Level Forecasting Performance}
\label{sec:appendix_economic_forecasting_country}
Table \ref{table:forecasting_gdp_by_country} shows the median relative improvement in forecasting performance of nominal gross domestic product on a country-level of our model over the IIASA model \cite{poledna2023economic}. The values on the left of the '/' correspond to model calibration using \npe{}, and those on the right correspond to \nre{}. We highlight country/horizon combinations for which the IIASA model performs better than our model in red.

\begin{table}[H]
\centering
\renewcommand{\arraystretch}{0.92}
\begin{tabularx}{\textwidth}{X|cccccc}
\toprule
\multirow{2}{*}{Country} & \multicolumn{6}{c}{Horizon} \\
 & 1 Quarter & 2 Quarters & 3 Quarters & 1 Year & 2 Years & 3 Years \\
\cmidrule(l){1-7}
AUS & 154 / 113 & 103 / 57 & 80 / 76 & 41 / 92 & 30 / 88 & 39 / 111 \\
AUT & 30 / 38 & 37 / 51 & 31 / 40 & 36 / 44 & 46 / 46 & 56 / 44 \\
BEL & 25 / 60 & 28 / 35 & 34 / 42 & 41 / 50 & 37 / 39 & 38 / 47 \\
CAN & 93 / 107 & 38 / 38 & 55 / 51 & 43 / 50 & 66 / 70 & 77 / 101 \\
CHE & 82 / 89 & 79 / 85 & 62 / 60 & 96 / 93 & 102 / 102 & 82 / 83 \\
CHL & 59 / 38 & 94 / 79 & 76 / 71 & 87 / 70 & 88 / 74 & 84 / 79 \\
COL & 101 / 104 & 130 / 101 & 103 / 93 & 90 / 91 & 75 / 83 & 91 / 83 \\
CRI & 108 / 211 & 150 / 206 & 179 / 166 & 174 / 163 & 141 / 143 & 152 / 132 \\
CZE & 1 / \textcolor{red}{-22} & \textcolor{red}{-14} / \textcolor{red}{-8} & 4 / 0 & 8 / 3 & 19 / 8 & 32 / 6 \\
DEU & 29 / 56 & \textcolor{red}{-21} / \textcolor{red}{-5} & \textcolor{red}{-7} / 2 & \textcolor{red}{-10} / \textcolor{red}{-8} & 21 / \textcolor{red}{-16} & 20 / \textcolor{red}{-7} \\
DNK & 107 / 109 & 104 / 109 & 106 / 103 & 103 / 103 & 85 / 97 & 73 / 77 \\
ESP & 40 / 39 & \textcolor{red}{-29} / 20 & \textcolor{red}{-36} / 30 & \textcolor{red}{-32} / 47 & 0 / 77 & 15 / 92 \\
EST & 77 / 78 & 56 / 56 & 73 / 74 & 78 / 80 & 89 / 90 & 99 / 98 \\
FIN & 98 / 111 & 49 / 50 & 51 / 47 & 72 / 70 & 94 / 97 & 92 / 92 \\
FRA & \textcolor{red}{-3} / \textcolor{red}{-19} & \textcolor{red}{-38} / 2 & \textcolor{red}{-33} / \textcolor{red}{-14} & \textcolor{red}{-35} / \textcolor{red}{-28} & \textcolor{red}{-34} / \textcolor{red}{-48} & \textcolor{red}{-27} / \textcolor{red}{-44} \\
GBR & 200 / 174 & 135 / 70 & 99 / 62 & 44 / 66 & 72 / 55 & 73 / 66 \\
GRC & 0 / 0 & 36 / 53 & 48 / 77 & 58 / 60 & 71 / 97 & 65 / 83 \\
HUN & 53 / 25 & 37 / 29 & 33 / 20 & 47 / 33 & 47 / 35 & 66 / 43 \\
IRL & 2 / 3 & 12 / 13 & 29 / 28 & 39 / 40 & 25 / 25 & 25 / 26 \\
ISL & 82 / 85 & 42 / 39 & 32 / 40 & 56 / 57 & 62 / 60 & 67 / 66 \\
ISR & 117 / 112 & 94 / 91 & 84 / 82 & 88 / 87 & 98 / 98 & 87 / 87 \\
ITA & 13 / 15 & \textcolor{red}{-87} / \textcolor{red}{-78} & \textcolor{red}{-85} / \textcolor{red}{-74} & \textcolor{red}{-84} / \textcolor{red}{-72} & \textcolor{red}{-77} / \textcolor{red}{-58} & \textcolor{red}{-73} / \textcolor{red}{-51} \\
JPN & 66 / 87 & 32 / \textcolor{red}{-1} & 18 / \textcolor{red}{-7} & 0 / \textcolor{red}{-26} & \textcolor{red}{-14} / \textcolor{red}{-90} & \textcolor{red}{-88} / \textcolor{red}{-92} \\
KOR & 88 / 110 & 83 / 143 & 74 / 161 & 56 / 162 & 99 / 230 & 149 / 225 \\
LTU & 188 / 188 & 217 / 198 & 216 / 208 & 176 / 204 & 149 / 146 & 152 / 130 \\
LUX & 164 / 149 & 197 / 181 & 163 / 172 & 155 / 150 & 198 / 197 & 206 / 213 \\
LVA & 184 / 185 & 205 / 259 & 215 / 229 & 235 / 259 & 269 / 249 & 277 / 263 \\
MEX & 24 / 24 & 81 / 79 & 85 / 78 & 106 / 95 & 112 / 103 & 111 / 109 \\
NLD & 4 / 3 & 13 / 13 & 32 / 32 & 35 / 35 & 38 / 38 & 36 / 36 \\
NOR & 47 / 48 & 41 / 41 & 46 / 46 & 52 / 53 & 37 / 39 & 27 / 28 \\
NZL & 2 / 7 & 21 / 22 & 32 / 37 & 54 / 58 & 91 / 95 & 78 / 79 \\
POL & 189 / 35 & 182 / 0 & 157 / 9 & 163 / 24 & 135 / 50 & 110 / 72 \\
PRT & \textcolor{red}{-9} / \textcolor{red}{-18} & \textcolor{red}{-14} / \textcolor{red}{-9} & \textcolor{red}{-8} / \textcolor{red}{-5} & 10 / 13 & 28 / 22 & 29 / 21 \\
SVK & 116 / 149 & 136 / 173 & 134 / 149 & 132 / 162 & 140 / 141 & 164 / 188 \\
SVN & 32 / 2 & 47 / 6 & 30 / 0 & 63 / 36 & 65 / 59 & 52 / 21 \\
SWE & 14 / 68 & 13 / 67 & 27 / 57 & 29 / 44 & 24 / 35 & 32 / 44 \\
TUR & 88 / 87 & 75 / 75 & 39 / 42 & \textcolor{red}{-17} / 21 & 3 / 13 & 9 / 19 \\
USA & 75 / 84 & 146 / 159 & 136 / 172 & 121 / 129 & 106 / 122 & 103 / 109 \\
\bottomrule
\end{tabularx}
\vspace{0.5cm}
\caption{Median relative improvements in RMSE of our model's growth rates of nominal GDP over the IIASA model \cite{poledna2023economic}. The values on the left of the '/' correspond to model calibration using \npe{}, and those on the right correspond to \nre{}. In both cases, the model was run for 20 initialisation times between 2013-Q1 and 2017-Q4.}
\label{table:forecasting_gdp_by_country}
\end{table}
\normalsize

\subsection{The Statistical Significance of our Improvements in Performance}\label{app:stat_sig}

In this section, we detail our calculations for assessing the statistical significance of the improvements in model performance over the baseline IIASA model. 

Let $n$ be the number of countries on which we test the performance of our model and the IIASA baseline. Let $X_i$, $i=1,\ldots,n$ be an indicator variable which takes value $1$ if our model outperforms the IIASA model according to some metric (e.g., Bayes factor, RMSE forecasting error etc.) on country $i$, and $0$ otherwise. We assume that the $X_i$ are independent and identically distributed Bernoulli random variables, each with common success probability $p$. Then, the number of successes
\begin{equation}
    T = \sum_{i=1}^n X_i \sim \text{Bin}(\cdot \mid n, p) =: g_{n,p}(\cdot),
\end{equation}
where $\text{Bin}(\cdot \mid n, p)$ denotes the Binomial distribution with $n$ trials and success probability $p$. 
We consider the following composite hypothesis test, in which we test
\begin{equation}
    H_0 : p \in \Theta_0 = [0,1/2]\quad \text{vs.}\quad H_1 : p \in \Theta_1 = (1/2, 1],
\end{equation}
with $\Theta := \Theta_0 \cup \Theta_1$. The null hypothesis $H_0$ corresponds to the case in which the IIASA model outperforms our model, while the alternative corresponds to our model outperforming the IIASA baseline. 

A common test statistic for composite null-composite alternative hypothesis tests is the generalised likelihood ratio,
\begin{equation}
    \lgrl{T} = -2\log\frac{\sup_{p\in\Theta_0}g_{n,p}(T)}{\sup_{p\in\Theta} g_{n,p}(T)}.
\end{equation}
The maximum likelihood estimate $\hat{p}_{\rm MLE}$ for $p$ is
\begin{equation}
    \hat{p}_{\rm MLE} = \arg \sup_{p\in\Theta} g_{n,p}(T) = \frac{T}{n}.
\end{equation}
Thus, we have that
\begin{equation}
    \lgrl{T} = -2\log\frac{\sup_{p\in\Theta_0}g_{n,p}(T)}{g_{n,\frac{T}{n}}(T)}.
\end{equation}
From the formula for the Binomial distribution, we have that
\begin{equation}
    \arg\sup_{p\in\Theta_0}g_{n,p}(T) = \begin{cases}
        T/n & \text{if}\quad T\leq n/2\\
        1/2& \text{if}\quad T> n/2,
    \end{cases}
\end{equation}
(since $p \mapsto \log g_{n,p}(T)$ is strictly concave in $p$ on $[0,1]$) which gives
\begin{equation}
    \lgrl{T} 
    = 
    \begin{cases}
        0 & \text{if}\quad T \leq n/2\\
        2\left(n\log{2} + T\log{\frac{T}{n}} + (n-T)\log\left(1 - \frac{T}{n}\right)\right) & \text{if}\quad T > n/2.
    \end{cases}
\end{equation}
Further, for a given $p$, we have that
\begin{equation}
    \lgrl{T} \mid p
    = 
    \begin{cases}
        0 & \text{with probability}\quad \sum_{k=1}^{n/2} g_{n,p}(k)\\
        2\left(n\log{2} + k\log{\frac{k}{n}} + (n-k)\log\left(1 - \frac{k}{n}\right)\right) & \text{with probability}\quad g_{n,p}(k),\ k > n/2.
    \end{cases}
\end{equation}
For this test to have a significance level of $\alpha$, we seek a $\tilde{\lambda}$ such that
\begin{equation}
    \sup_{p\in\Theta_0} \mathbb{P}\left(\lgrl{T} \geq \tilde{\lambda} \mid p\right) \leq \alpha.
\end{equation}
(That is, we seek the value $\tilde{\lambda} \in (0, \infty)$ such that values for the test statistic that are more extreme than $\tilde{\lambda}$ occur no more than $(100\alpha)\%$ of the time under the null hypothesis.) Since $\lgrl{k}$ is non-decreasing in $k$, the probability with which $\lgrl{k} \geq \tilde{\lambda}$ is the probability with which $k \geq \tilde{k}$ for some $\tilde{k}$ corresponding to $\tilde{\lambda}$. It is therefore sufficient to find $\tilde{k}$ such that
\begin{equation}
    \sup_{p \in \Theta_0} \mathbb{P}(k \geq \tilde{k} \mid p) \leq \alpha
\end{equation}
for our chosen significance level $\alpha$. We also have that $\mathbb{P}(k \geq \tilde{k} \mid \frac{1}{2}) > \mathbb{P}(k \geq \tilde{k} \mid p)$ for any $0 \leq p < 1/2$, so in particular it is sufficient to find a $\tilde{k}$ such that
\begin{equation}
    \mathbb{P}\left(k \geq \tilde{k}\ \bigg\vert\ \frac{1}{2}\right) \leq \alpha.
\end{equation}
For $n = 38$ countries, we can easily obtain numerically the fact that the lowest such $\tilde{k}$ is $\tilde{k} = 25$, giving
\begin{equation}
    \mathbb{P}\left(k \geq 25\ \bigg\vert\ \frac{1}{2}\right) \approx 0.036 \leq 0.05.
\end{equation}
Thus, a significance level of $\alpha = 0.05$ corresponds to succeeding in outperforming a baseline model in 25 of the 38 countries, and a significance level of $\alpha = 0.005$ corresponds to 28 out of 38 countries. Similarly, treating all 760 entries of Tables \ref{table:bayes_factors1} \& \ref{table:bayes_factors2} and all 228 entries of Table \ref{table:forecasting_gdp_by_country} as independent, we require that our model outperforms the baseline on, respectively, 402 and 126 for a significance level of $0.05$, or 415 and 133 for a significance level of $0.005$.

From the results presented in both 
\ref{sec:appendix_bayes_factors} and 
Table \ref{table:forecasting_gdp_by_country} of \ref{sec:appendix_economic_forecasting_country}, we see that our improvement in predictive performance is statistically significant at the $0.005$ significance level under this hypothesis test across all prediction horizons when the model is calibrated with either \npe{} or \nre{}. In particular, for the Bayes factor computations, it is noteworthy that our chosen uniform prior is not entirely suitable, given that a uniform prior places equal weight on regions of the parameter space that produce economic collapse in our model, which we reasonably believe to be implausible behaviour. It is, therefore, likely that adjusting the prior place greater weight on more sensible economic outcomes than economic collapse will result in even more favourable Bayes factors for our model.

\end{document}